\documentclass[12pt,preprint]{aastex}  % for e-submission to ApJ - one column

\usepackage{graphicx}

\def\Ms{{\rm M_\odot}}
\def\h2{H$_2$}

\begin{document}

\title{Population III star formation in a $\Lambda$CDM universe, \
II:  Effects of a photodissociating background}

\author{Brian W. O'Shea\altaffilmark{1} \& Michael L. Norman\altaffilmark{2}}

\altaffiltext{1}{Theoretical Astrophysics (T-6), MS B227, Los Alamos National
Laboratory, Los Alamos, NM 87545; bwoshea@lanl.gov}

\altaffiltext{2}{Center for Astrophysics and Space Sciences,
University of California at San Diego, La Jolla, CA 
92093; mlnorman@ucsd.edu}

%%%%%%%%%%%%%%%%%%%%%%%%%%%%%%%%%%%%%%%%%%%%%%%%%%%%%%%%%%%%%%%%%%%%%%%%%%%%
\begin{abstract}
We examine aspects of primordial star formation in the presence of a molecular 
hydrogen-dissociating ultraviolet background. We compare a set of AMR hydrodynamic 
cosmological simulations using a single cosmological realization but with a range 
of ultraviolet background strengths in the Lyman-Werner band. This allows us to 
study the effects of Lyman-Werner radiation on suppressing \h2 cooling at low 
densities as well as the high-density evolution of the collapsing cloud core in a 
self-consistent cosmological framework. We find that the addition of a 
photodissociating background results in a delay of the collapse of high density 
gas at the center of the most massive halo in the simulation and, as a result, 
an increase in the virial mass of this halo at the onset of baryon collapse.  
We find that, contrary to previous results, Population III star formation is 
not suppressed for J$_{21} \geq 0.1$, but occurs even with backgrounds as high 
as J$_{21} = 1$.  We find that \h2 cooling leads to collapse despite the 
depressed core molecular hydrogen fractions due to the elevated \h2 cooling 
rates at $T=2-5 \times 10^3$ K.  We observe a relationship between the strength 
of the photodissociating background and the rate of accretion onto the evolving 
protostellar cloud core, with higher LW background fluxes resulting in higher 
accretion rates.  Finally, we find that the collapsing cloud cores in our 
simulations do not fragment at densities below $n \sim 10^{10}$~cm$^{-3}$ 
regardless of the strength of the LW background, suggesting that Population 
III stars forming in halos with T$_{vir} \sim 10^4$~K may still form in isolation.
\end{abstract}


%%%%%%%%%%%%%%%%%%%%%%%%%%%%%%%%%%%%%%%%%%%%%%%%%%%%%%%%%%%%%%%%%%%%%%%%%%%%
\section{Introduction}\label{intro}

The study of the formation of Population III stars in a cosmological context 
via high-resolution simulations is becoming a mature discipline, with multiple 
groups finding similar results
\markcite{2002ApJ...564...23B, 2004NewA....9..353B, ABN02, oshea07a, 2006ApJ...652....6Y, 
2006astro.ph.10174G}({Bromm}, {Coppi}, \&  {Larson} 2002; {Bromm} \& {Loeb} 2004; {Abel}, {Bryan}, \& {Norman} 2002; {O'Shea} \& {Norman} 2007; {Yoshida} {et~al.} 2006b; {Gao} {et~al.} 2006). However, all of these calculations operate under a common 
fundamental assumption -- namely, the absence of an ultraviolet background.  
Given the abundance of halos in which Population III stars would form and the 
general consensus that these stars must be massive, one expects 
that, though only a small fraction of the volume of the universe would be
ionized, a significant background of ultraviolet radiation in the Lyman-Werner 
(LW) band (11.18-13.6 eV), which is capable of photodissociating molecular
hydrogen, would be present for the formation of the bulk of Population III stars
 \markcite{2001ApJ...546..635O,2003ApJ...592..645Y,2001ApJ...548..509M,2005ApJ...629..615W}({Omukai} 2001; {Yoshida} {et~al.} 2003; {Machacek}, {Bryan}, \&  {Abel} 2001; {Wise} \& {Abel} 2005).
Given that atomic hydrogen is optically thin to this radiation and that it easily 
destroys molecular hydrogen, this could have a significant impact on the formation
of primordial stars.  \markcite{2001ApJ...548..509M}{Machacek} {et~al.} (2001) used cosmological AMR simulations to
study the formation of primordial stars in the presence of a soft UV background, 
and later including an x-ray background \markcite{2003MNRAS.338..273M}({Machacek}, {Bryan}, \&  {Abel} 2003).
They found that the LW background delays the formation
of Population III stars and shifts halo formation to higher masses.  However, their calculations
were not of sufficiently high resolution to give any significant information about
the formation of the protostellar cloud at the core of each collapsing cosmological halo.
 \markcite{2003ApJ...592..645Y}{Yoshida} {et~al.} (2003) performed a suite of large SPH simulations of high redshift 
structure formation, including two calculations with a soft UV background with 
$J_{21} = 0.01$ and $0.1$, where $J_{21}$ is the mean intensity of the UV background 
in the LW band in units of $10^{-21}$~erg s$^{-1}$~cm$^{-2}$~Hz$^{-1}$~sr$^{-1}$.  They 
found that gas cooling is suppressed in the higher LW background case, and predicted that 
star formation would not occur. On the other hand, \markcite{2002ApJ...569..558O}{Oh} \& {Haiman} (2002) showed 
analytically that primordial gas in halos with 
T$_{vir} \ga 10^4$~K can still collapse to high densities in the presence of a strong 
molecular hydrogen-dissociating background via atomic hydrogen line radiation, and can 
eventually form H$_2$. They referred to these objects as ``second generation objects" in 
the sense that although still of primordial composition, their formation pathway is
different from the first stars. 

An issue that needs to be addressed concerns the mode of primordial star formation 
in halos with T$_{vir} \sim 10^4$~K and above.  It is apparent from recent studies using large 
numbers of simulations that smaller halos, with masses of $\sim 10^5-10^6$~M$_\odot$, 
appear to always form primordial stars in isolation~\markcite{2006astro.ph.10174G,oshea07a}({Gao} {et~al.} 2006; {O'Shea} \& {Norman} 2007).  
Is this still the case in halos whose masses are an order of magnitude or more 
larger?  Does a single primordial star form, or several?  This is particularly 
relevant given that many semi-analytic models of structure formation that follow the 
chemical evolution of structures at high redshift \markcite{2005ApJ...633.1031S,2006MNRAS.369..825S}({Scannapieco} {et~al.} 2005; {Schneider} {et~al.} 2006)
suggest that halos of this scale which have not been enriched by metals (and thus will form
primordial stars) exist up to at least $z \sim 5$.  If this is true, there is expected 
to be a strong photodissociating background at that time~\markcite{2003ApJ...592..645Y,2005ApJ...629..615W}({Yoshida} {et~al.} 2003; {Wise} \& {Abel} 2005).

This paper is the second in a series.  In \markcite{oshea07a}{O'Shea} \& {Norman} (2007) (hereafter referred to as Paper I), we
examined several aspects of Population III star formation in a $\Lambda$CDM universe, in the 
absence of an ultraviolet background which photodissociates molecular hydrogen.
In this paper, we study Population III star formation in a single cosmological realization 
varying the strength of the LW
background, with the goal of investigating the effect that this background has on the 
evolution and properties of gas in the halo core (at radii $\ll 1$~pc) and the ultimate 
fate of the gas at the center of
T$_{vir} \sim 10^4$~K halos.  We use a single cosmological realization,
varying the strength of the photodissociating background over the range
suggested by~\markcite{2005ApJ...629..615W}{Wise} \& {Abel} (2005), and look for the cutoff in Population III star formation
suggested by~\markcite{2003ApJ...592..645Y}{Yoshida} {et~al.} (2003).

In agreement with the findings of \markcite{2001ApJ...548..509M}({Machacek} {et~al.} 2001), we find that the addition of a 
photodissociating background 
results in a delay of the collapse of high density gas at the center of the most
massive halo in the simulation and, as a result, an increase in the virial mass of this 
halo at the onset of baryon collapse.  We find that, contrary to the results 
of~\markcite{2003ApJ...592..645Y}{Yoshida} {et~al.} (2003), star formation is not suppressed for J$_{21} \geq 0.1$, 
but occurs even with backgrounds as high as J$_{21} = 1$.
We find that \h2 cooling leads to collapse despite the depressed halo core molecular hydrogen 
fractions f$_{H2} \sim 10^{-6}$ by two multiplicative effects: (1) the elevated \h2 cooling 
rates per molecule at $T=2-5 \times 10^3$ K, and (2) time. We find that halo core collapse occurs in the usual 
way once the gas in the halo core has become dense enough that the cooling time becomes much less than
the Hubble time. 
We also observe a relationship between the strength
of the photodissociating background and the rate of accretion onto the evolving 
protostellar cloud core, with higher LW background fluxes resulting in higher 
accretion rates. This is a simple consequence of the suppression of molecular 
hydrogen formation (and thus suppression of cooling) by the photodissociating background, as well as
the higher virial temperatures of the more massive
halos at the epoch of collapse.  This may have implications for the range of Population III stellar 
masses. Finally, we find that the collapsing halo cores in our simulations do not 
fragment at densities below $n \sim 10^{10}$~cm$^{-3}$ regardless of the strength of the 
soft UV background, suggesting that Population III stars forming in halos with
T$_{vir} \sim 10^4$~K may still form in isolation.

The organization of this paper is as follows.  In Section~\ref{methodology} we provide a 
description of Enzo, the code used to perform the calculations in this paper and of
the simulation setup.  The results from our simulations are presented in 
Sections~\ref{results.meanprop} through \ref{results.rephalos}: 
Section~\ref{results.meanprop} discusses some of the mean halo properties observed in 
the calculations, Section~\ref{results.evolution} discusses the evolution
of the halo core prior to collapse,
Section~\ref{results.collapse} compares spherically-averaged halo 
properties for all simulations at the epoch of collapse, 
Section~\ref{results.fixred} discusses a variety of halo properties at a fixed 
redshift, and Section~\ref{results.rephalos} compares the evolution of two representative 
simulations taken from our ensemble.
In Section~\ref{sect.issues} we discuss neglected physics and possible numerical issues, 
and in Section~\ref{discuss} we discuss some of the results presented in this work and 
their implications.  Finally, we present a summary of the main results in Section~\ref{summary}.

%%%%%%%%%%%%%%%%%%%%%%%%%%%%%%%%%%%%%%%%%%%%%%%%%%%%%%%%%%%%%%%%%%%%%%%%%%%%
\section{Methodology}\label{methodology}

%---------------------------------------------------------------------------
\subsection{The Enzo code}\label{enzocode}

`Enzo'\footnote{http://lca.ucsd.edu/portal/software/enzo/} is a publicly available, extensively tested 
adaptive mesh refinement (AMR)
cosmology code developed by Greg Bryan and others \markcite{bryan97,bryan99,norman99,oshea04,
2005ApJS..160....1O}({Bryan} \& {Norman} 1997a, 1997b; {Norman} \& {Bryan} 1999; {O'Shea} {et~al.} 2004, 2005b).
The specifics of the Enzo code are described in detail in these papers (and references therein),
but we present a brief description here for clarity.

The Enzo code couples an N-body particle-mesh (PM) solver \markcite{Efstathiou85, Hockney88}({Efstathiou} {et~al.} 1985; {Hockney} \& {Eastwood} 1988) 
used to follow the evolution of a collisionless dark
matter component with an Eulerian AMR method for ideal gas dynamics by \markcite{Berger89}{Berger} \& {Colella} (1989), 
which allows high dynamic range in gravitational physics and hydrodynamics in an 
expanding universe.  This AMR method (referred to as \textit{structured} AMR) utilizes
an adaptive hierarchy of grid patches at varying levels of resolution.  Each
rectangular grid patch (referred to as a ``grid'') covers some region of space in its
\textit{parent grid} which requires higher resolution, and can itself become the 
parent grid to an even more highly resolved \textit{child grid}.  Enzo's implementation
of structured AMR places no fundamental restrictions on the number of grids at a 
given level of refinement, or on the number of levels of refinement.  However, owing 
to limited computational resources it is practical to institute a maximum level of 
refinement, $\ell_{max}$.  Additionally, the Enzo AMR implementation allows arbitrary 
integer ratios of parent
and child grid resolution, though in general for cosmological simulations (including the 
work described in this paper) a refinement ratio of 2 is used.

Since the addition of more highly refined grids is adaptive, the conditions for refinement 
must be specified.  In Enzo, the criteria for refinement can be set by the user to be
a combination of any or all of the following:  baryon or dark matter overdensity
threshold, minimum resolution of the local Jeans length, local density gradients,
local pressure gradients, local energy gradients, shocks, and cooling time.
A cell reaching
any or all of the user-specified criteria will then be flagged for refinement.  Once all 
cells of a given level have been examined, rectangular solid boundaries are determined which 
minimally 
encompass the flagged cells on that level.  A refined grid 
patch is then introduced within each such bounding 
volume, and the results are interpolated to a higher level of resolution.

In Enzo, resolution of the equations being solved is adaptive in time as well as in
space.  The timestep in Enzo is satisfied on a level-by-level basis by finding the
largest timestep such that the Courant condition (and an analogous condition for 
the dark matter particles) is satisfied by every cell on that level.  All cells
on a given level are advanced using the same timestep.  Once a level $L$ has been
advanced in time $\Delta t_L$, all grids at level $L+1$ are 
advanced, using the same criteria for timestep calculations described above, until they
reach the same physical time as the grids at level $L$.  At this point grids at level
$L+1$ exchange baryon flux information with their parent grids, providing a more 
accurate solution on level $L$.  Cells at level $L+1$ are then examined to see 
if they should be refined or de-refined, and the entire grid hierarchy is rebuilt 
at that level (including all more highly refined levels).  The timestepping and 
hierarchy rebuilding processes are repeated recursively on every level to the 
maximum existing grid level in the simulation.

Two different hydrodynamic methods are implemented in Enzo: the piecewise parabolic
method (PPM) \markcite{Woodward84}({Woodward} \& {Colella} 1984), which was extended to cosmology by 
\markcite{Bryan95}{Bryan} {et~al.} (1995), and the hydrodynamic method used in the ZEUS magnetohydrodynamics code
\markcite{stone92a,stone92b}({Stone} \& {Norman} 1992a, 1992b).  We direct the interested reader to the papers describing 
both of these methods for more information, and note that PPM is the preferred choice
of hydro method since it is higher-order-accurate and is based on a technique that 
does not require artificial viscosity, which smoothes shocks and can smear out 
features in the hydrodynamic flow.

The chemical and cooling properties of primordial (metal-free) gas are followed 
using the method of \markcite{abel97}{Abel} {et~al.} (1997) and \markcite{anninos97}{Anninos} {et~al.} (1997).  
This method follows the non-equilibrium evolution of a 
gas of primordial composition with 9 total species:  
$H$, $H^+$, $He$, $He^+$, $He^{++}$, $H^-$, $H_2^+$, $H_2$, and $e^-$.  The code 
also calculates 
radiative heating and cooling following atomic line excitation, recombination,
collisional excitation, free-free transitions, molecular line cooling, and Compton
scattering of the cosmic microwave background, as well as any of
approximately a dozen different models for a metagalactic ultraviolet background that heat
the gas via photoionization and/or photodissociation.  We model the cooling processes
detailed in~\markcite{abel97}{Abel} {et~al.} (1997), but use the~\markcite{1998A&A...335..403G}{Galli} \& {Palla} (1998) molecular
hydrogen cooling function.  The multispecies rate equations are solved out of
equilibrium to properly model situations where, e.g., the cooling time of the gas
is much shorter than the hydrogen recombination time.  
A total of 9 kinetic equations are solved, including 29 kinetic and radiative 
processes, for the 9 species mentioned above.  
The chemical reaction equation network is technically challenging to solve due to 
the huge range of reaction time scales involved; the characteristic creation
and destruction time scales of the various species and reactions can differ by 
many orders of magnitude.  As a result, the set of rate equations is extremely 
stiff, and an explicit scheme for integration of the rate equations can be 
costly if small enough timestep are taken to keep the network
stable.  This makes an implicit scheme preferable for such a set of 
equations, and Enzo solves the rate equations using a method based on a backwards 
differencing formula (BDF) in order to provide a stable and accurate solution.
 
It is important to note the regime in which this chemistry model is valid.  According to 
\markcite{abel97}{Abel} {et~al.} (1997) and \markcite{anninos97}{Anninos} {et~al.} (1997), the reaction network is valid for temperatures
between $10^0 - 10^8$ K.  The original model discussed in these two references is only
valid up to n$_H \sim 10^4$~cm$^{-3}$.  However, addition of the 3-body H$_2$ formation
process allows correct modeling of the gas chemistry up until the point where
collisionally induced emission from molecular hydrogen becomes an important
cooling processes, which occurs at $n_H \sim 10^{14}$~cm$^{-3}$.  We do not
include heating by molecular hydrogen formation, which will be significant at 
densities of $\sim 10^8$~cm$^{-3}$ and above, and may affect temperature evolution at
these high densities.  A further concern is that the optically thin approximation
for radiative cooling breaks down beginning at n$_H \simeq 10^{10} - 10^{12}$~cm$^{-3}$.
Beyond this point, 
modifications to the cooling function that take into account the non-negligible
opacity of the gas to line radiation from molecular hydrogen must be made, as 
discussed by \markcite{ripamonti04}{Ripamonti} \& {Abel} (2004).  Even with these modifications, a more correct 
description of the cooling of gas of primordial composition at high densities will 
require some form of radiation transport, which will greatly 
increase the cost of the simulations.

%---------------------------------------------------------------------------
\subsection{Simulation setup}\label{sect.simsetup}

The simulations discussed in this paper are set up in a similar way 
to those in Paper I.  A dark matter-only
calculation with $128^3$ particles in a three-dimensional simulation volume 
which is $0.6$~h$^{-1}$~Mpc (comoving) on a side
is set up at $z=99$ assuming a ``concordance''
cosmological model with no baryons:  $\Omega_m = \Omega_{DM} = 0.3$, 
$\Omega_b = 0.0$, $\Omega_\Lambda = 0.7$, $h=0.7$ (in units of 100 km/s/Mpc), 
$\sigma_8 = 0.9$, and using an Eisenstein \& Hu power spectrum \markcite{eishu99}({Eisenstein} \& {Hu} 1999)
with a spectral index of $n = 1$.  The cold dark matter cosmological model
is assumed.  This calculation is then evolved to 
$z=15$ using a maximum of four levels of adaptive mesh refinement, 
refining on a dark matter overdensity of 8.0.  At $z=15$, the 
Hop halo finding algorithm \markcite{eishut98}({Eisenstein} \& {Hut} 1998) is used to find the most massive 
halo in the simulation.

At this point, we generate a new set of initial conditions which contain the same large-scale
power as the dark matter-only calculation, but include both dark matter and baryons 
such that the Lagrangian volume in which the halo in the dark matter-only calculation formed
is resolved at high spatial and mass resolution using a series of static 
nested grids, with a $128^3$ 
root grid and three static nested grids, for an overall effective root grid size of $1024^3$
cells.
The highest resolution grid  is $256^3$ grid cells, and corresponds
to a volume $150$~h$^{-1}$ comoving kpc on a side.
The dark matter particles in the highest
resolution grid are 14.48~h$^{-1}$~M$_\odot$ and the spatial resolution
of the highest resolution grid is 586~h$^{-1}$ parsecs (comoving). 
Previous work shows that this particle mass resolution is adequate 
to fully resolve the collapse of the halo \markcite{ABN02,oshea07a}({Abel} {et~al.} 2002; {O'Shea} \& {Norman} 2007).

All simulations are performed using the adaptive
mesh cosmology code Enzo, which is described  in Section~\ref{enzocode}.  
The simulations are started at $z=99$ and allowed to evolve until the collapse
of the gas within the center of the most massive halo, assuming the presence of 
an unevolving soft UV background with intensities in the Lyman-Werner band
of J$_{LW} = 0, 10^{-24}, 10^{-23.5}, 10^{-22.5}, 10^{-22}, 10^{-21.67}, 10^{-21.33},$ 
and~$10^{-21}$ ~erg s$^{-1}$~cm$^{-2}$~Hz$^{-1}$~sr$^{-1}$.  
This range covers a much wider range of parameter space than the results 
described in \markcite{2001ApJ...548..509M}{Machacek} {et~al.} (2001), and completely encompasses the range of possible J$_{LW}$ values
suggested by \markcite{2005ApJ...629..615W}{Wise} \& {Abel} (2005), for a wide range of mean Population III stellar masses.
Note that many publications use F$_{LW}$ instead of J$_{LW}$:  F$_{LW}$ has units of erg~s$^{-1}$~cm$^{-2}$~Hz$^{-1}$,
and thus F$_{LW} = 4 \pi$~J$_{LW}$.  We will use J$_{LW}$ throughout this paper, and for convenience 
express values in 
units of J$_{21}$, where $J_{21}$ is the mean intensity of the UV background in the LW band in units of 
$10^{-21}$~erg s$^{-1}$~cm$^{-2}$~Hz$^{-1}$~sr$^{-1}$.

The equations of hydrodynamics
are solved using the PPM method with a dual energy formulation, which is required 
to adequately resolve the thermal properties of gas in high-Mach flows.  The nonequilibrium 
chemical
evolution and optically thin radiative cooling of the primordial gas is 
modeled as described in Section~\ref{enzocode}, following 9 
separate species including molecular hydrogen (but excluding deuterium), with an initial
electron fraction of $2.35 \times 10^{-4}$ (which is roughly consistent with
\markcite{1968ApJ...153....1P}{Peebles} (1968) for an $\Omega_m = 0.3$, $\Omega_b = 0.04$
universe).  Note that the initial electron fraction in the 
calculation is relatively unimportant to the molecular hydrogen formation rates
in halo cores, as the electron fraction at the center of a given halo is 
controlled primarily
by mergers and the shock formed by accretion of gas onto the halo. 

Adaptive
mesh refinement is used throughout the innermost high resolution region 
such that cells are refined by factors of two along each 
axis, with a maximum of 22 total levels of refinement.  This corresponds to a 
maximum spatial resolution of 115~h$^{-1}$ AU (comoving)
at the finest level of resolution, with an overall spatial dynamical range of
$5.37 \times 10^8$.  To avoid effects due to the finite size of the dark matter
particles, the dark matter density is smoothed on a comoving scale of $\sim 0.5$~pc
(which corresponds to $\simeq$ 0.03 proper pc at $z \simeq 18$).
This is reasonable because at that scale in all of our calculations the gravitational
potential in the halo of interest is completely dominated by the baryons.

Grid cells are adaptively refined based upon several criteria.  We refine on
baryon and dark matter overdensities in cells of 4.0 and 8.0, respectively.
This corresponds to a maximum mass of gas or dark matter per cell (on the
most highly refined static grid) of M$_{max} = 12.78$ and $166.16$~M$_\odot$,
respectively.  In addition, the \emph{MinimumMassForRefinementLevelExponent} 
parameter is set to $-0.2$ for both the dark matter and baryon overdensity
refinement criteria, meaning that the mass required to refine to a higher
level decreases as a function of increasing level, as:

\begin{equation}
M_{ref}(L) = M_{max} \times 2^{-0.2 L}
\label{eqn-refine}
\end{equation}

where L is the current level of refinement.  The negative exponent means 
that the mass resolution in the calculations is 
super-Lagrangian -- for example,
M$_{ref}(L=20) = 0.0625$~M$_{ref}(L=0)$.
In addition to refining on baryon and dark matter overdensity, these simulations include
refinement criteria which ensure that shocks are always well-resolved, that the 
cooling time in a given cell is always longer
than the sound crossing time of that cell, and that the Jeans length is always
resolved by at least 16 cells.  This last criterion guarantees that the Truelove
criterion \markcite{truelove97}({Truelove} {et~al.} 1997) is always resolved by a factor of four more cells 
in each dimension than is strictly necessary, ensuring that no artificial fragmentation 
will take place.

%%%%%%%%%%%%%%%%%%%%%%%%%%%%%%%%%%%%%%%%%%%%%%%%%%%%%%%%%%%%%%%%%%%%%%%%%%%%
\section{Mean halo properties at collapse}\label{results.meanprop}

\begin{figure}
\begin{center}
%%% set textwidth to 0.45 for apj, 0.9 for one-column MS format
\includegraphics[width=0.9\textwidth]{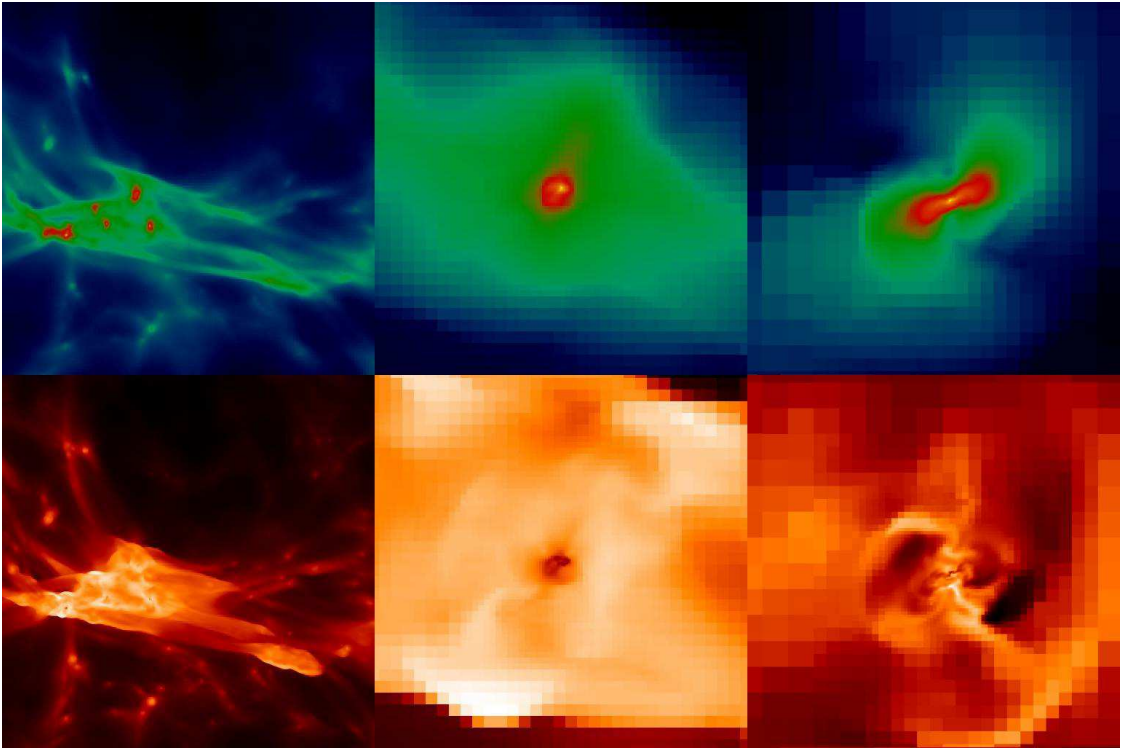}
\end{center}
\caption{
Projections of baryon density and temperature at the final output of the J$_{21} = 0$ simulation
($z = 24.119$; M$_{vir} = 5.68 \times 10^{5}$~M$_\odot$).
Top row:  Projected log baryon density.  Bottom row: Projected, mass-weighted log baryon temperature.
Left column: Region 2.13 kpc (proper) across and deep.
Middle column: Region 133.3 pc (proper) across and deep.
Right column:  Region 0.130 pc (proper) across and deep.
The middle and right column are zoomed images focusing on the high baryon density core which forms in the center of the halo.
}
\label{fig.image_jlw_0}
\end{figure}

\begin{figure}
\begin{center}
\includegraphics[width=0.9\textwidth]{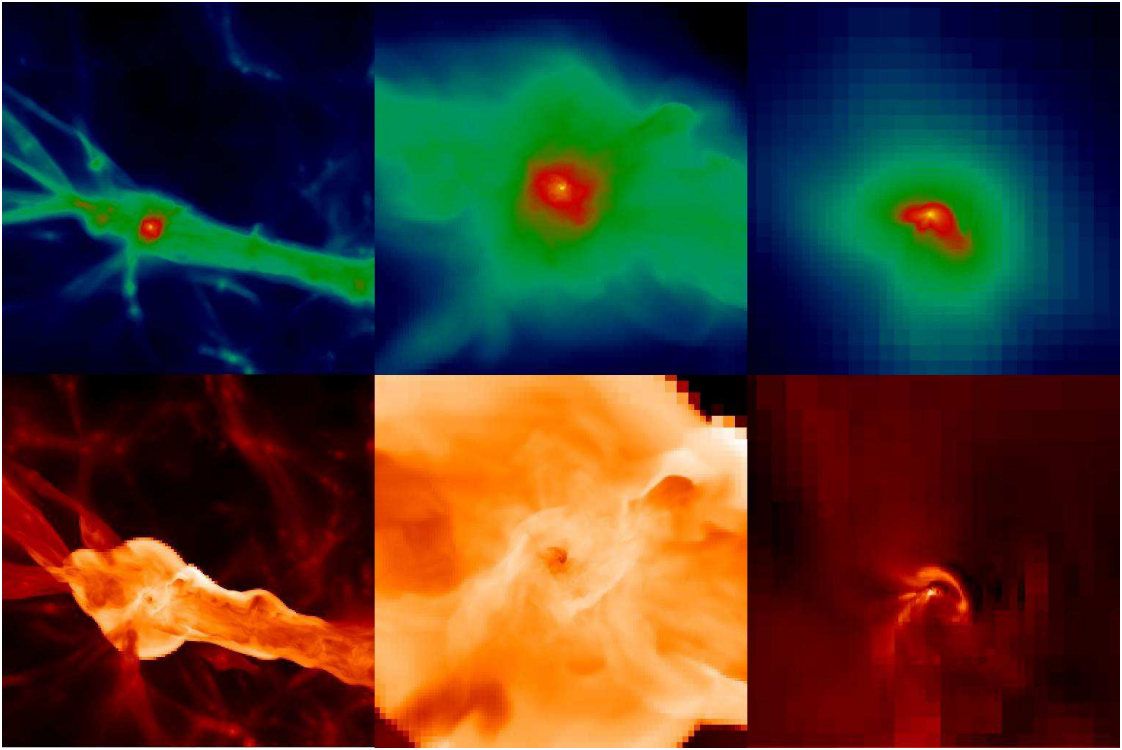}
\end{center}
\caption{
Projections of baryon density and temperature at the final output of the J$_{21} = 1$ simulation
($z = 17.322$; M$_{vir} = 1.260 \times 10^{7}$~M$_\odot$).
Top row:  Projected log baryon density.  Bottom row: Projected, mass-weighted log baryon temperature.
Left column: Region 2.13 kpc (proper) across and deep.
Middle column: Region 533 pc (proper) across and deep.
Right column:  Region 0.260 pc (proper) across and deep.
The middle and right column are zoomed images focusing on the high baryon density core which forms in the center of the halo.
}
\label{fig.image_jlw_1em21}
\end{figure}

Figures~\ref{fig.image_jlw_0} and~\ref{fig.image_jlw_1em21} show projections of
baryon density and temperature for the J$_{21} = 0$ and J$_{21} = 1$ 
calculations, respectively, at the epoch at which each calculation collapses,
defined as the redshift
at which the baryon number density reaches $\simeq 10^{10}$~cm$^{-3}$.
Note that due to the rapid evolution of gas at high density, the
``collapse redshift'' depends very weakly on the exact choice of
density threshold.
The calculations are started from the same set of initial conditions, and 
the J$_{21} = 1$ calculation is clearly a later stage in the evolution
of the J$_{21} = 0$ run.  The satellite halos surrounding the primary halo which 
collapses at z=24.12 in the J$_{21} = 0$ run have merged by z=17.32 when the halo in the J$_{21} = 1$ case collapses (cf. Fig. \ref{fig.merger}). With virial masses of $5.68 \times 10^5 \Ms$ and $1.26 \times 10^7 \Ms$, respectively, the latter halo is more than 20 times as massive as the former.  Though quite different in mass, both halos exhibit
similar morphologies -- they are extremely centrally-concentrated, and only a single
condensed object (that is to say, a primordial protostellar cloud core) is visible in the
highest-resolution panel in each image.  The existence of a single cloud core
is common to all calculations discussed in this paper.

\clearpage
%%%%%%%%%%%%%%%%%%% FIGURE %%%%%%%%%%%%%%
%%%%%%%%%%%%%%%%%%%%%%%%%%%%%%%%%%%%%%%%%

%%%%%%%%%%%%%%%%%%% FIGURE %%%%%%%%%%%%%%
%%%%%%%%%%%%%%%%%%%%%%%%%%%%%%%%%%%%%%%%%
\clearpage

\begin{figure}
\begin{center}
\includegraphics[width=0.9\textwidth]{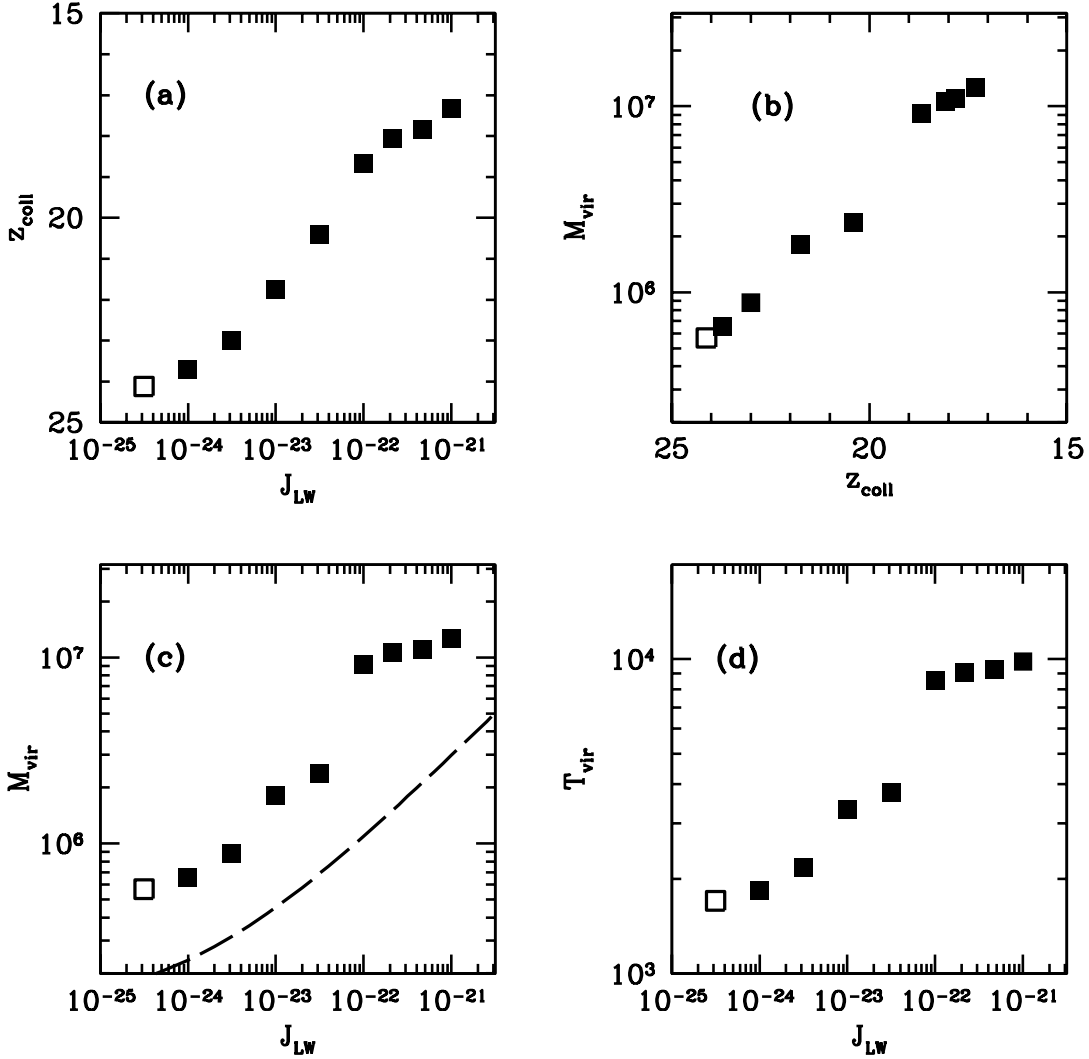}
\end{center}
\caption{
Mean halo quantities for several simulations with the same cosmic 
realization but a range of Lyman-Werner molecular hydrogen
photodissociating flux
backgrounds.  
Panel (a):  J$_{LW}$ vs. halo collapse redshift.
Panel (b): halo virial mass vs. halo collapse redshift
Panel (c): halo virial mass vs. J$_{LW}$
Panel (d): halo virial temperature vs. J$_{LW}$
The J$_{21} = 0$ ``control'' result are shown as an open square 
(and is at log J$_{LW} = -24.5$ in the panels which
are a function of J$_{LW}$).
In the bottom left panel, the dashed line corresponds to 
the fitting function for threshold mass
from \markcite{2001ApJ...548..509M}{Machacek} {et~al.} (2001), Eqtn. 8.}
\label{fig.halovals1}
\end{figure}

Figure~\ref{fig.halovals1} shows several mean halo quantities for all simulations
discussed in this paper, including halo collapse redshift as a function 
of J$_{LW}$, virial mass as a function of J$_{LW}$, virial temperature as a function
of J$_{LW}$, and the virial mass as a function of collapse redshift.  Simulations where the
soft UV background is turned on are shown by solid squares, and the ``control'' J$_{21} = 0$
calculation is shown by an open square.  In plots where J$_{LW}$ is shown on the x-axis, the
J$_{21} = 0$ simulation is placed at log$_{10} J_{LW} = -24.5$.

Figure~\ref{fig.halovals1} shows that there is a clear relationship between the LW
intensity and the collapse redshift and virial mass of the halo.  A larger LW
intensity results in a later collapse time and larger virial mass because the halo must be hotter
in order to have a cooling time which is less than a Hubble time despite the depressed \h2 abundance in the high-density gas at the halo's center. 
This is discussed further in Section~\ref{results.evolution}  The final mass of the halos in the simulations with 
J$_{21} = 1$ is approximately a factor of 20
higher than that in the ``control'' simulation.  This is in qualitative agreement with \markcite{2001ApJ...548..509M}{Machacek} {et~al.} (2001),
who suggest that there is a ``minimum halo mass'' as a function of the strength of the LW background of the form

\begin{equation}
M_{TH}(M_\odot) = 1.25 \times 10^5 + 8.7 \times 10^5 ~\left( \frac{F_{LW}}{10^{-21}} \right)^{0.47}, F_{LW} \leq 10^{-21}
\label{eqn-mach-mthresh}
\end{equation}

This threshold mass is plotted in panel (c) of Figure~\ref{fig.halovals1}.
Eq. \ref{eqn-mach-mthresh} is only strictly valid over the range $0 \leq F_{LW}=4\pi J_{LW} \leq 10^{-21}$, because this was the range simulated by \markcite{2001ApJ...548..509M}{Machacek} {et~al.} (2001). We see that our points for $J_{21} \leq 0.1$ parallel the threshold curve but at a mass approximately four times higher. This difference can be ascribed to two factors. First,
the threshold mass is the lowest possible halo mass that can collapse, derived from a statistical sample. Since we study only one realization which focuses on the most massive halo in the box, its mass is bound to be higher than the statistical minimum. Second, \markcite{2001ApJ...548..509M}{Machacek} {et~al.} (2001)'s criterion for cooling catches halos at an earlier stage of evolution compared to our data points, which give the halos' virial masses at the time of 
central baryon collapse (the ``collapse redshift'').   

 Examination of panels (b) and (c) of Figure~\ref{fig.halovals1} show that there is some sort 
of ``phase change'' between J$_{21} = 10^{-1.5}$
and $10^{-1}$ causing the halo mass at the time of collapse to increase
steeply and non-monotonically. 
The collapse redshift steadily
decreases as a function of UV increasing background strength 
(i.e. collapse of gas at the center of the halo is delayed). 
However, panel (c) shows that the virial mass steadily increases with increasing
UV background strength until J$_{21} = 10^{-1.5}$, at which point there is a jump in 
mass by more than a factor of four, above which the halo mass increases only 
slightly with increasing UV background strength. That something interesting should happen at these LW intensities is consistent with the findings of~\markcite{2003ApJ...592..645Y}{Yoshida} {et~al.} (2003) who found that \h2 cooling is strongly suppressed at J$_{21}=0.1$. They predicted that star formation would be inhibited since the equilibrium \h2 fraction is below the critical fraction for cooling derived by \markcite{1997ApJ...474....1T}{Tegmark} {et~al.} (1997). Contrary to these predictions, we find collapse not only at J$_{21}=0.1$, but also at values as large as J$_{21}=1$. We note that highest value
considered by \markcite{2001ApJ...548..509M}{Machacek} {et~al.} (2001) was J$_{21}= 0.0796$, and the highest simulated by~\markcite{2003ApJ...592..645Y}{Yoshida} {et~al.} (2003) was J$_{21}=0.1$. Therefore our cases J$_{21} > 0.1$ have not been examined before, and certainly not at the resolution of our simulations.   

How is the gas in the center of these halos able to cool and collapse in such high radiation backgrounds? This is analyzed in some detail in the next section. A hint is provided in (d) of Figure~\ref{fig.halovals1}, which plots the halo virial temperature as a function of J$_{LW}$. At background strengths above 
J$_{21} = 10^{-1.5}$, the virial temperature is consistently approximately $10^4$~K, 
which is roughly the temperature at which atomic hydrogen cooling is effective. However, an examination of the radial temperature profiles shows that high-density gas in the center of the halo never reaches these temperatures, but is more typically 2000 K. Atomic line cooling is unimportant at such low temperatures; however the \h2 cooling rate per particle is roughly 100 times as large at 2000K as at 500K. \h2 cooling still operates in the halo centers despite low \h2 abundance due to the higher cooling rates and long evolutionary timescales. In the rest of this section we merely present additional mean properties of the halo and its central region at collapse, and defer discussion of the relevant timescales to Section~\ref{results.evolution}. 

\clearpage
%%%%%%%%%%%%%%%%%%% FIGURE %%%%%%%%%%%%%%
%%%%%%%%%%%%%%%%%%%%%%%%%%%%%%%%%%%%%%%%%
\clearpage

\begin{figure}
\begin{center}
\includegraphics[width=0.9\textwidth]{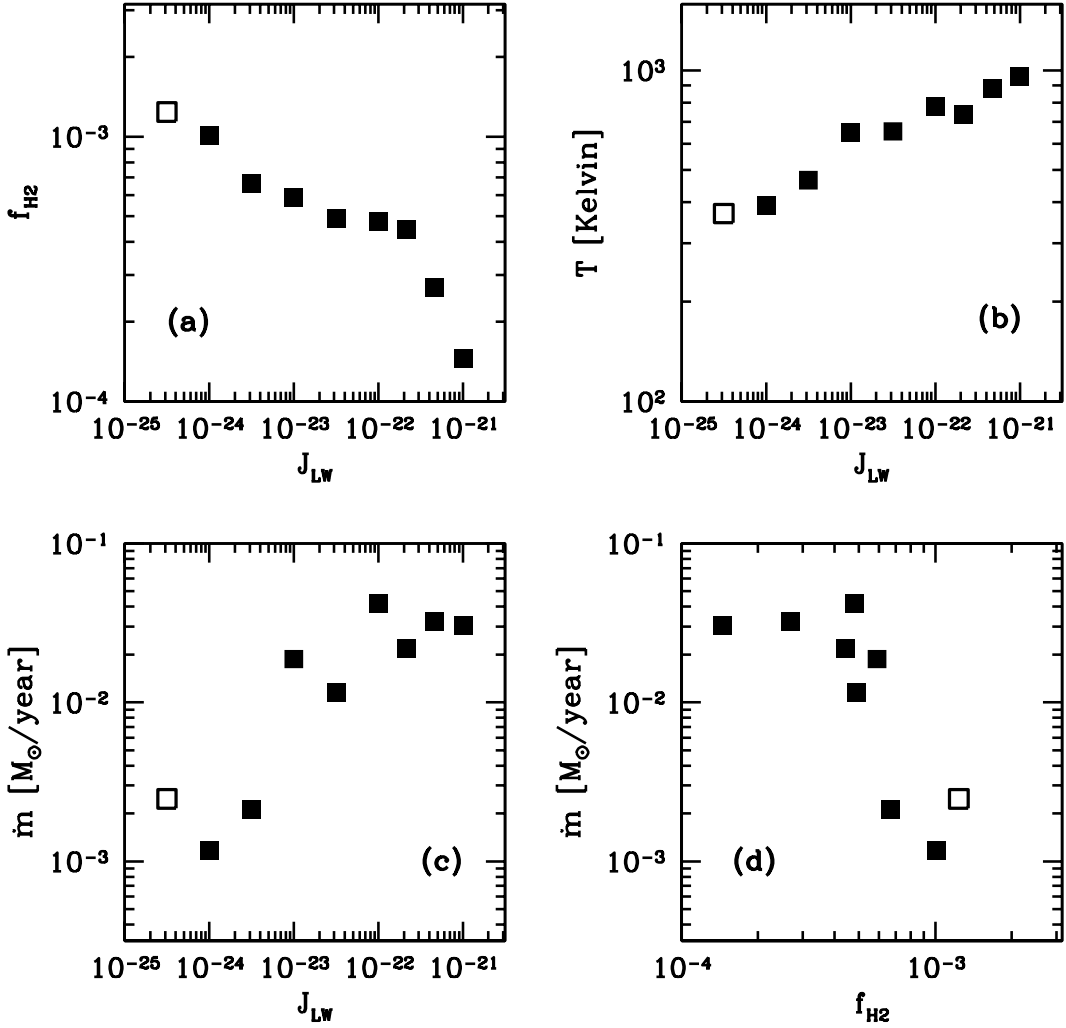}
\end{center}
\caption{
Mean quantities within the central $100 \Ms$ core for several simulations with the same cosmic 
realization but a range of LW background intensities.  
Panel (a):  J$_{LW}$ vs. baryon core temperature.
Panel (b):  J$_{LW}$ vs. baryon core $H_2$ fraction.
Panel (c): J$_{LW}$ vs. instantaneous accretion rate.
Panel (d): Baryon core $H_2$ fraction vs. accretion rate.
The J$_{21} = 0$ ``control'' result are shown as an open square 
(and is at log J$_{LW} = -24.5$ in the panels which
are a function of J$_{LW}$).  All ``core'' values are spherically-averaged and 
measured at the mass shell where
$100$~M$_\odot$ of gas is enclosed.
}
\label{fig.halovals2}
\end{figure}

Figure~\ref{fig.halovals2} shows several properties of the halo core at the epoch of
collapse, including the core temperature, molecular hydrogen fraction, and spherically-averaged
accretion rate as a function of ultraviolet background strength, and the spherically-averaged
accretion rate as a function of the molecular hydrogen fraction.  All ``core'' values are
spherically-averaged and measured at the mass shell where $100$~M$_\odot$ of gas
is enclosed.   Panel (a) shows that the core H$_2$ fraction decreases significantly with 
increasing FUV flux, with a corresponding increase in the core temperature (Panel (b)).  This
relationship is similar to that noted in~\markcite{oshea07a}{O'Shea} \& {Norman} (2007), where the amount of molecular
hydrogen at densities of $10^4 - 10^8$~cm$^{-3}$ varies between simulations, and correlates
strongly with baryon temperature.  Given that the accretion of gas onto the protostellar
cloud is subsonic, this results in a strong relationship
between the soft UV background flux and accretion rate onto the protostellar cloud, as shown by
Panel (c), with accretion rates varying
by more than a factor of 30 between the J$_{21} = 0$ and $1$ cases.  Panel (d) shows
the strength of the correlation between core H$_2$ fraction and accretion rate.  Note that
the values discussed above are insensitive to the exact definition of the halo ``core.''

\clearpage
%%%%%%%%%%%%%%%%%%% FIGURE %%%%%%%%%%%%%%
%%%%%%%%%%%%%%%%%%%%%%%%%%%%%%%%%%%%%%%%%
\clearpage

\begin{figure}
\begin{center}
\includegraphics[width=0.9\textwidth]{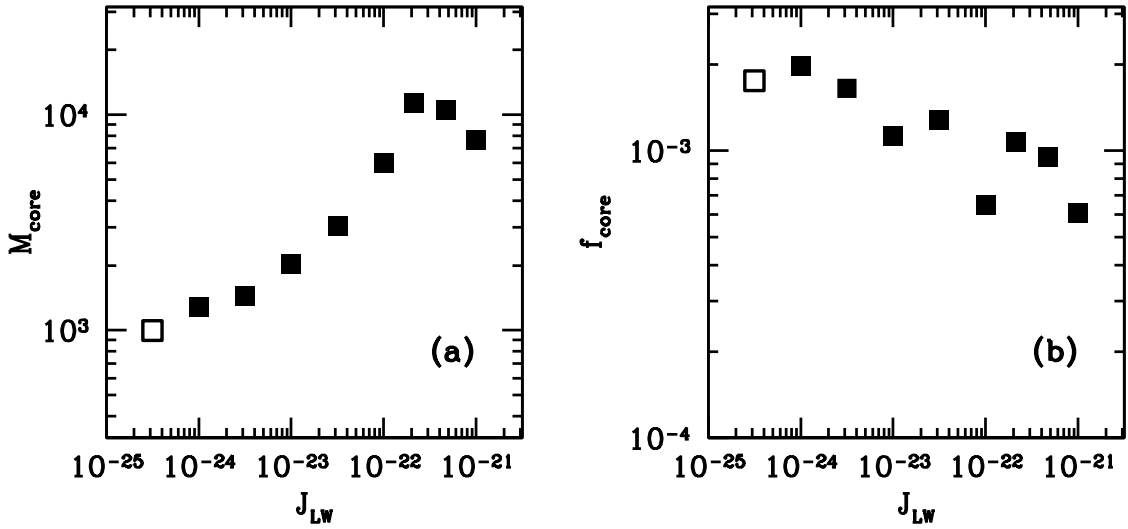}
\end{center}
\caption{
Halo core mass (Panel (a)) and core mass fraction (Panel (b)) for several 
simulations with the same cosmic 
realization but a range of LW background intensities.  
The ``core mass'' is defined as the mass of gas in the halo which has a density 
of n$_H = 10^4$~cm$^{-3}$ or above.  The ``core mass fraction'' is defined as the ``core
mass'' divided by the virial mass of the halo at the epoch of collapse.
The J$_{LW} = 0$ ``control'' result are shown as an open square 
(and is at log J$_{LW} = -24.5$ in the panels which
are a function of J$_{LW}$).
}
\label{fig.halovals3}
\end{figure}

Figure~\ref{fig.halovals3} shows the halo core mass and mass fraction at the epoch of collapse
(note that the definition of ``core'' is somewhat different than in the previous figure).
The halo ``core mass'' is defined as being all gas at a density of n$_H = 10^4$~cm$^{-3}$ or above,
and the core mass fraction is defined as the core mass divided by the virial mass of the halo at the epoch
of collapse.  This choice of minimum density ensures that we have captured the entirety of the quasistatically contracting
analog of a galactic molecular cloud core, and for practical purposes is comparable to the gas fraction of
cold, dense gas discussed in~\markcite{2001ApJ...548..509M}{Machacek} {et~al.} (2001). 
The size of the halo core increases in absolute terms as the strength of the photodissociating
background is increased, from $\simeq 10^3$~M$_\odot$ until J$_{21} = 0.1$, where it reaches $\simeq 11,000$~M$_\odot$.
Above this value of J$_{21}$, the core mass then begins to decrease, reaching $\simeq 7600$~M$_\odot$ at J$_{21} = 1$.
The core mass fraction generally decreases (though with some noise in the relationship) from 
$\simeq 1.7 \times 10^{-3}$ at J$_{21} = 0$ to $\simeq 6.96 \times 10^{-4}$ at J$_{21} = 1$, though the total
overall change is less than a factor of 2.  

\clearpage
%%%%%%%%%%%%%%%%%%% FIGURE %%%%%%%%%%%%%%
%%%%%%%%%%%%%%%%%%%%%%%%%%%%%%%%%%%%%%%%%
\clearpage

\section{Central evolution prior to collapse}\label{results.evolution}

\begin{figure}
\begin{center}
\includegraphics[width=0.9\textwidth]{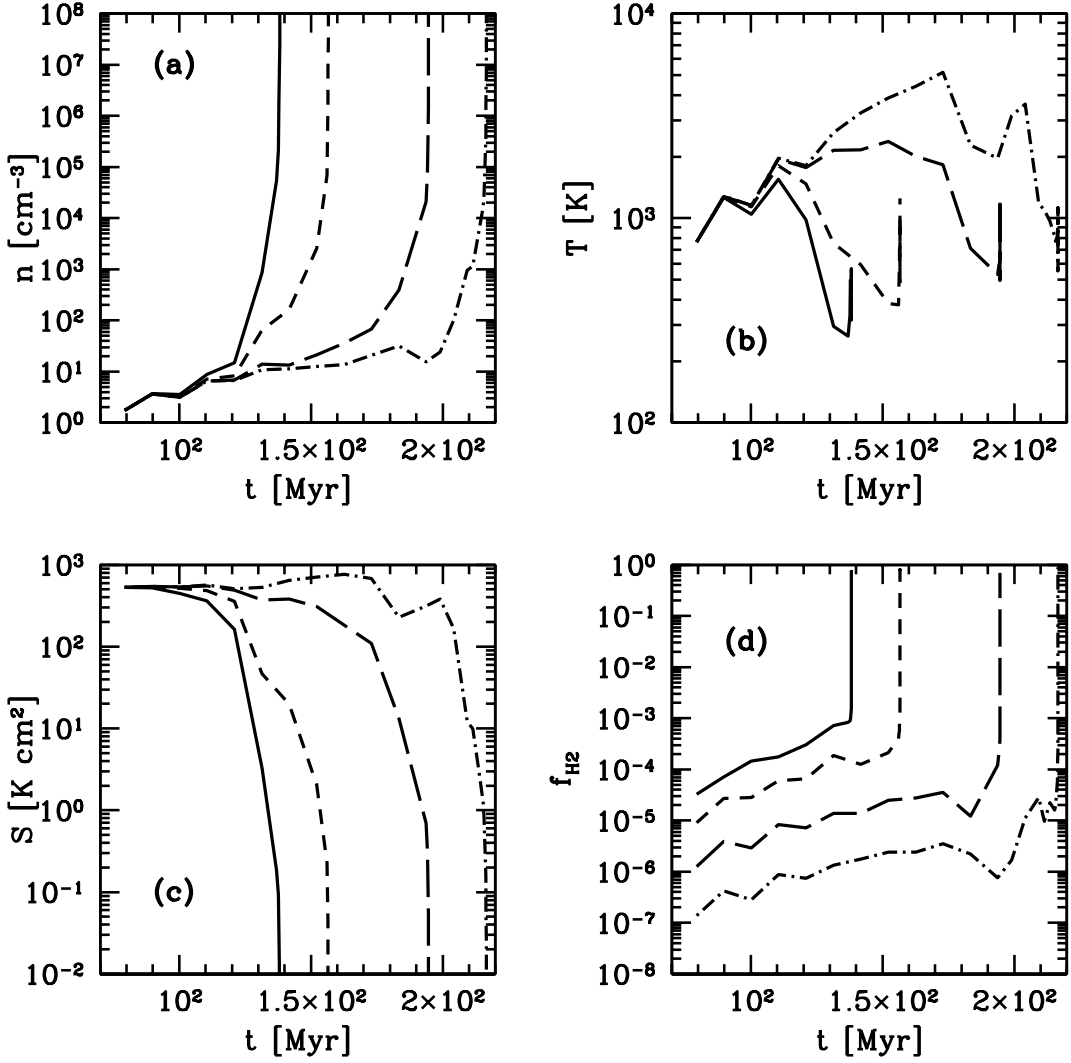}
\end{center}
\caption{
Evolution of the central density (Panel (a)), 
temperature (Panel (b)), entropy (Panel (c)), 
and \h2 fraction (Panel (d))as a function 
of time for the cases $J_{21}=10^{-3}$ (solid), $10^{-2}$ (short-dashed), $10^{-1}$ 
(long-dashed) and $1$ (dot-dashed). 
}
\label{fig.evolution}
\end{figure}

We now examine in more detail the evolution of the conditions in the cloud center 
leading to collapse. In Figure~\ref{fig.evolution} we plot the central density, 
temperature, entropy $S \equiv T/n_H^{2/3}$, and \h2 fraction versus time for 
the calculations with $J_{21}=10^{-3}$ (solid line), $10^{-2}$ (short-dashed line), 
$10^{-1}$ (long-dashed line) and $0$ (dot-dashed line).
In these plots, central values of temperature and density are defined as being
the values of these quantities in the cell with the highest baryon density.
 It is most instructive to compare 
the two extreme cases, J$_{21}=10^{-3}$, which very closely resembles the 
J$_{21}=0$ case, and the J$_{21}=1$ case, which according to~\markcite{2003ApJ...592..645Y}{Yoshida} {et~al.} (2003) 
should not collapse at all.  In panel (a) we see the central density increase modestly 
from $\sim$ 2 cm$^{-3}$ to $\sim$ 20 cm$^{-3}$ over 40 Myr, and then increase rapidly 
thereafter, reaching $10^8$ cm$^{-3}$ in a scant additional 15 Myr. This evolution is 
driven by \h2 cooling, as can be seen from the entropy evolution in panel (c). As 
expected, runaway cooling occurs when the \h2 fraction reaches $\sim 2 \times 10^{-4}$ 
(panel d), consistent with the \markcite{1997ApJ...474....1T}{Tegmark} {et~al.} (1997) analysis. By contrast the 
J$_{21}=1$ case requires 120 Myr for the central density to increases from $\sim$ 2 
cm$^{-3}$ to $\sim$ 20 cm$^{-3}$. As shown in panel (c), the first 50 Myr of this 
contraction is adiabatic, followed by an increase in entropy due to some heating 
event (mergers). The fact that central entropy is not decreasing for 90 Myr is indicative 
of the fact that \h2 cooling is unimportant over this interval due to extremely low 
equilibrium \h2 fractions (f$_{H2} < 10^{-6}$; (panel d)). 

\begin{figure}
\begin{center}
\includegraphics[width=0.9\textwidth]{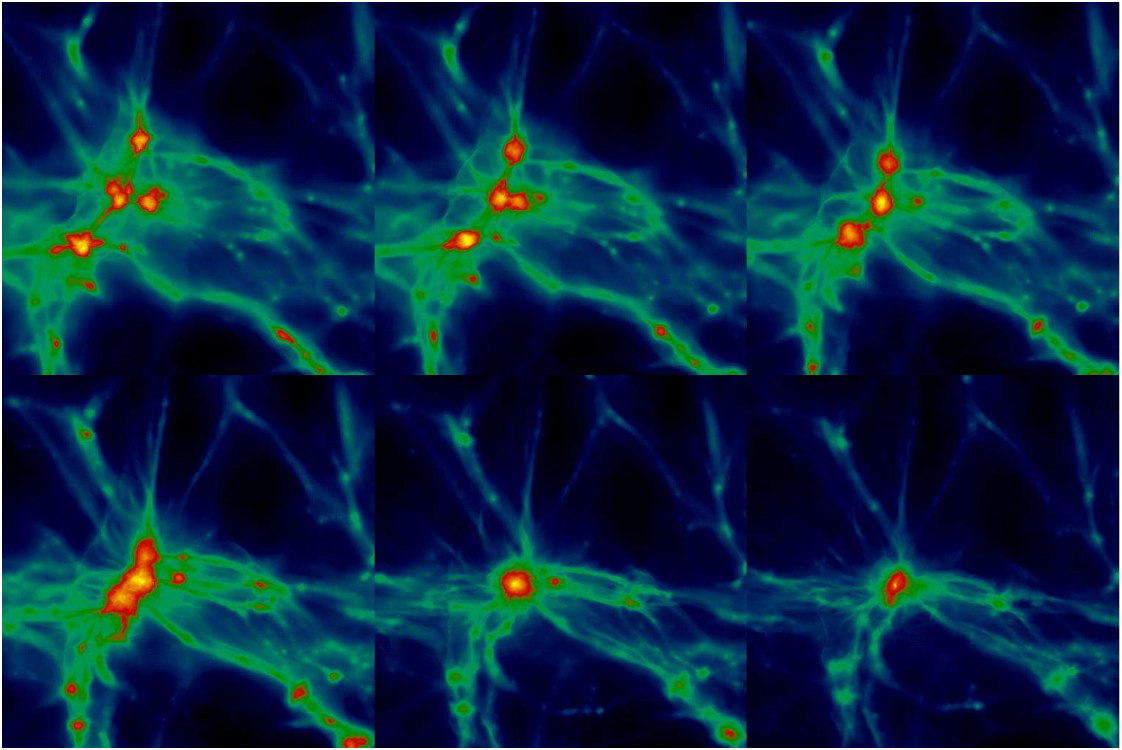}
\end{center}
\caption{
Merger history of the $M_{vir}=1.26 \times 10^7 \Ms$, T$_{vir}=10^4$~K halo that collapses 
at z=17.32 with a Lyman-Werner background flux $J_{21}=1$. In comic strip order, the redshifts are 
z=22, 21, 20, 19, 18, and 17.32. Field of view is 53.571 comoving kpc (2.678 proper 
kpc at z = 19). Logarithm of the projected baryon density (column density) is displayed, with color table scaled 
to the maximum and minimum values in each image.  
}
\label{fig.merger}
\end{figure}

However, something interesting happens in the J$_{21} = 1$ calculation 
at $t \approx 170$ Myr. Prior to that time the 
central temperature has crept up to 5000 K due to the increased virial mass. The cooling 
rate per \h2 molecule is about 3 orders of magnitude higher at 5000K than at 500K -- a 
typical temperature at the halo center in low UV background evolutions. We believe this high temperature 
is due to a combination of adiabatic heating as the potential well deepens, and merger-induced
 shock heating. As shown in Figure \ref{fig.merger}, three large halos merge between $z=20$ and $z=18$, 
quadrupling the halo mass (Figure \ref{fig.halovals1}, panel (b)). The elevated cooling rates 
cool the gas at the halo center to $T \approx 2000K$, allowing the gas density to increase slightly. 
The two effects reduce the central entropy from $S \approx$ 700 K cm$^2$ to $\approx$ 250 K cm$^2$. 
This is followed by another heating event at t=190 Myr, presumably due to another merger, 
followed by a second more catastrophic cooling event. At t=200 Myr, the halo core collapses, 
driving the central density higher, temperature and entropy lower, and \h2 fraction higher. 

\begin{figure}
\begin{center}
\includegraphics[width=0.9\textwidth]{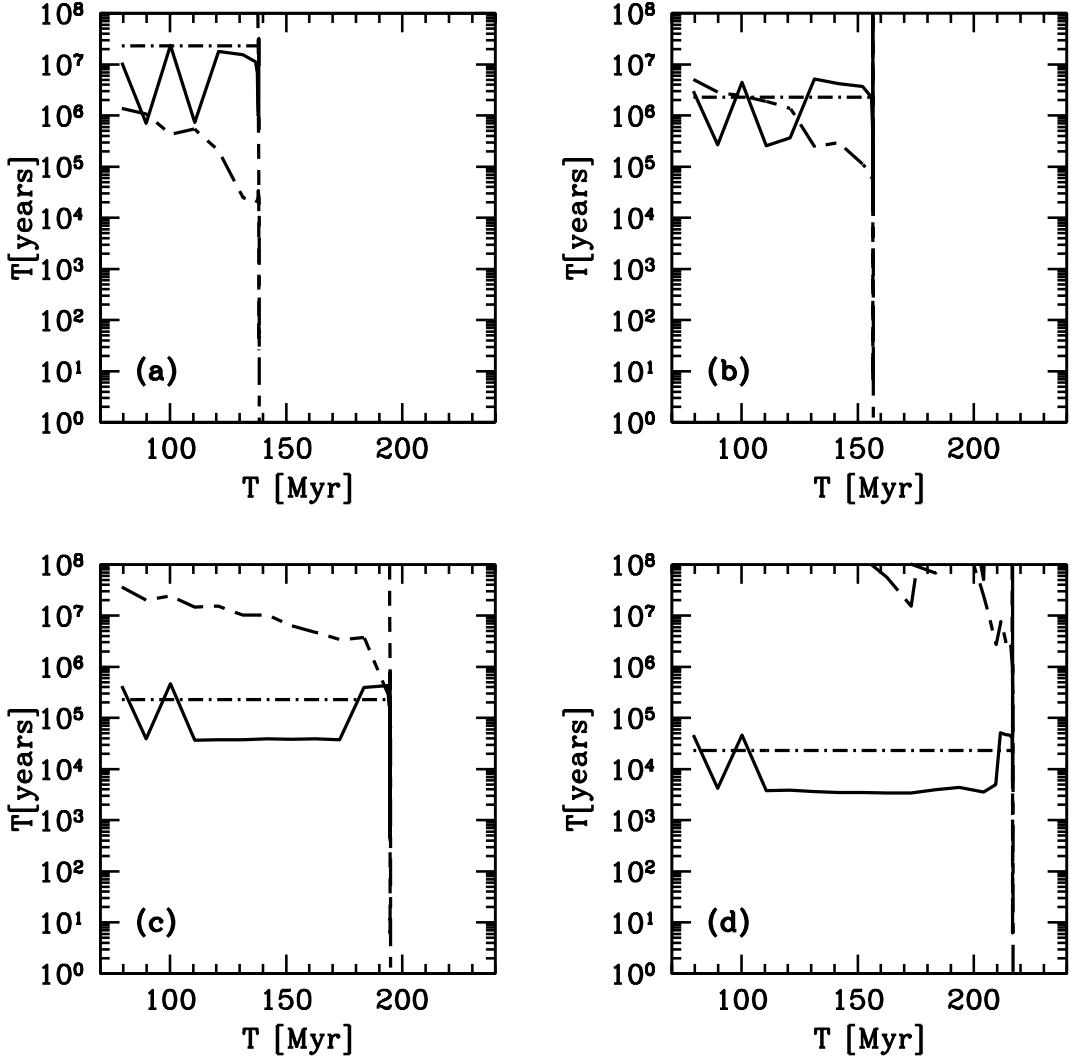}
\end{center}
\caption{
Evolution of timescales in the central zone for the cases $J_{21}=10^{-3}, 10^{-2}, 
10^{-1}$ and $1$ (Panels (a)--(d), respectively): 
2-body \h2 formation time (solid); 3-body \h2 formation time (short-dashed); 
\h2 collisional dissociate time (long-dashed); 
\h2 photodissociation time (dot-dashed); 
\h2 cooling 
time (short-long-dashed).  
}
\label{fig.timescales}
\end{figure}

Why does this collapse occur? This is analyzed in Figure \ref{fig.timescales}, where we plot 
the evolution of the important timescales in the center of the halo. In the J$_{21}=10^{-3}$ 
case, the 2-body \h2 formation timescale is always roughly an order of magnitude smaller than the 
photodissociation timescale. Consequently, the \h2 abundance is out of equilibrium and grows 
steadily with time (Figure \ref{fig.evolution} panel d). The \h2 cooling time in the central 
zone ($\leq 1$ Myr) is far less than the Hubble time ($\sim 100$ Myr), and before too long 
runaway collapse occurs. At higher values of J$_{LW}$, the 2-body \h2 formation time always 
hovers around the photodissociation timescale, regardless of how long that timescale is. This tells 
us that \h2 is in equilibrium: the 2-body formations balance the photo-destructions by the 
LW background. The equilibrium \h2 fraction is given by 

\begin{equation}
f_{H2} \approx \frac{k_{H^-}n_e}{k_{diss}}
\label{eqn-fh2}
\end{equation}

where k$_{H^-}$ is the rate coefficient for the formation of H$^-$ (the limiting reaction 
for H$_2$ formation at the temperature and density range considered here), $n_e$ is the 
electron density, and k$_{diss}=1.1 \times 10^8 F_{LW} s^{-1}$ is 
the \h2 photodissociation rate due to the Solomon process~\markcite{abel97,2003ApJ...592..645Y}({Abel} {et~al.} 1997; {Yoshida} {et~al.} 2003). 
Using T $=2000$~K and $n_e=10^{-4}$, we get $f_{H2}=1.46\times 
10^{-6}~\left( \frac{F_{LW}}{10^{-21}} \right)^{-1}$, which is in good agreement with 
the \h2 fractions we see in Figure \ref{fig.evolution}. Since 
$t_{H2}\equiv n_{H2}$/ \.{n}$_{H2} \propto f_{H2}$, the smaller $f_{H2}$ is, the shorter 
its formation time. 

Now let us consider the \h2 cooling time evolution (short-long-dashed line). It is far lass than a 
Hubble time for J$_{21} \leq 10^{-1}$, ensuring that these halo cores will eventually 
cool and collapse. In the J$_{21}=1$ case, the cooling time drops below the Hubble time for the 
first time at t=170 Myr, and again at t=190 Myr where it begins a steady decrease toward 
$10^6$ yr. Despite the low equilibrium \h2 fraction, if we wait long enough the inexorable 
press of time eventually establishes the condition $t_{cool} < t_{Hubble}$ and we get 
runaway collapse, even for J$_{21}=1$. Note that we get catastrophic cooling because we are 
in the density regime ($n<10^4$ cm$^{-3}$) where \h2 cooling is proportional to the square 
of the gas density, not linear in the gas density~\markcite{2000ApJ...540...39A,ABN02}({Abel}, {Bryan}, \&  {Norman} 2000; {Abel} {et~al.} 2002).

\clearpage
%%%%%%%%%%%%%%%%%%% FIGURE %%%%%%%%%%%%%%
%%%%%%%%%%%%%%%%%%%%%%%%%%%%%%%%%%%%%%%%%

%%%%%%%%%%%%%%%%%%% FIGURE %%%%%%%%%%%%%%
%%%%%%%%%%%%%%%%%%%%%%%%%%%%%%%%%%%%%%%%%

%%%%%%%%%%%%%%%%%%% FIGURE %%%%%%%%%%%%%%
%%%%%%%%%%%%%%%%%%%%%%%%%%%%%%%%%%%%%%%%%
\clearpage

\section{Halo properties at the epoch of collapse}\label{results.collapse}

\begin{figure}
\begin{center}
\includegraphics[width=0.3\textwidth]{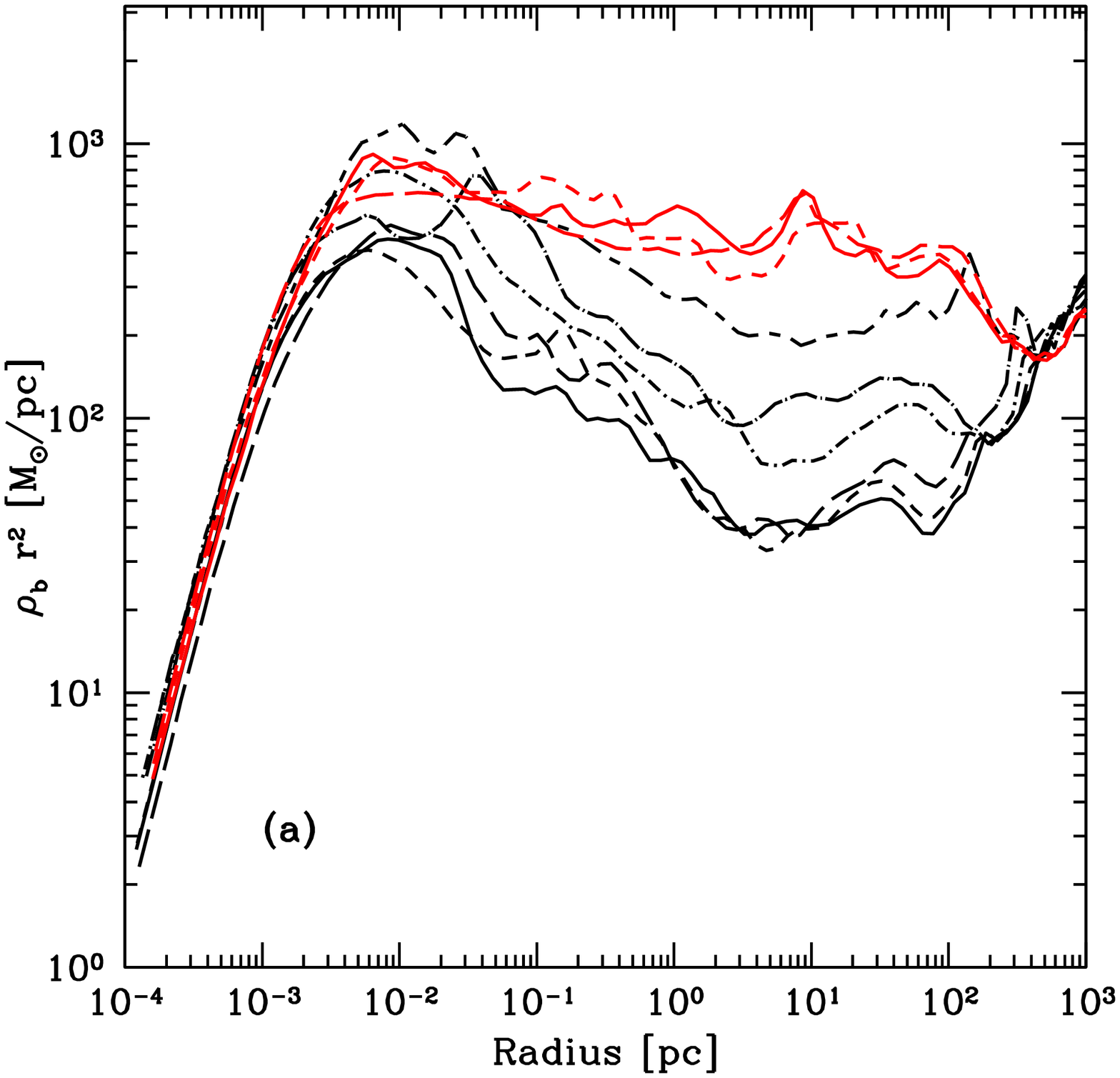}
\includegraphics[width=0.3\textwidth]{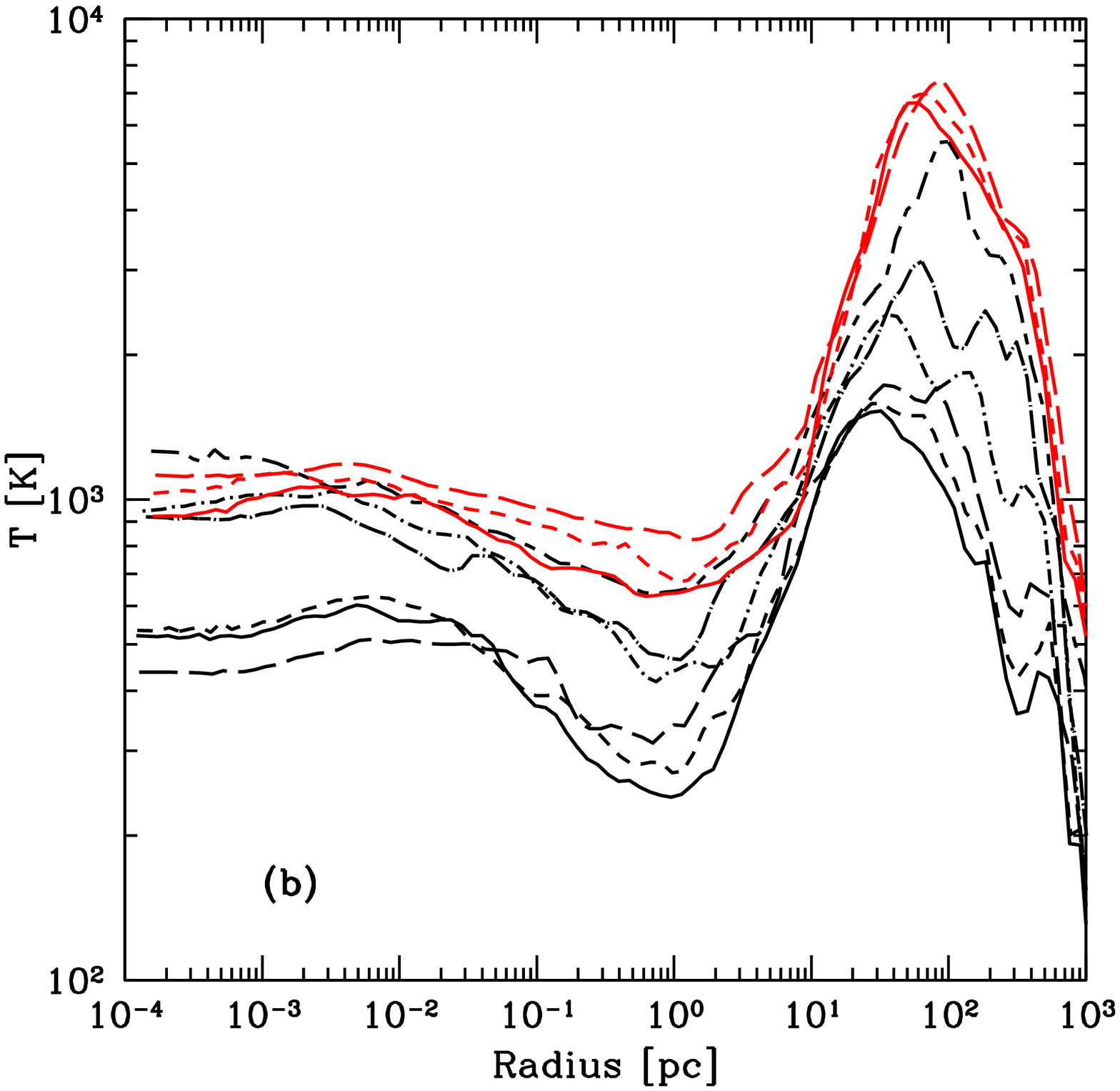}
\includegraphics[width=0.3\textwidth]{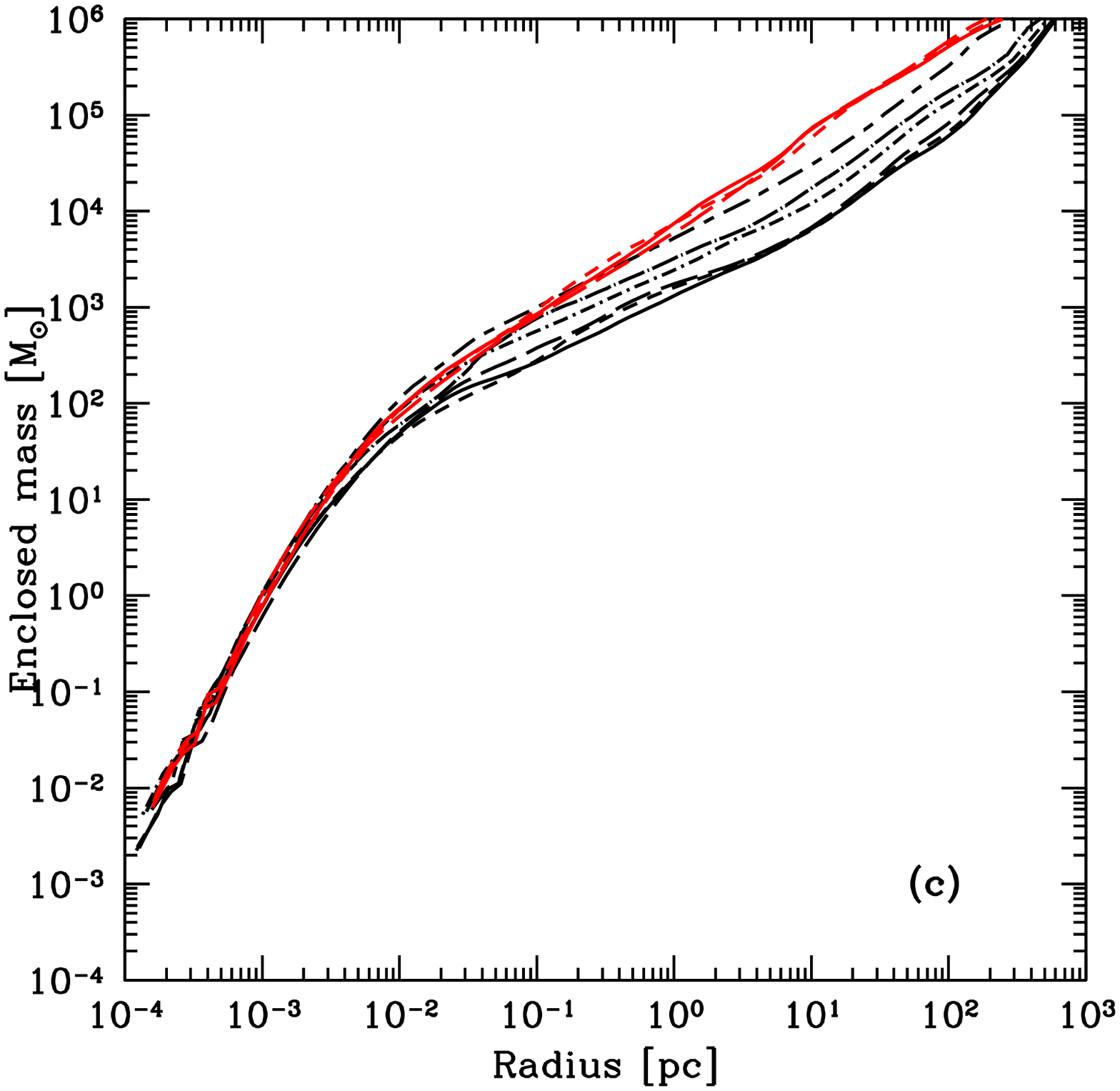}
\includegraphics[width=0.3\textwidth]{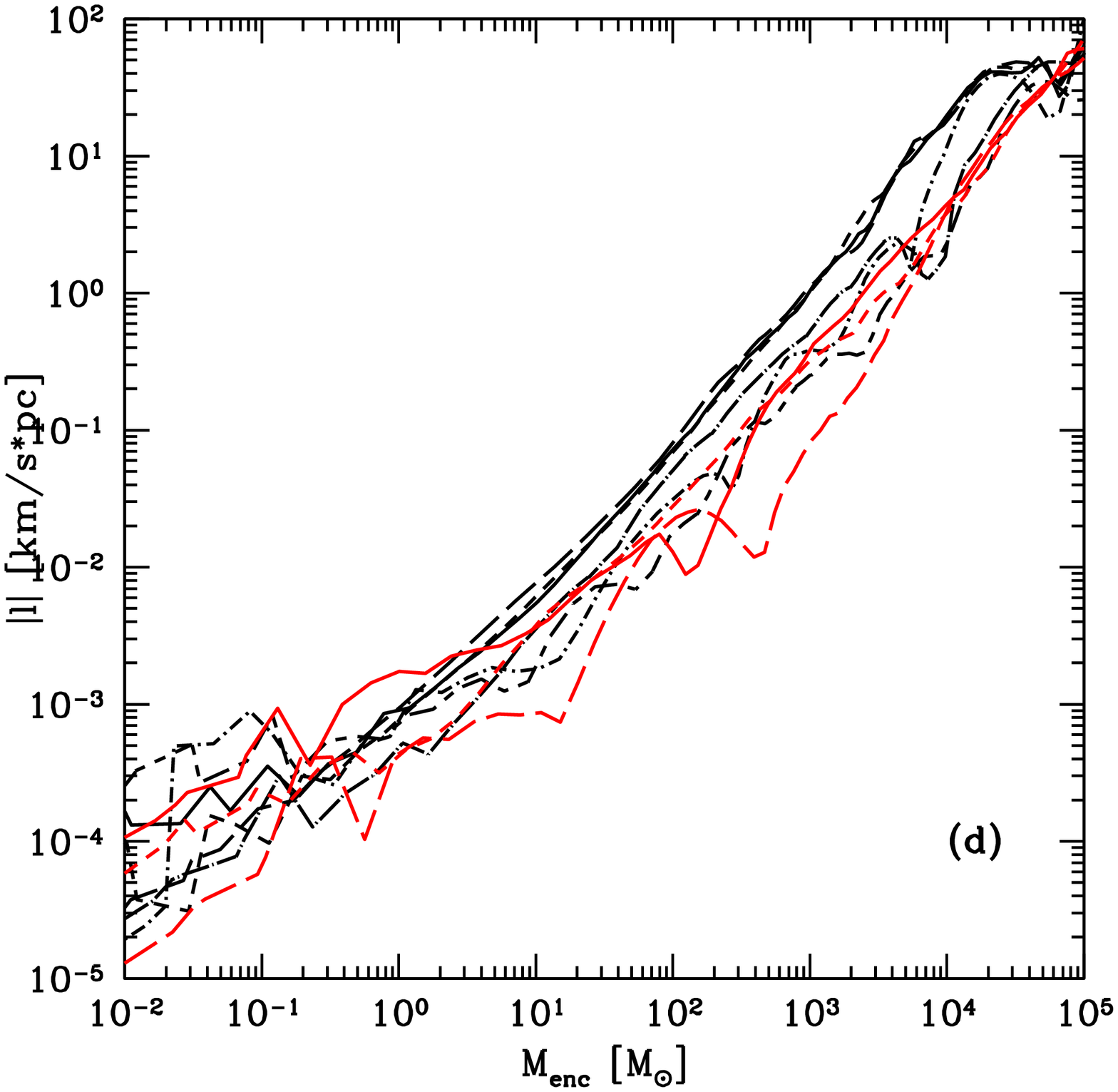}
\includegraphics[width=0.3\textwidth]{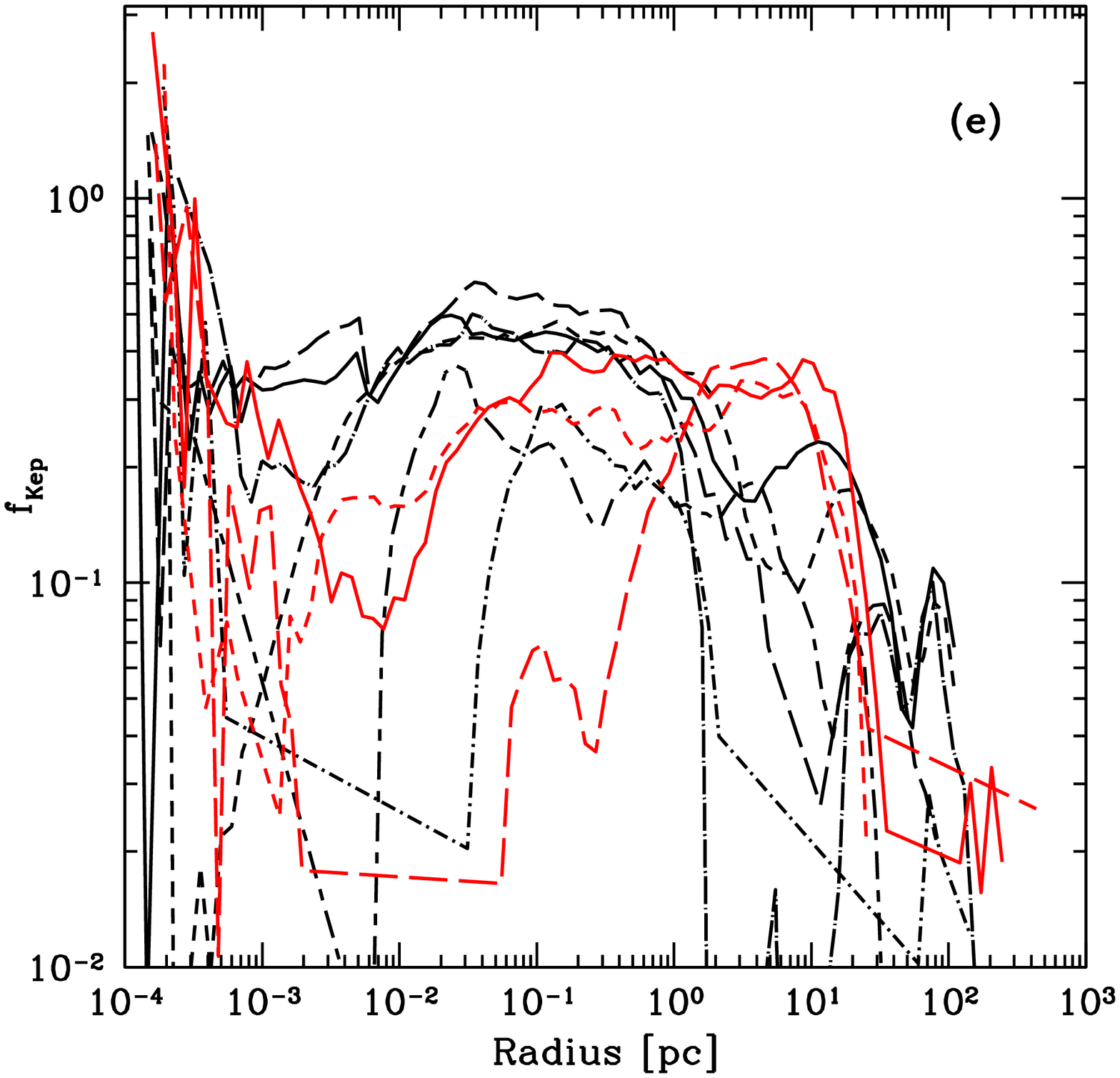}
\includegraphics[width=0.3\textwidth]{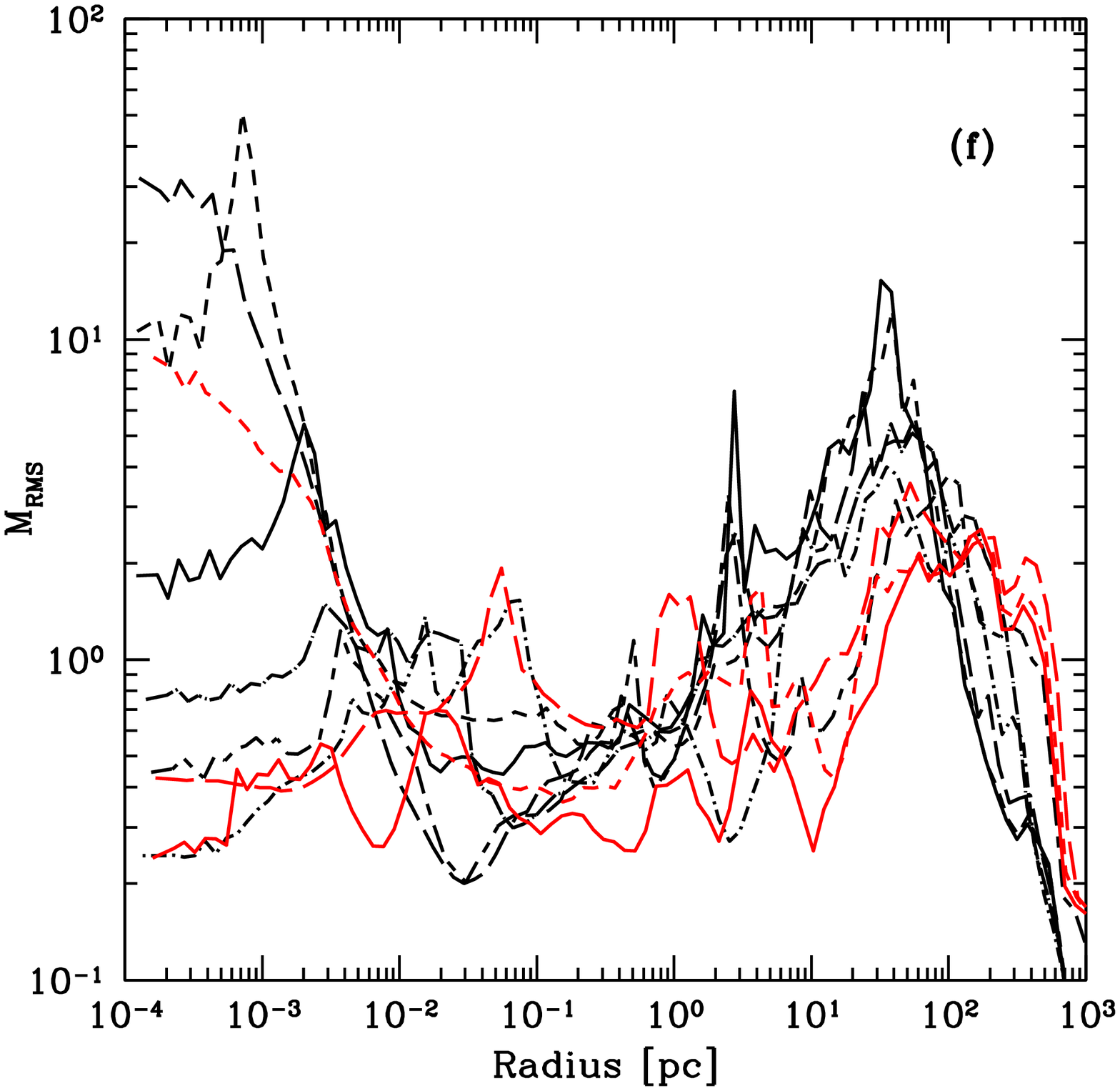}
\end{center}
\caption{
Evolution of several spherically-averaged baryon quantities for simulations
with all values of J$_{21}$.
Panel (a):  $\rho_b r^2$ as a function of radius.
Panel (b):  Temperature as a function of radius.
Panel (c):  Enclosed gas mass as a function of radius.
Panel (d):  Specific angular momentum as a function of enclosed mass.
Panel (e):  Keplerian velocity fraction as a function of radius.
Panel (f):  Gas RMS Mach number as a function of radius.
All quantities except enclosed gas mass are mass-weighted, and
all simulations are shown at the point where the maximum
number density is approximately $10^{10}$~cm$^{-3}$.
Line types and weights correspond to simulations, as follows.
Black solid line: J$_{21} = 0$.
Black short-dashed line: J$_{21} = 10^{-3}$.
Black long-dashed line: J$_{21} = 10^{-2.5}$.
Black dot short-dashed line: J$_{21} = 10^{-2}$.
Black dot long-dashed line: J$_{21} = 10^{-1.5}$.
Black short dashed-long dashed line: J$_{21} = 10^{-1}$.
Red solid line: J$_{21} = 10^{-0.67}$.
Red short-dashed line: J$_{21} = 10^{-0.33}$.
Red long-dashed line: J$_{21} = 1$.
}
\label{fig.radprof_zcoll.1}
\end{figure}

Figure~\ref{fig.radprof_zcoll.1} shows radial profiles of baryon number density, temperature, 
enclosed mass, circular velocity, and RMS Mach number as a function of radius, as well as 
specific angular momentum as a function of enclosed mass, for all simulations discussed in 
this paper.  To facilitate comparison, results from the output of each simulation where the peak
baryon number density at the center of the halo is approximately $10^{10}$ cm$^{-3}$ are shown, in 
order to capture each halo at a similar evolutionary stage, rather than at a fixed
point in time.  Panel (a) shows that all of the simulations have similar density profiles.  
The scatter in number density at a given radius is readily explained by variation in halo 
mass -- halos which collapse at later times (due to higher UV background strength) are more 
massive, and thus have higher overall baryon densities.  This is shown in another way in 
Panel (c), in a  plot of enclosed mass as a function of radius which shows that all halos 
have very similar profiles at $r \la 10^{-2}$ pc, but a significant variation at larger 
radii which is related to halo mass.  The plot of temperature as a function of radius in 
Panel (b) shows an interesting trend -- as the LW background is increased, the overall halo 
temperature as well as the halo core (where ``core'' is roughly defined as gas within 
$\sim 1$ pc of the halo center) temperature go up.  In the outskirts of the halo, where 
baryon densities are low and thus cooling times are long, this is due primarily to the 
increase in halo mass, with the peak temperature corresponding approximately to the halo 
virial temperature.  In the halo core, however, this temperature is correlated more strongly 
with the H$_2$ fraction, and thus the UV background, as suggested by 
Figure~\ref{fig.halovals2}.  The plots of specific angular momentum as a function of enclosed 
mass and Keplerian velocity fraction (defined as the local circular
velocity divided by the Keplerian orbital velocity due to mass within that radius) 
as a function of radius in panels (d) and (e) show some mild trends 
trends.  It appears that halos in simulations with stronger UV backgrounds tend to have 
less specific angular momentum at a given mass shell, and also tend to have a lower
Keplerian velocity fraction at a given radius than gas in halos which form in the 
presence of a lower UV background -- indeed, the halo which forms in the J$_{21} = 1$
calculation has the least angular momentum of all of the simulations, and the lowest
Keplerian velocity fraction out to $\sim 1$~pc.  We speculate that this is due to
transport and segregation of angular momentum by turbulence -- halos whose collapse
is delayed have more time for these processes to act within the halo core, resulting 
in less angular momentum overall when the halo ultimately collapses (note that this will
be discussed more fully in a later paper).
The plot of RMS Mach number as a function of radius
in panel (f) also 
shows no obvious trend, though at radii below $\sim 10^{-2}$ pc, simulations with a smaller 
UV background generally have higher RMS Mach numbers.  This is due primarily to the gas 
being colder, and not the gas velocities being higher.

\clearpage
%%%%%%%%%%%%%%%%%%% FIGURE %%%%%%%%%%%%%%
%%%%%%%%%%%%%%%%%%%%%%%%%%%%%%%%%%%%%%%%%
\clearpage

\begin{figure}
\begin{center}
\includegraphics[width=0.3\textwidth]{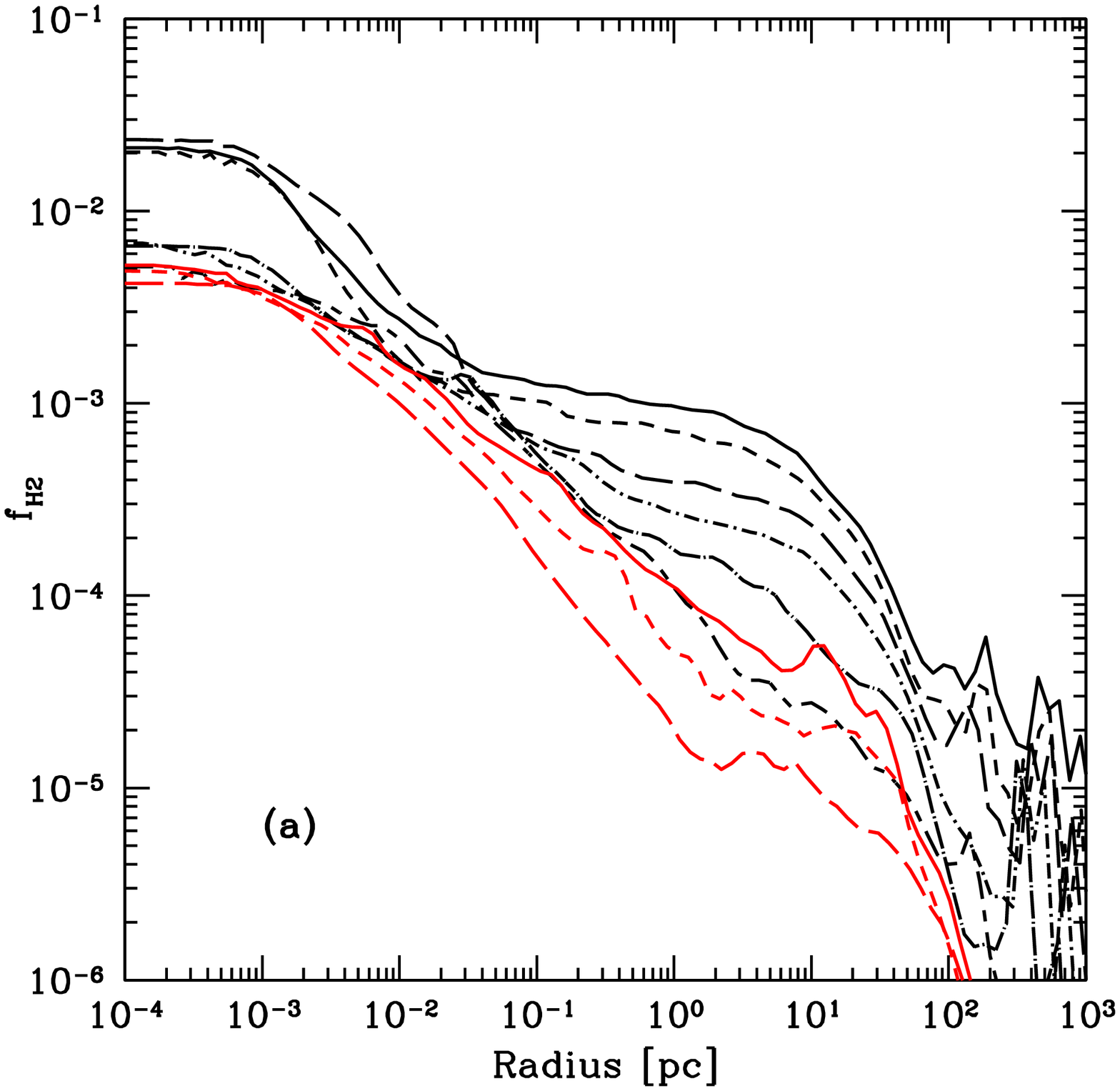}
\includegraphics[width=0.3\textwidth]{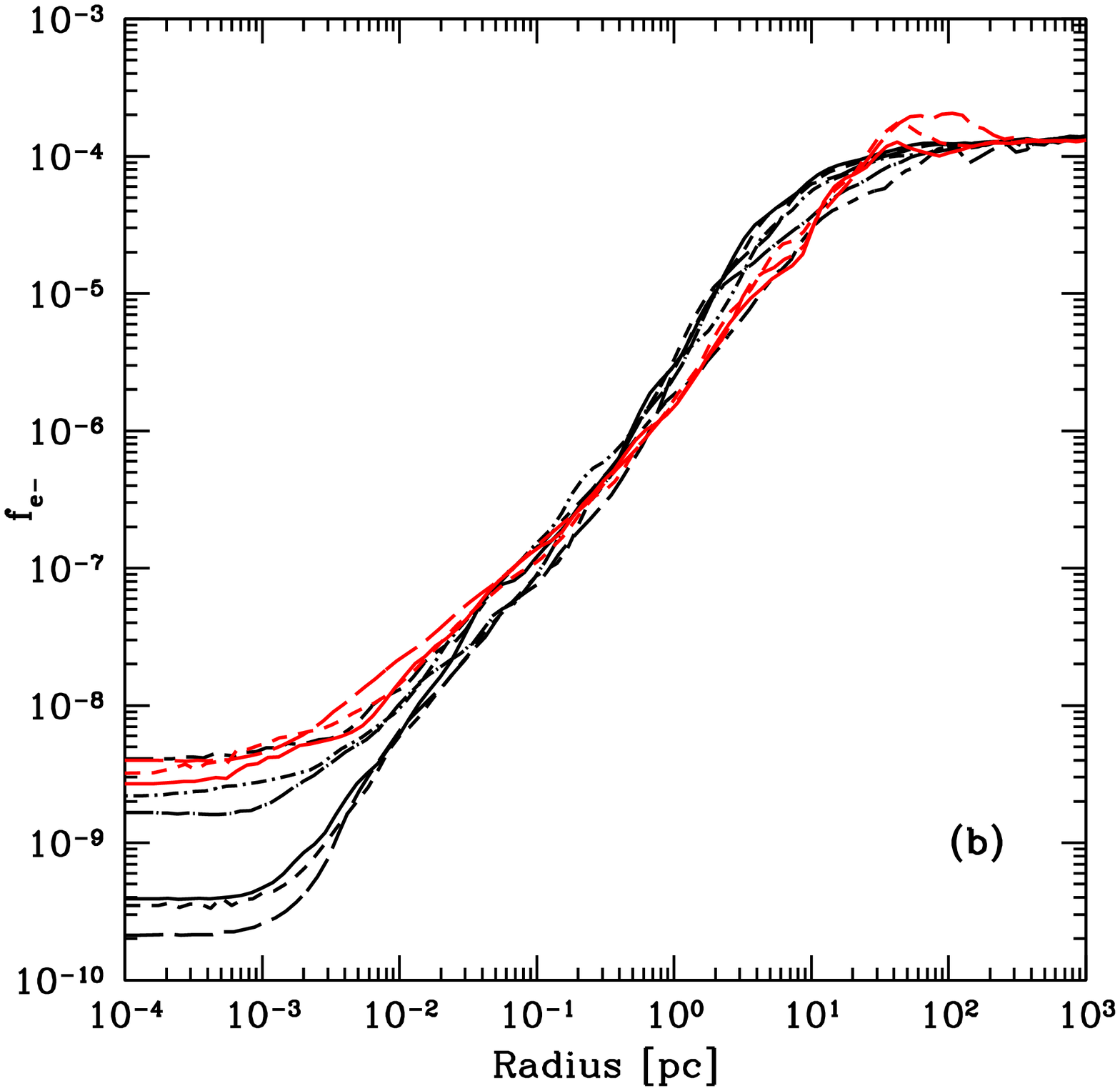}
\includegraphics[width=0.3\textwidth]{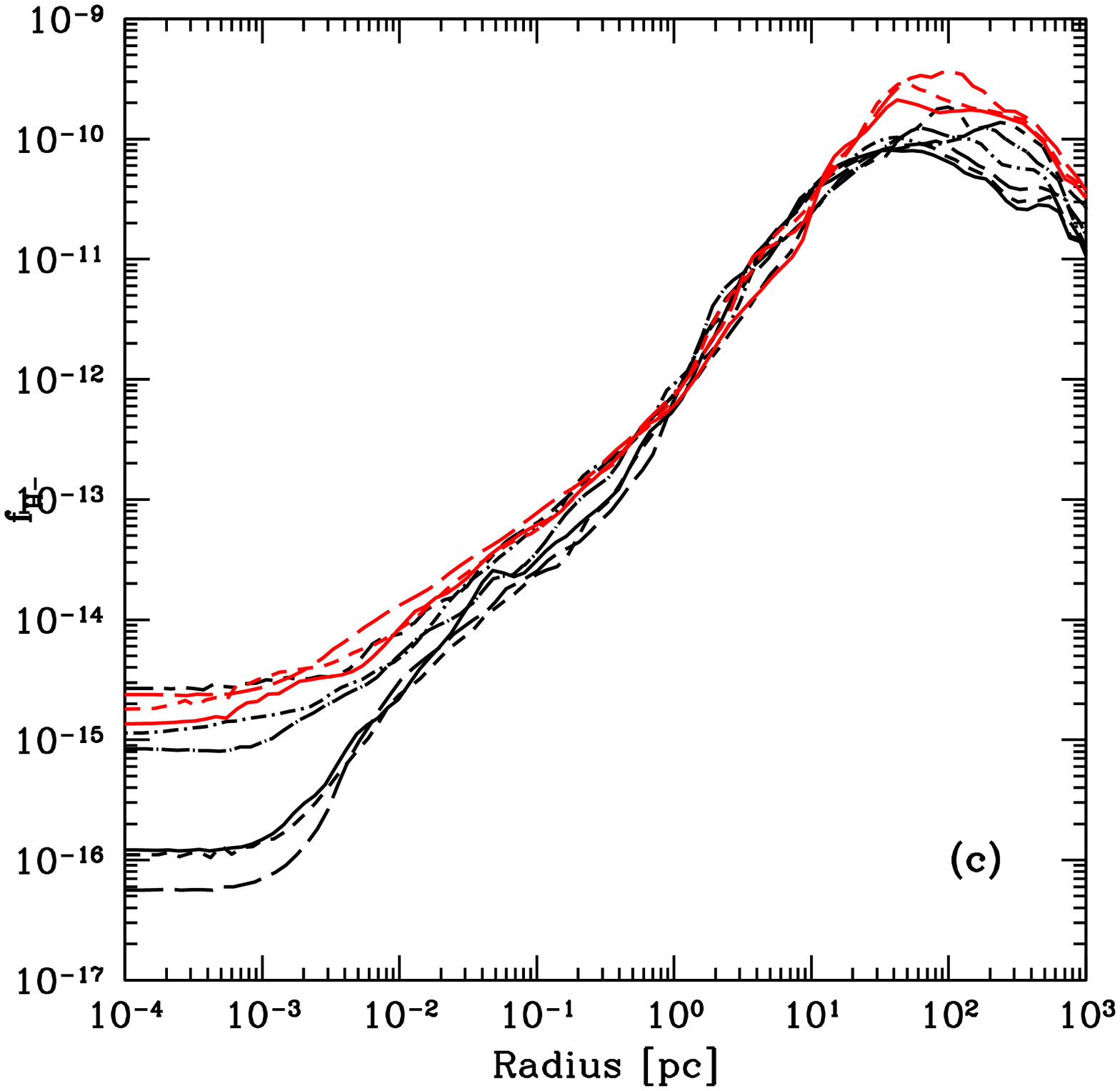}
\includegraphics[width=0.3\textwidth]{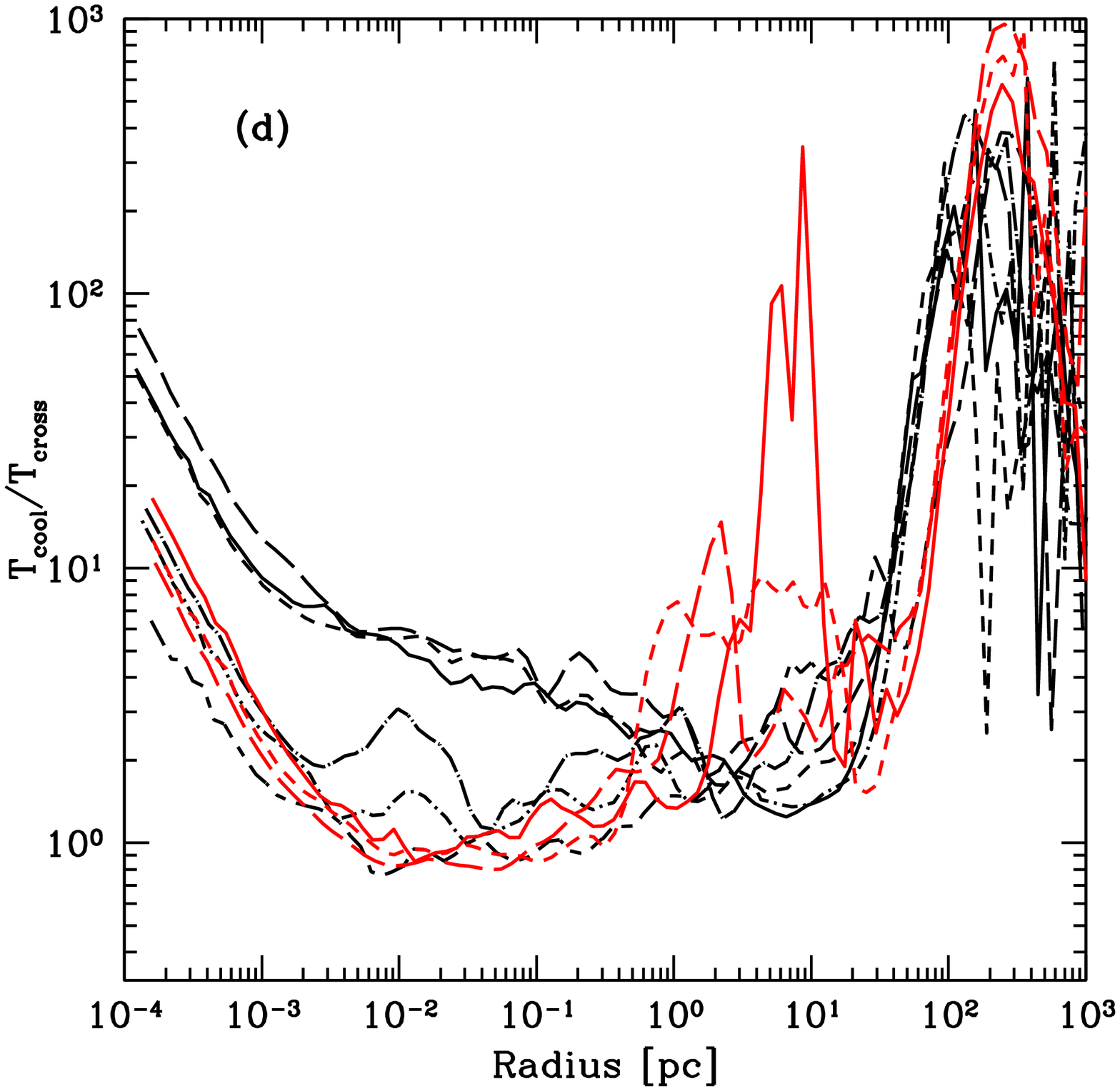}
\includegraphics[width=0.3\textwidth]{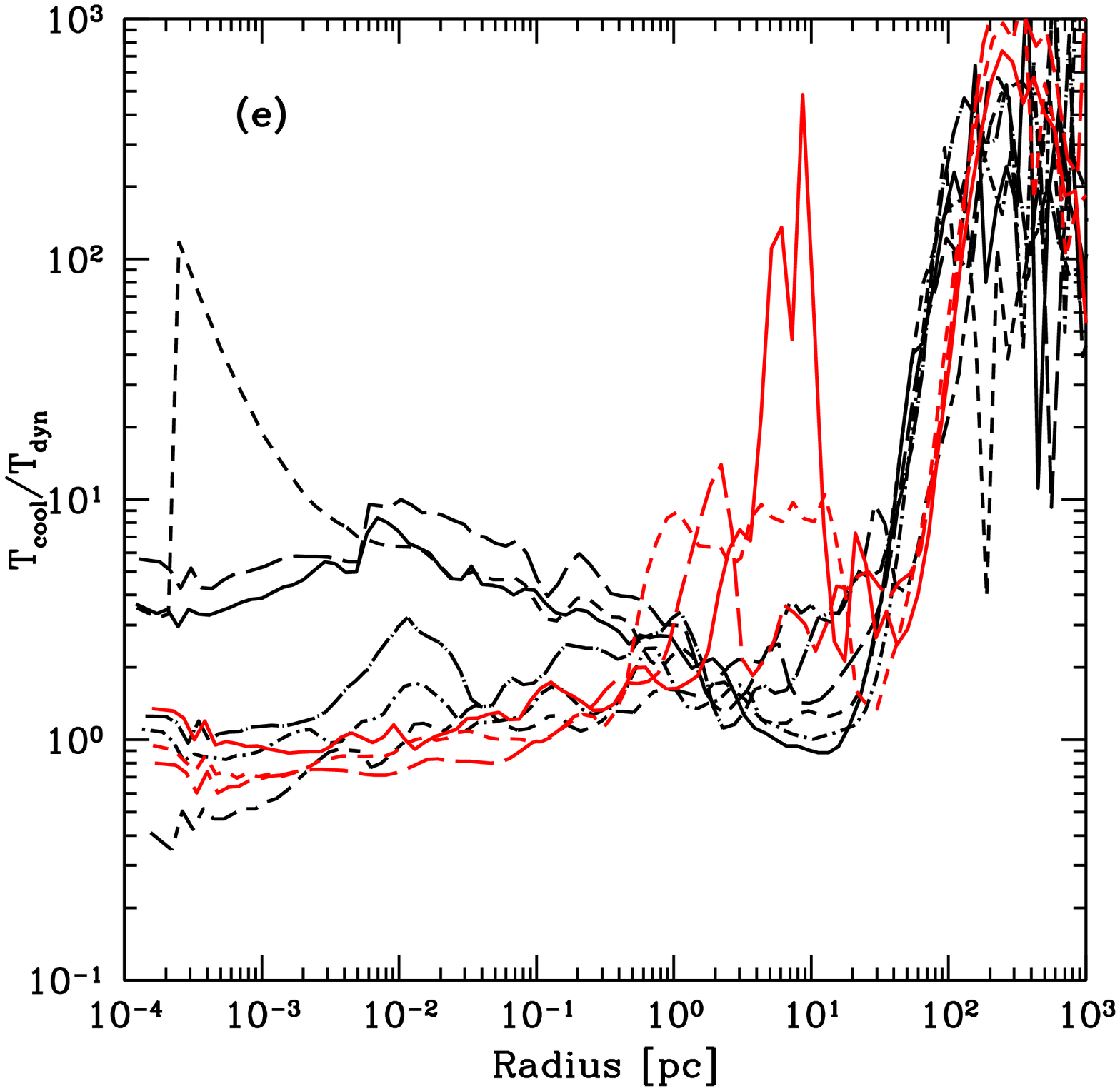}
\includegraphics[width=0.3\textwidth]{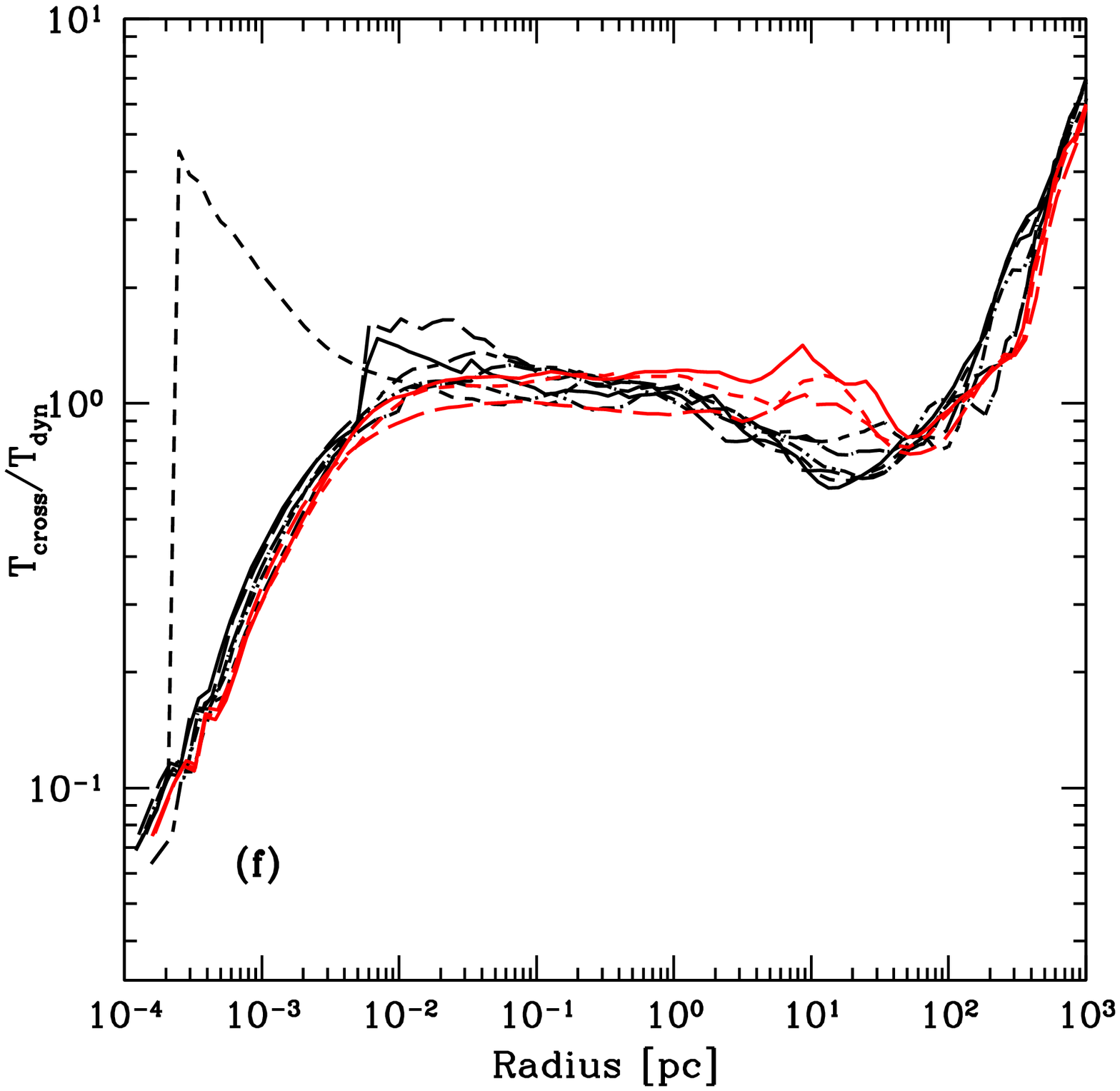}
\end{center}
\caption{
Several spherically-averaged baryon quantities for simulations
with all values of J$_{21}$.
Panel (a):  Molecular hydrogen fraction as a function of radius.
Panel (b):  Electron fraction as a function of radius.
Panel (c):  $H^-$ fraction as a function of radius.
Panel (d):  Ratio of gas cooling time to sound crossing time as a function of radius.
Panel (e):  Ratio of gas cooling time to dynamical time as a function of radius.
Panel (f):  Ratio of gas sound crossing time to dynamical time as a function of radius.
All quantities except enclosed gas mass are mass-weighted, and
all simulations are shown at the point where the maximum
number density is approximately $10^{10}$~cm$^{-3}$.
Line types and weights correspond to those in Figure~\ref{fig.radprof_zcoll.1}.
}
\label{fig.radprof_zcoll.2}
\end{figure}

Figure~\ref{fig.radprof_zcoll.2} shows the H$_2$ fraction, electron fraction, H$^-$ fraction, 
ratio of cooling time to sound crossing time, ratio of cooling time to dynamical time, and ratio
of sound crossing time to dynamical time as a function of radius, for all
simulations.  Outputs and line types correspond to those in Figure~\ref{fig.radprof_zcoll.1}.
There is a clear relationship at all radii between H$_2$ fraction and UV background strength 
in panel (a) -- as
the FUV background is increased, the overall H$_2$ fraction decreases.  This difference
is most noticeable at radii of $0.1 - 10$ pc, but is maintained at larger and smaller radii.
There is a ``kink'' at approximately $10^{-2}$ pc where the values are all quite similar.  This
corresponds to a baryon number density of $\sim 10^8$~cm$^{-3}$ in all simulations, which is
where 3-body H$_2$ formation begins to occur.  Panels (b) and (c) show that
  e$^-$ and $H^-$ fractions track each other,
which is to be expected -- the local electron fraction controls the amount of H$^-$ which
can be produced, which is the limiting reaction in the dominant mode of H$_2$ formation for
n $\la 10^8$ cm$^{-3}$ -- and there is a general trend with increasing electron and H$^-$ fractions with
increasing UV background flux.  

The plots of the ratio of cooling time to sound crossing time, ratio of cooling time to 
dynamical time, and ratio of sound crossing time to dynamical time as a function of radius
shown in panels (d), (e) and (f)  
display some interesting trends.  The three simulations with the lowest UV background strengths 
have a much longer cooling time than the other calculations, which are all grouped roughly 
together with no discernible pattern.  This agrees well with the plot of the ratio of cooling
time to dynamical time, where a similar trend is observed.  The difference between the three 
simulations with the lowest UV background strengths and the others is due to the somewhat higher 
H$_2$ fraction in the halo core of these calculations.  The gas temperature is $\sim 200$~K at 
the center of the halo cores in these calculations, and the gas cannot cool further, which results 
in a very long cooling time.  The sound crossing time is also increased, but not as significantly.  
All simulations have approximately the same ratio of sound crossing to dynamical times at all radii.  
Given that the cooling time is typically longer than both the sound crossing time and dynamical time,
one can infer that the collapse of the halo is occurring quasi-statically for all simulations.

\clearpage
%%%%%%%%%%%%%%%%%%% FIGURE %%%%%%%%%%%%%%
%%%%%%%%%%%%%%%%%%%%%%%%%%%%%%%%%%%%%%%%%
\clearpage

\begin{figure}
\begin{center}
\includegraphics[width=0.45\textwidth]{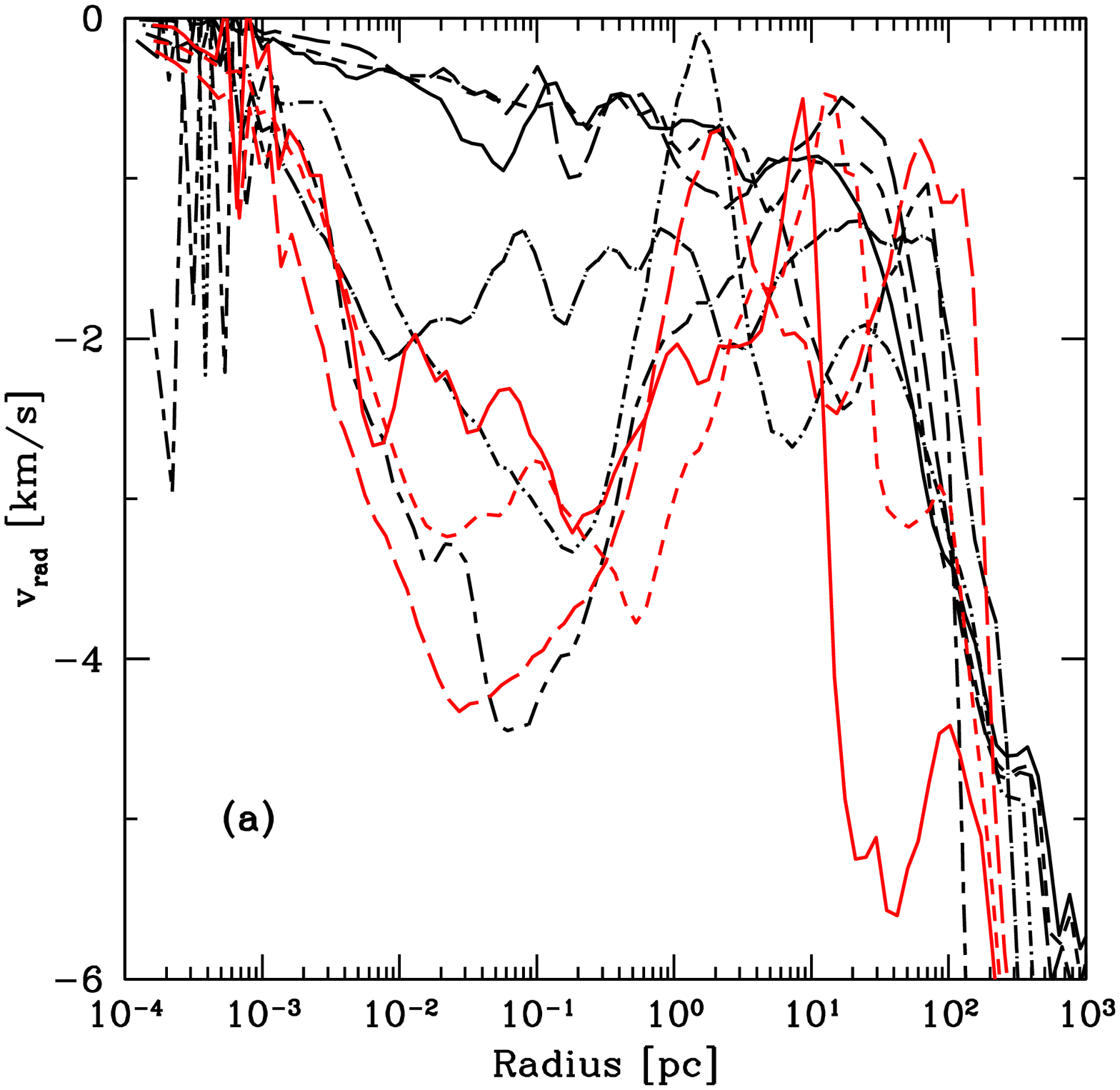}
\includegraphics[width=0.45\textwidth]{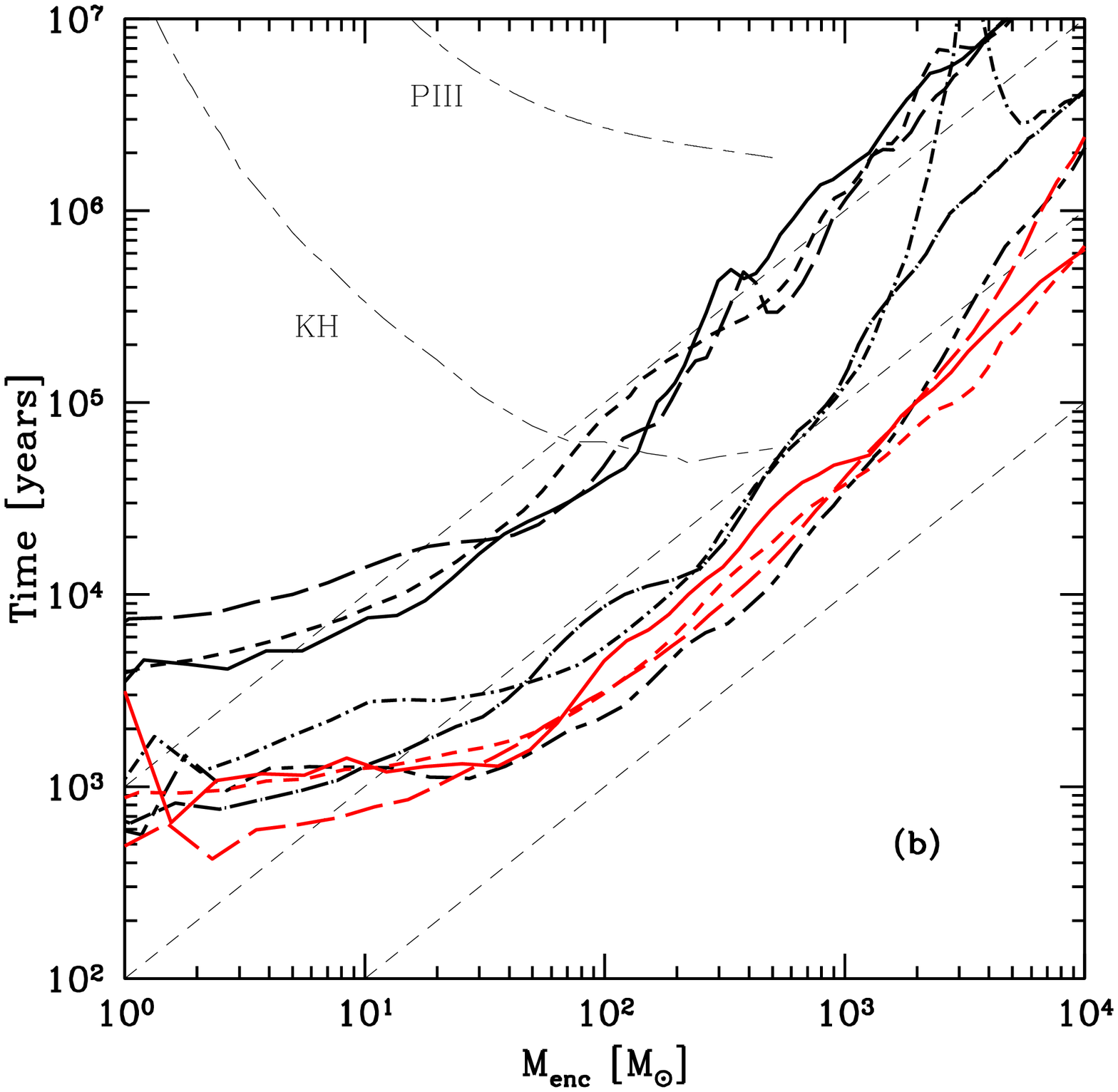}
\includegraphics[width=0.45\textwidth]{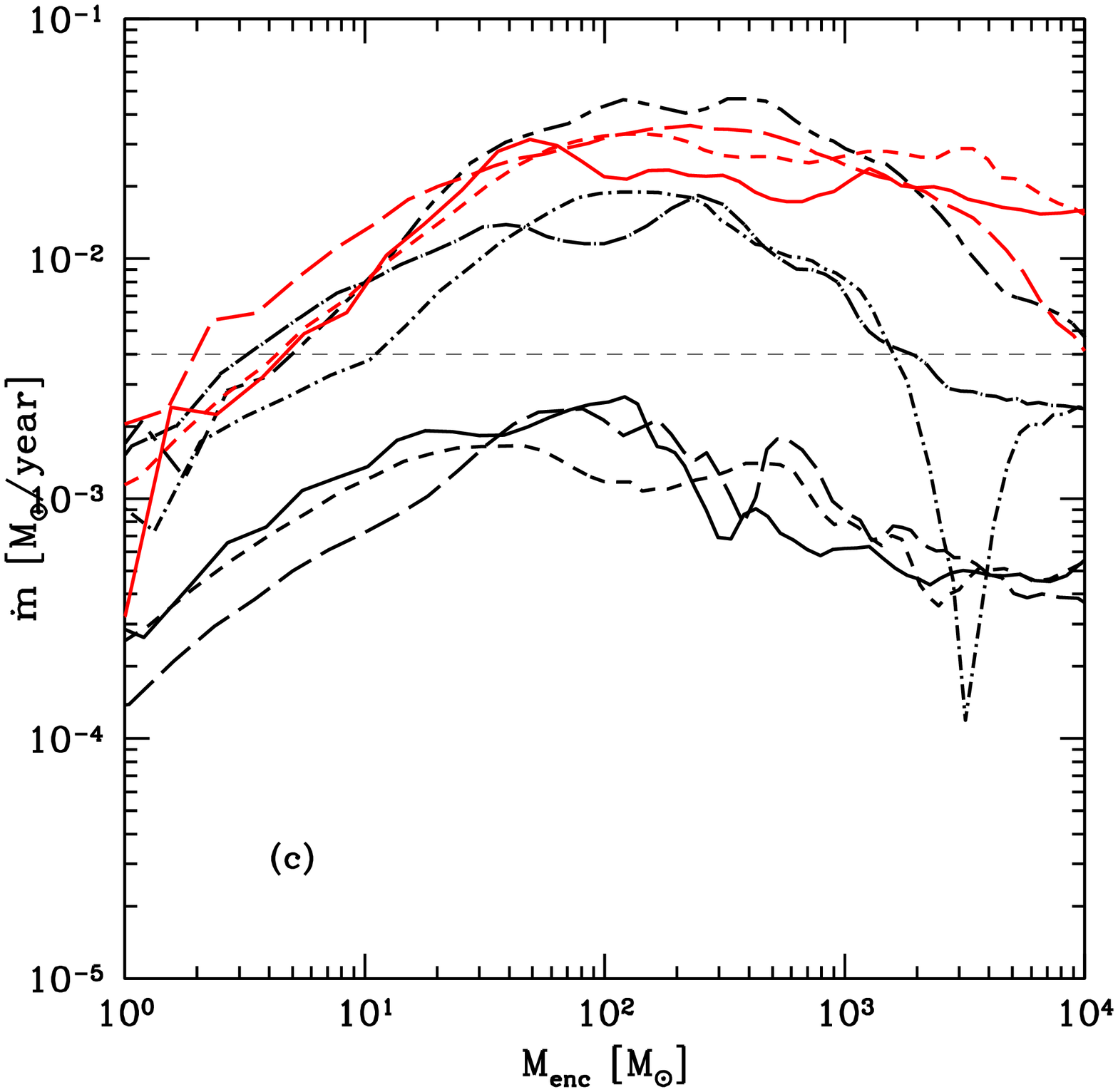}
\includegraphics[width=0.45\textwidth]{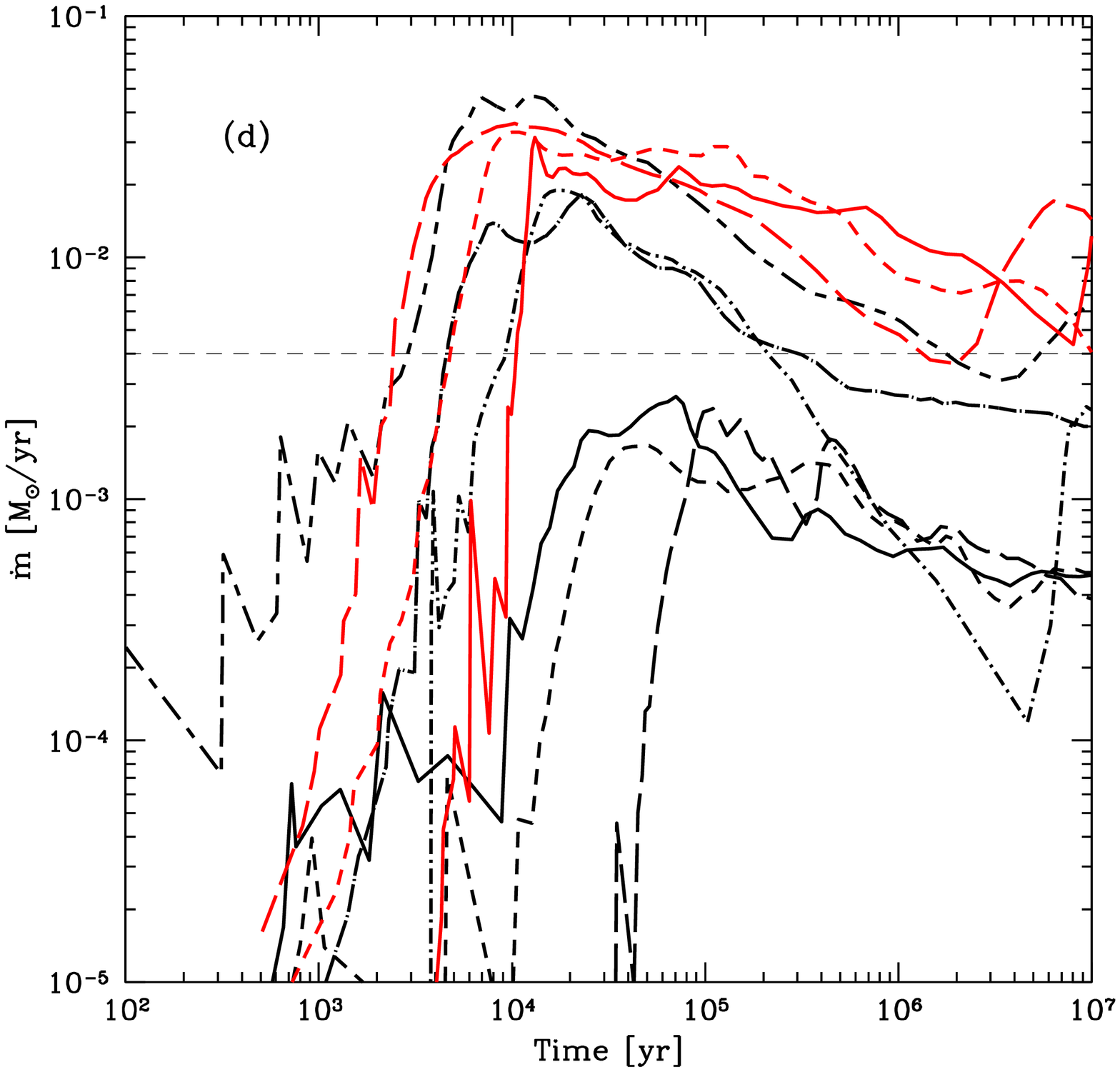}
\end{center}
\caption{
Several spherically-averaged baryon quantities for simulations
with all values of J$_{21}$.
Panel (a):  Radial velocity as a function of radius.
Panel (b):  Accretion time as a function of enclosed baryon mass.
Panel (c):  Instantaneous accretion rate as a function of enclosed baryon mass.
Panel (d):  Estimated accretion rate as a function of time.
All quantities except enclosed gas mass are mass-weighted, and
all simulations are shown at the point where the maximum
number density is approximately $10^{10}$~cm$^{-3}$.
Line types and weights correspond to those in Figure~\ref{fig.radprof_zcoll.1}.
The upper and lower light short-long-dashed curves which extend from the
upper left corner of Panel (b) correspond to 
the main sequence lifetime of a massive Population III star of that mass and
the Kelvin-Helmholtz timescale of a Population III timescale with a given 
luminosity and
radius.  All values are taken from~\markcite{2002A&A...382...28S}{Schaerer} (2002).
The three light diagonal short-dashed lines which extend
from bottom left to top right in panel (b) correspond to masses
accreted using constant accretion rates of (from top to bottom)
$\dot{m} = 10^{-3}$, $10^{-2}$, and $10^{-1}$ M$_\odot/$yr.
The light horizontal short-dashed line in panels (c) and 
(d) correspond to the ``critical'' accretion rate of~\markcite{2003ApJ...589..677O}{Omukai} \& {Palla} (2003),
$\dot{m} \simeq 4 \times 10^{-3}$~M$_\odot/$yr.
}
\label{fig.radprof_zcoll.3}
\end{figure}

Figure~\ref{fig.radprof_zcoll.3} shows radial velocity as a function of radius,
accretion time (defined as $M_{b,enc}/\dot{M}$) as a function of enclosed baryon mass
$\dot{M}$ as a function of enclosed baryon mass and $\dot{M}$ as a function of time
for all simulations.  Output time and lines correspond to those in 
Figures~\ref{fig.radprof_zcoll.1} and~\ref{fig.radprof_zcoll.2}.
Panel (a) shows a clear relationship between the strength of the soft UV background
and the radial velocity, with the calculations that have higher UV backgrounds
also having higher infall velocities.  This is easily understood by the quasistatic
contraction of the gas, which takes place at or below the local sound speed in the gas, 
as discussed by \markcite{ABN02}{Abel} {et~al.} (2002) and \markcite{oshea07a}{O'Shea} \& {Norman} (2007).  Given that the halo temperatures
are systematically higher in calculations where the UV background is stronger, the sound
speed is higher, and thus the rate at which the gas at the center of the halo contracts is higher.  This leads to 
lower accretion times (Panel (b)) and higher accretion rates (Panel (c)), with the accretion 
rates for the simulations with the strongest UV backgrounds being higher than those with 
a low (or no) UV background.  Panel (d) shows that the time evolution of the accretion 
rate in all simulations is qualitatively similar, but varies strongly in absolute magnitude 
and in the time at which the accretion rate onto the evolving protostellar cloud peaks, 
suggesting that the stars forming in these simulations may have very different evolutionary 
histories.

\clearpage
%%%%%%%%%%%%%%%%%%% FIGURE %%%%%%%%%%%%%%
%%%%%%%%%%%%%%%%%%%%%%%%%%%%%%%%%%%%%%%%%
\clearpage

\section{Halo properties at fixed redshift}\label{results.fixred}

\begin{figure}
\begin{center}
\includegraphics[width=0.45\textwidth]{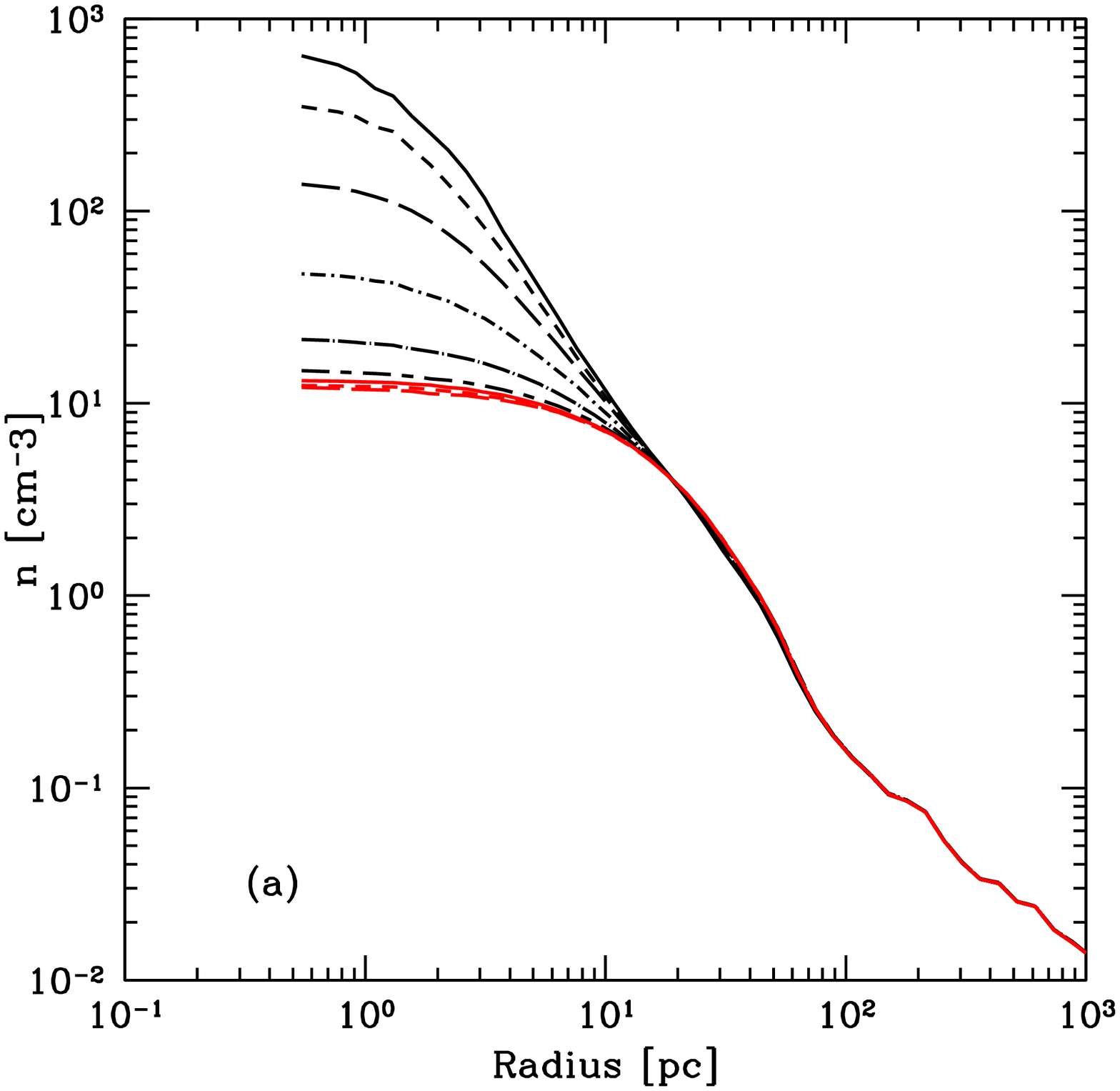}
\includegraphics[width=0.45\textwidth]{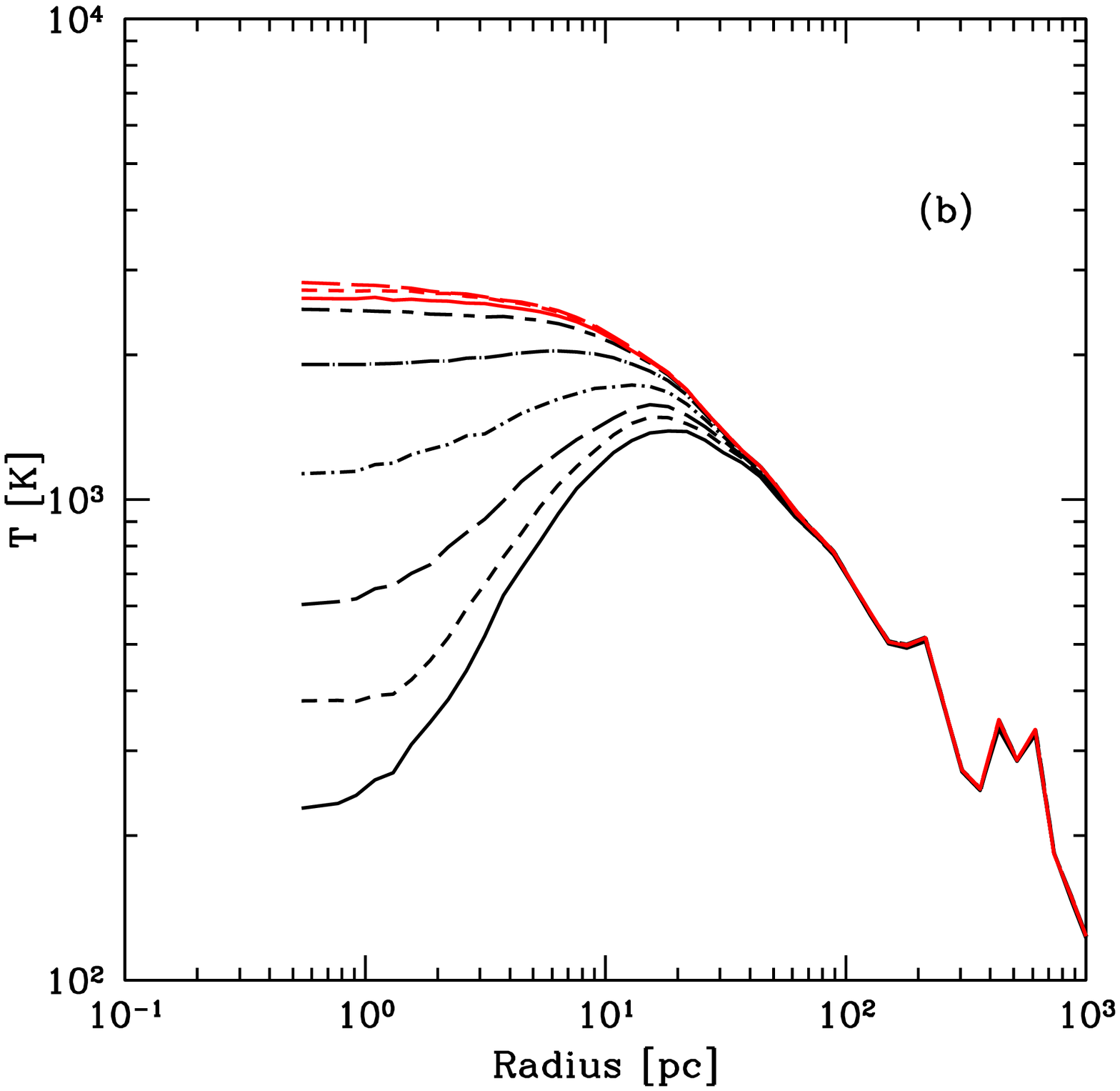}
\includegraphics[width=0.45\textwidth]{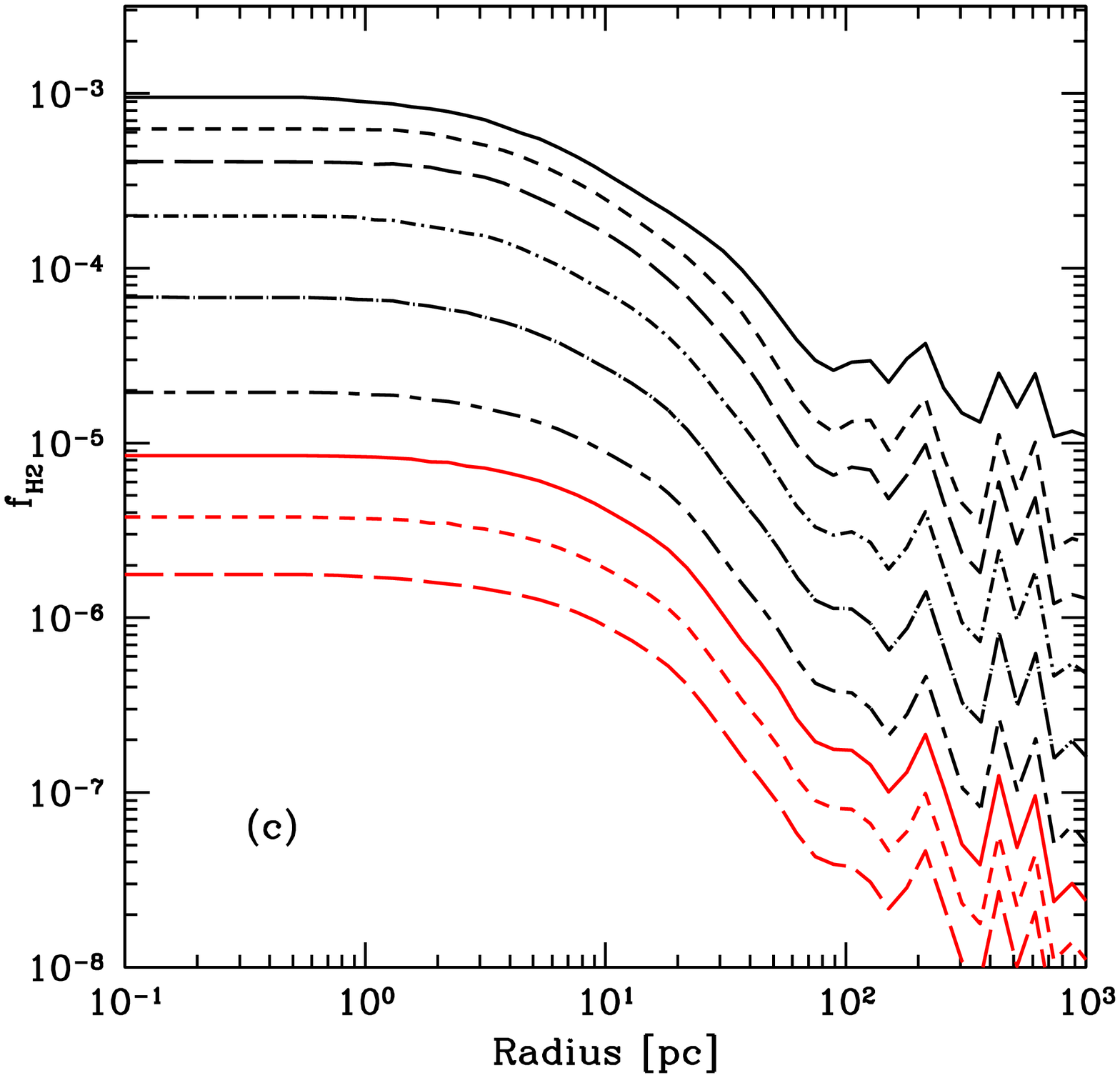}
\includegraphics[width=0.45\textwidth]{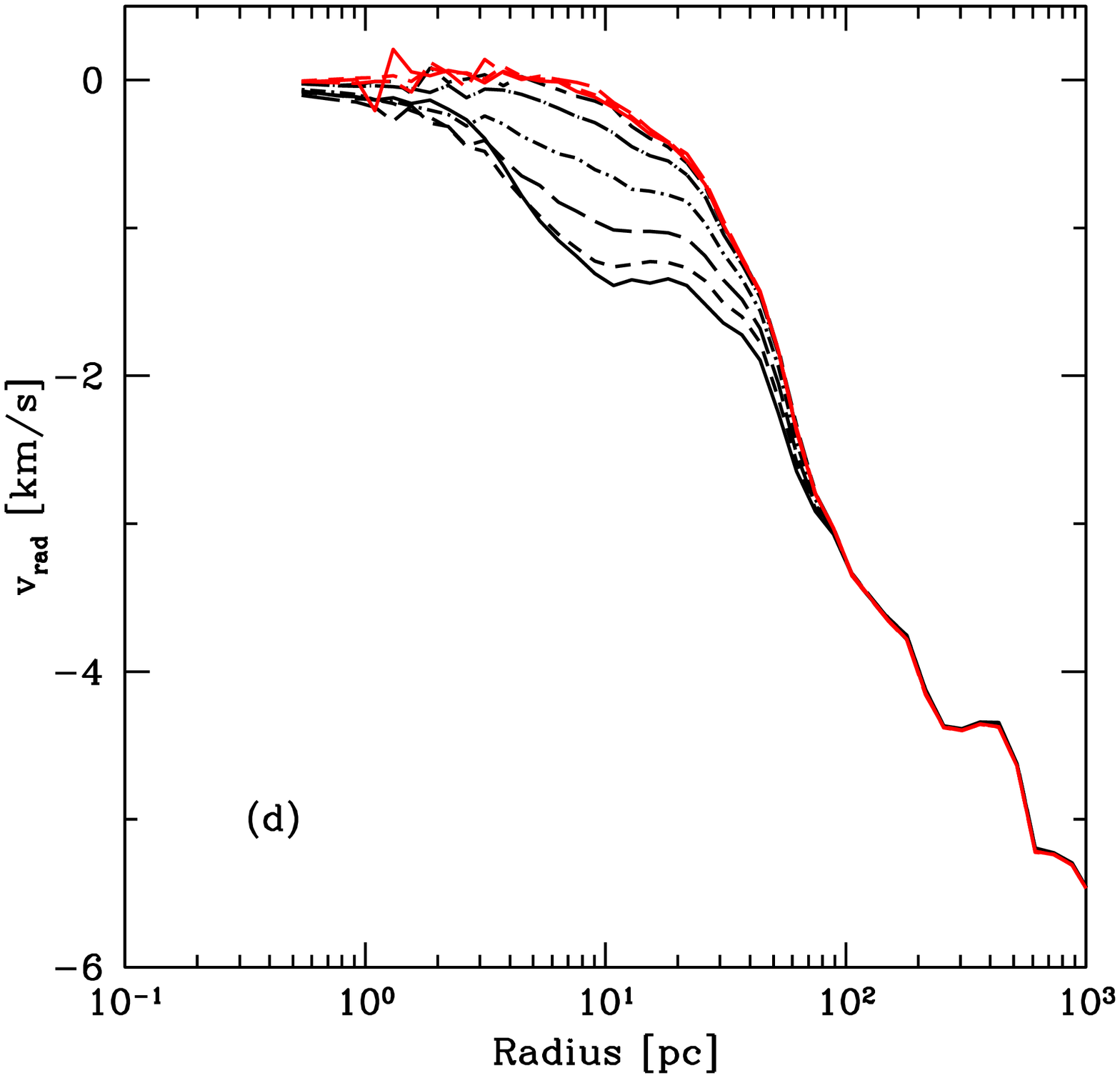}
\end{center}
\caption{
Several spherically-averaged baryon quantities for simulations
with all values of J$_{21}$.
Panel (a):  Number density as a function of radius.
Panel (b):  Temperature as a function of radius.
Panel (c):  $H_2$ fraction as a function of radius.
Panel (d):  Radial velocity as a function of radius.
All quantities except enclosed gas mass are mass-weighted, and
all simulations are shown at $z=25$ (shortly before the J$_{21} = 0$ 
simulation collapses to high density.
Line types and weights correspond to simulations, as follows.
Black solid line: J$_{21} = 0.0$.
Black short-dashed line: J$_{21} = 10^{-3}$.
Black long-dashed line: J$_{21} = 10^{-2.5}$.
Black dot short-dashed line: J$_{21} = 10^{-2}$.
Black dot long-dashed line: J$_{21} = 10^{-1.5}$.
Black short dashed-long dashed line: J$_{21} = 10^{-1}$.
Red solid line: J$_{21} = 10^{-0.67}$.
Red short-dashed line: J$_{21} = 10^{-0.33}$.
Red long-dashed line: J$_{21} = 1$.
}
\label{fig.radprof_fixred.1}
\end{figure}

Figure~\ref{fig.radprof_fixred.1} shows the status of several
spherically-averaged quantities for halos from all simulations at a fixed
redshift, $z=25$, which is shortly before the baryons in the most massive halo
in the J$_{21} = 0$ simulation collapses to high
density.  This is at an early point in the evolution of the majority of these halos, and 
several conclusions can be drawn from the plots of number density, temperature, H$_2$ fraction,
and radial velocity as a function of radius.  The molecular hydrogen fraction declines monotonically
with increasing UV background strength, in agreement with an estimate for equilibrium H$_2$ fraction at low
($n \ll 10^8$~cm$^{-3}$) densities:

\begin{equation}
f_{H2} = \frac{k_{H^-}n_e}{k_{LW}}
\label{eqn-fh2}
\end{equation}

where k$_{H^-}$ is the rate coefficient for the formation of H$^-$ (the limiting reaction 
for H$_2$ formation at the temperature and density range considered here) and k$_{LW}$ is 
the rate coefficient for H$_2$ in the LW band~\markcite{abel97,2003ApJ...592..645Y}({Abel} {et~al.} 1997; {Yoshida} {et~al.} 2003). 
 Halo cores at higher central densities are departing from the approximations used in the above
 estimate and thus have values of the H$_2$ fraction that are not quite the equilibrium 
values.  Halo cores which have reached core H$_2$ fractions higher than $\sim 10^{-4}$ 
have been able to cool to significantly below the virial temperature (since the cooling 
time is much less than the Hubble time at that value of the H$_2$ fraction), allowing 
the baryon density to increase.  The effect of cooling can also be seen in the plot of 
radial velocity as a function of radius, where the core regions of halos that have cooled are 
contracting (albeit very slowly), while those which are still at the virial temperature 
are not showing significant signs of contraction within $\sim 20$ pc.  All halos have 
some inflow at large radii, due to the infall of gas from adjoining filaments.

\clearpage
%%%%%%%%%%%%%%%%%%% FIGURE %%%%%%%%%%%%%%
%%%%%%%%%%%%%%%%%%%%%%%%%%%%%%%%%%%%%%%%%
\clearpage

\section{Evolution of two representative halos}\label{results.rephalos}

In this section we compare the evolution of two representative simulations.
We choose the calculations which correspond to J$_{21} = 0$ and $1$.  
These two simulations represent
the extremes in halo evolution -- the J$_{21} = 0$ calculation is an
example of the evolution of a ``standard'' halo, as discussed in previous 
literature, and the J$_{21} = 1$ simulation is the most extreme example
of a Population III protostar forming in a T$_{vir} \sim 10^4$~K halo
available from our suite of calculations.  As described in Section~\ref{sect.simsetup},
both calculations start from the same initial conditions, with only the strength of
the LW background being different.

Projections of the log baryon density and temperature for the J$_{21} = 0$ and $1$
simulations are shown in Figures~\ref{fig.image_jlw_0} and~\ref{fig.image_jlw_1em21}, 
respectively.  These projections are at the epoch of collapse, which is $z=24.1$~$(17.3)$
for the J$_{21} = 0$~$(1)$ simulation.  The halo masses are significantly
different: $5.68 \times 10^5$ and $1.26 \times 10^7$~M$_\odot$ for the two calculations,
reflecting the rapid pace of mergers at that redshift, and the amount of structure evolution
that takes place over that relatively short time period ($\sim 8.2 \times 10^7$ years).  This
can clearly be seen in the left column of both figures, when the large
amount of structure apparent in Figure~\ref{fig.image_jlw_0} has merged into the main 
halo by the time of protostellar cloud formation in the J$_{21} = 1$ case, as seen
in Figure~\ref{fig.image_jlw_1em21}.

Despite the differences in evolutionary states, the halos themselves are quite similar in 
Figures~\ref{fig.image_jlw_0} and~\ref{fig.image_jlw_1em21}.  Both calculations show halos
that are strongly centrally concentrated, as shown by the center column in both figures.  
This column shows projections of the halos which are centered on the baryon peak and are 
approximately one virial radius across (and thus are scaled differently in both images).
No fragmentation of the halo core is visible.  The rightmost panel in both calculations shows
the center of the halo, where the protostellar cloud is evolving.  There is still no evidence for
fragmentation up to a central baryon density of $n \sim 10^{10}$~cm$^{-3}$.

\begin{figure}
\begin{center}
\includegraphics[width=0.3\textwidth]{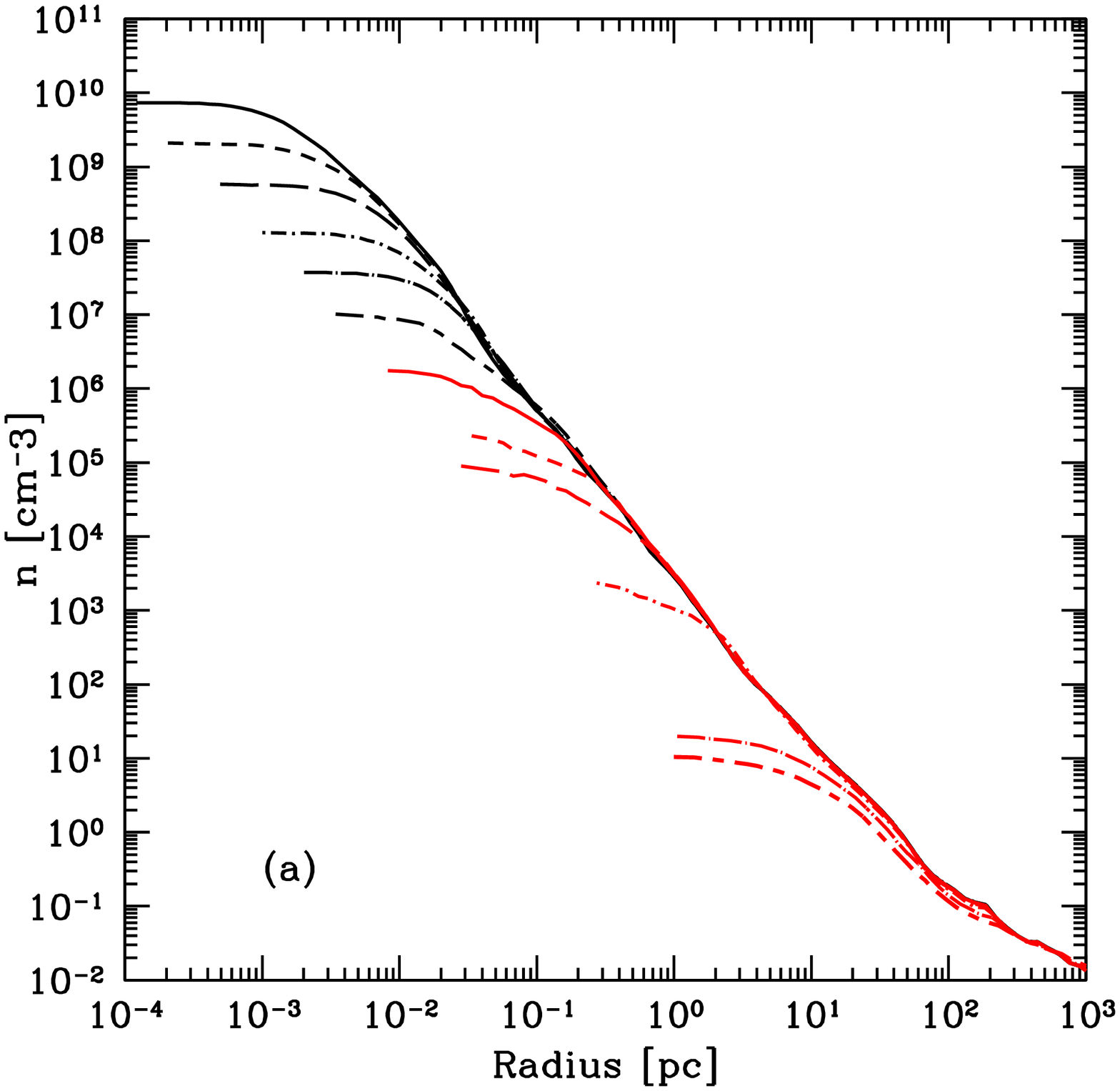}
\includegraphics[width=0.3\textwidth]{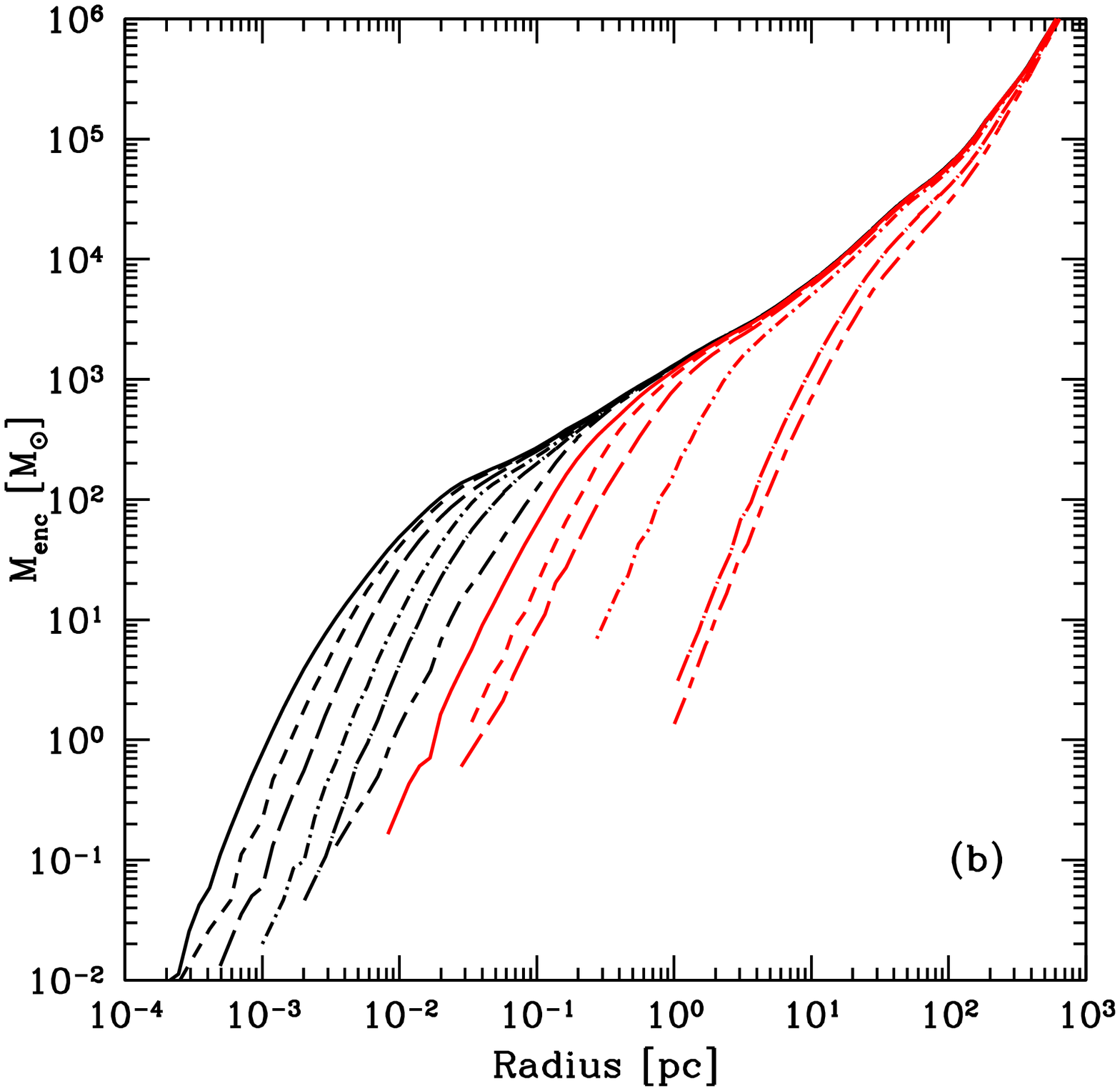}
\includegraphics[width=0.3\textwidth]{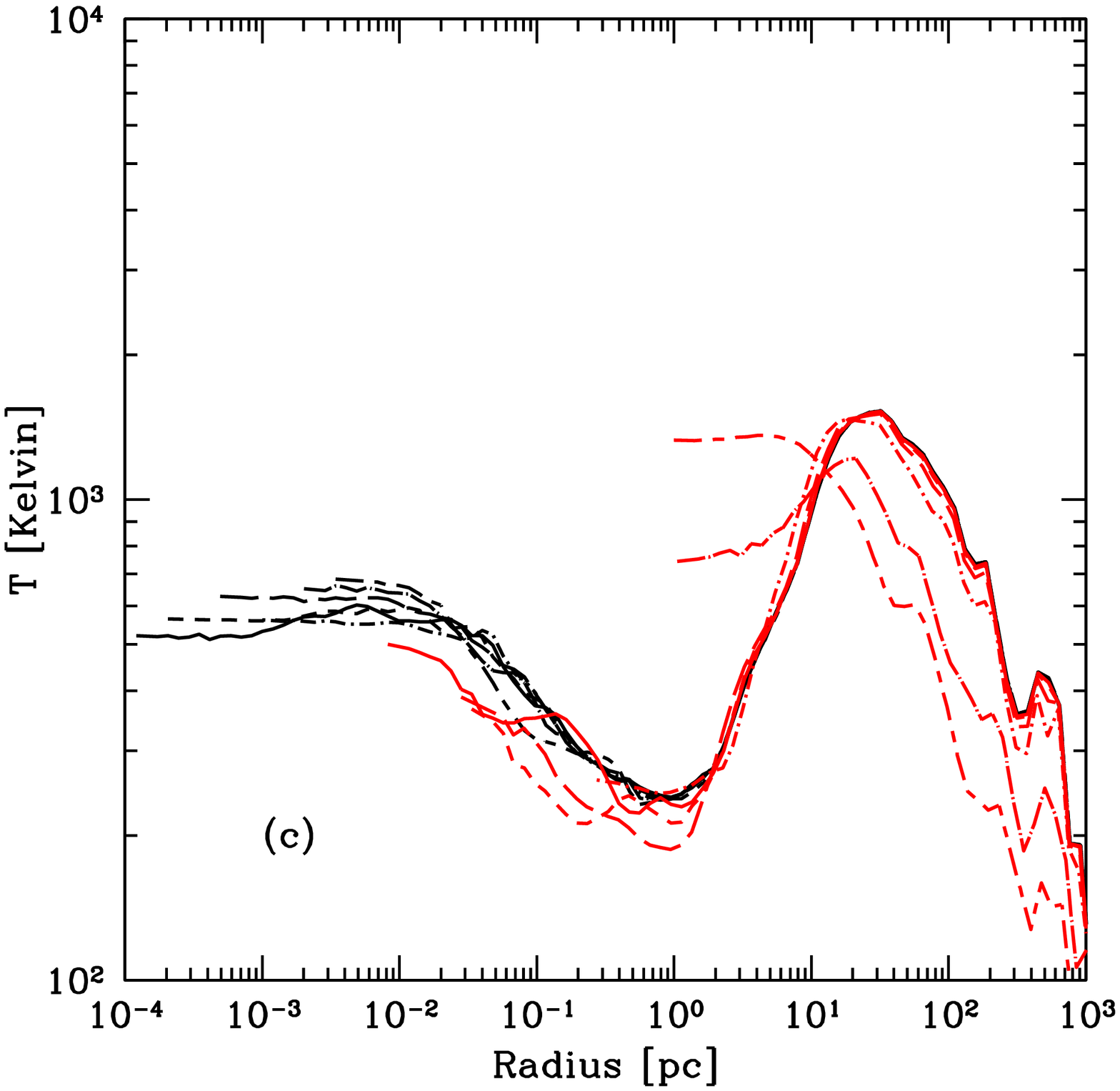}
\includegraphics[width=0.3\textwidth]{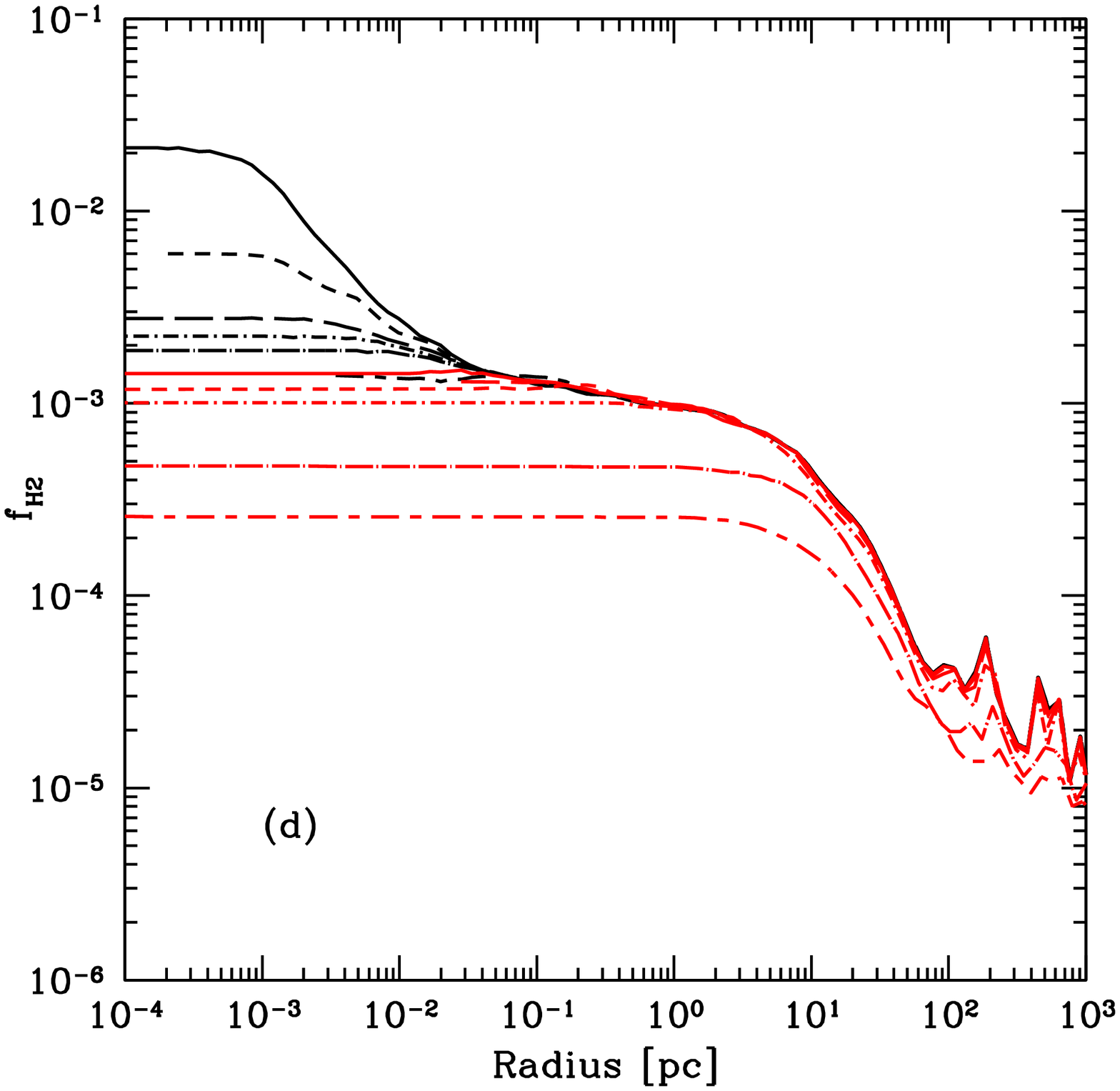}
\includegraphics[width=0.3\textwidth]{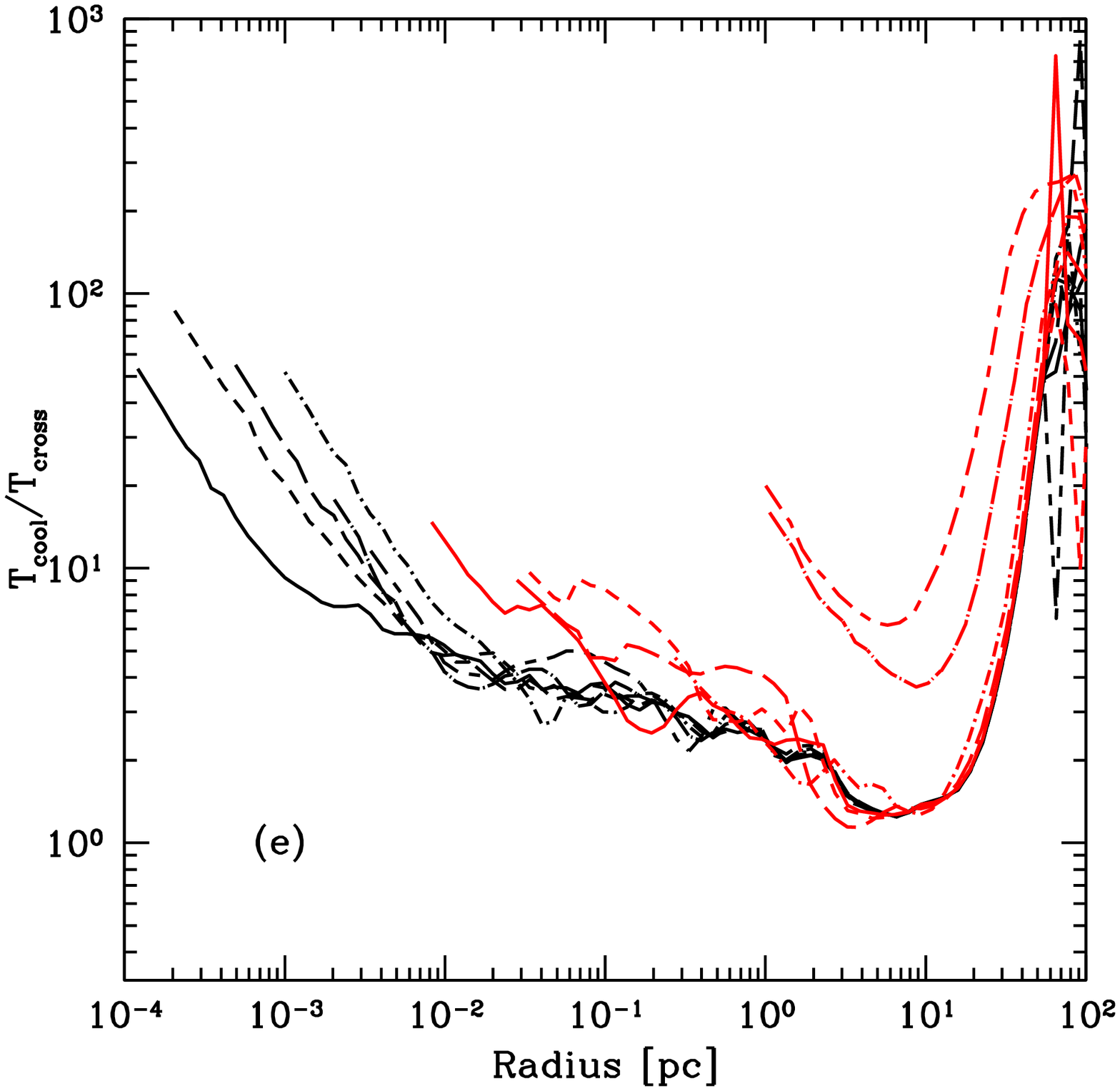}
\includegraphics[width=0.3\textwidth]{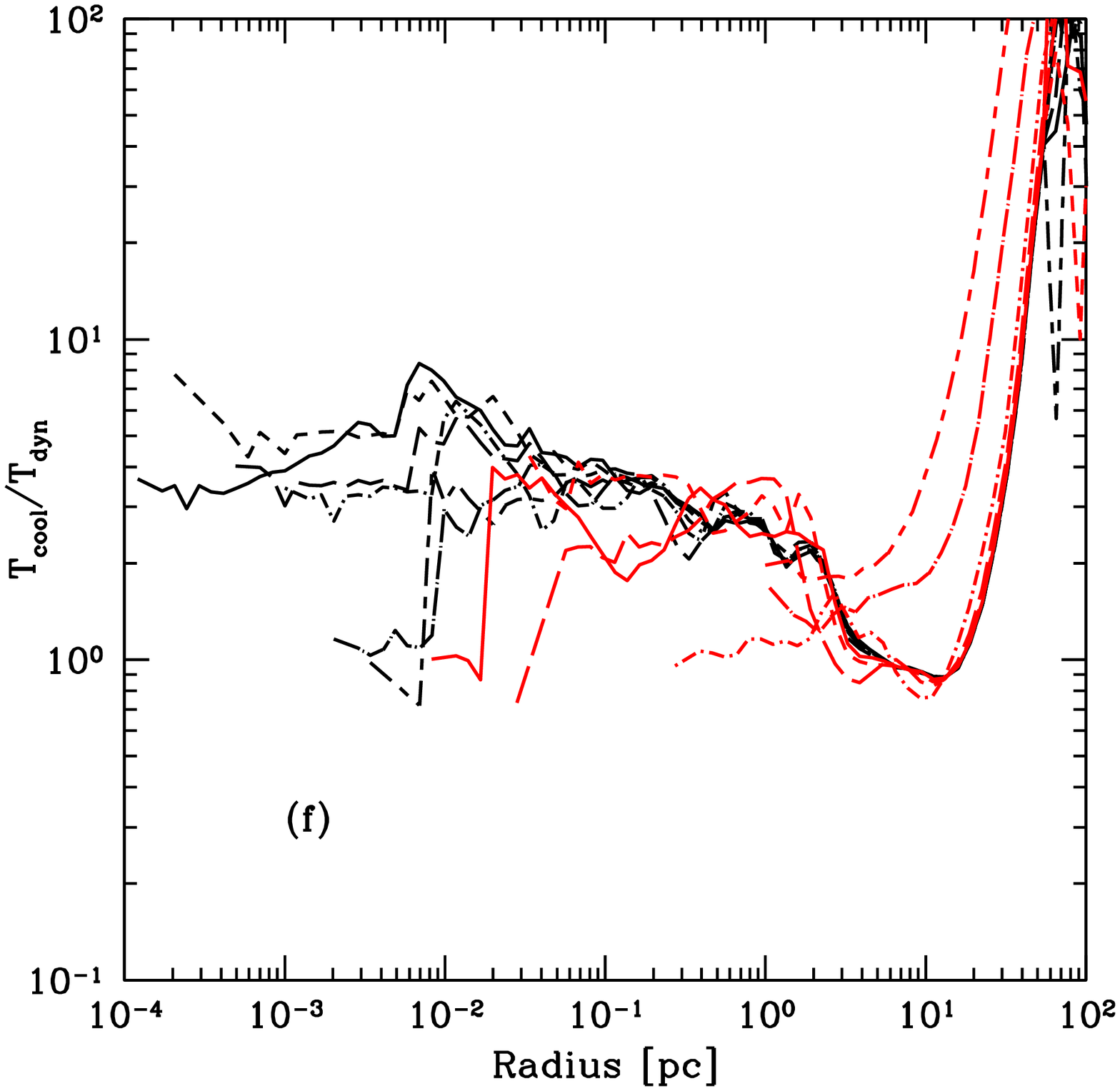}
\end{center}
\caption{
Evolution of several spherically-averaged baryon quantities as a function
of time for the J$_{21} = 0$ simulation.
Panel (a):  Number density as a function of radius.
Panel (b):  Enclosed gas mass as a function of radius.
Panel (c):  Temperature as a function of radius.
Panel (d):  $H_2$ fraction as a function of radius.
Panel (e):  Ratio of cooling time to sound crossing time as a function of radius.
Panel (f):  Ratio of cooling time to dynamical time as a function of radius.
All quantities except enclosed gas mass are mass-weighted.
Line types and weights correspond to different times, as follows.
Red short dashed-long dashed line:  $z=27.678$, $t=1.1088 \times 10^8$ years.
Red dot long-dashed line:  $z=26.010$, $\Delta t = 1.0428 \times 10^7$ years. 
Red dot short-dashed line:  $z = 24.565$, $\Delta t = 1.0428 \times 10^7$ years.
Red long-dashed line:  $z = 24.273$, $\Delta t = 2.291 \times 10^6$ years.
Red short-dashed line:  $z = 24.171$, $\Delta t = 8.0959 \times 10^5$ years.
Red solid line:  $z =  24.148101$, $\Delta t = 1.9034 \times 10^5$ years.
Black short dashed-long dashed line:  $z = 24.132556$, $\Delta t = 1.2528 \times 10^5$ years.
Black dot long-dashed line:  $z = 24.126839$, $\Delta t = 4.6131 \times 10^4$ years.
Black dot short-dashed line:  $z = 24.124178$, $\Delta t =  2.1483 \times 10^4$ years.
Black long-dashed line:  $z = 24.121760$, $\Delta t = 1.9517 \times 10^4$ years.
Black short-dashed line:  $z = 24.120224$, $\Delta t = 1.2404 \times 10^4$ years.
Black solid line:  $z = 24.119543$, $\Delta t = 5.5024 \times 10^3$ years.
}
\label{fig.radprof_jlw0.1}
\end{figure}

\begin{figure}
\begin{center}
\includegraphics[width=0.3\textwidth]{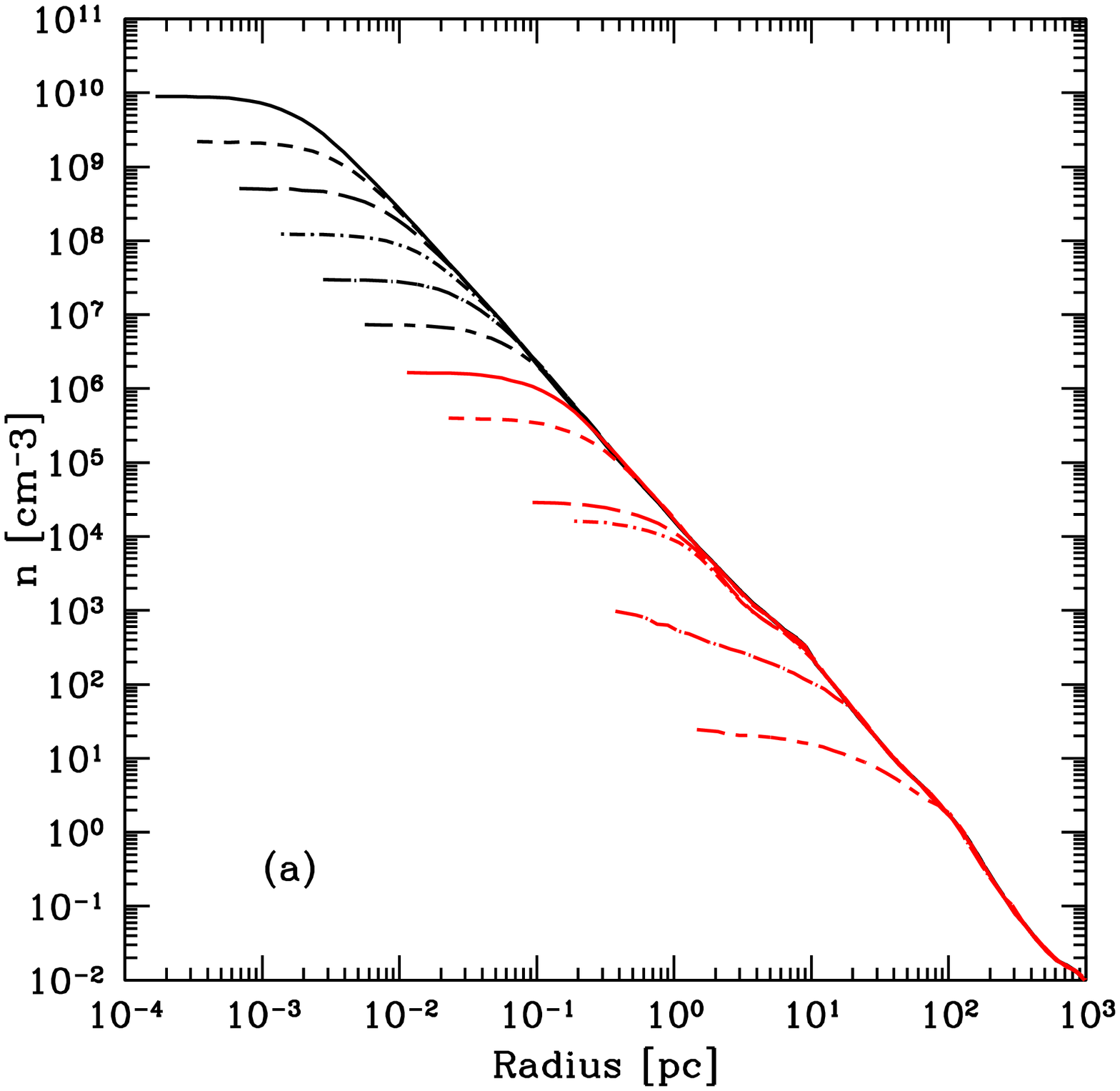}
\includegraphics[width=0.3\textwidth]{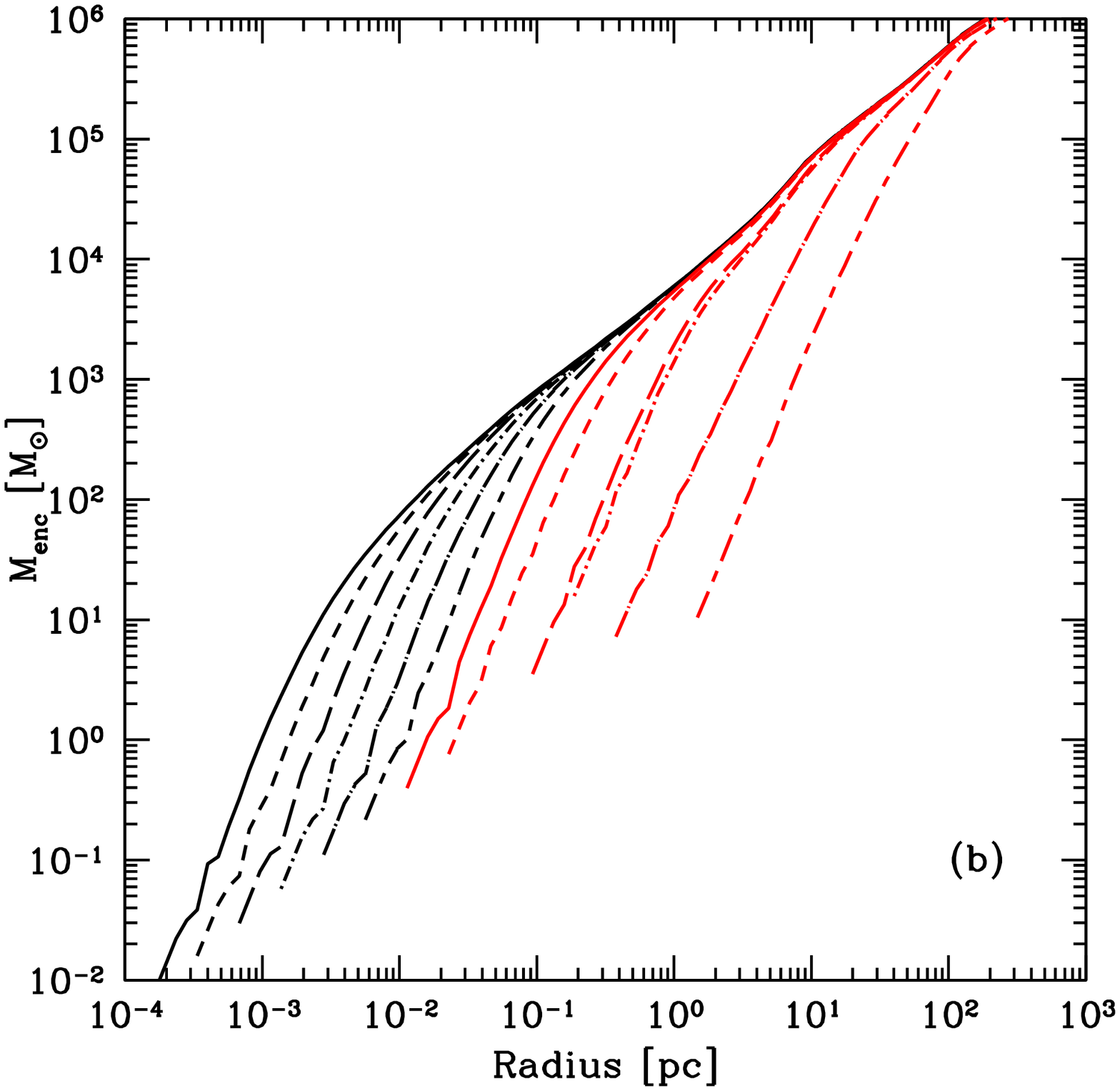}
\includegraphics[width=0.3\textwidth]{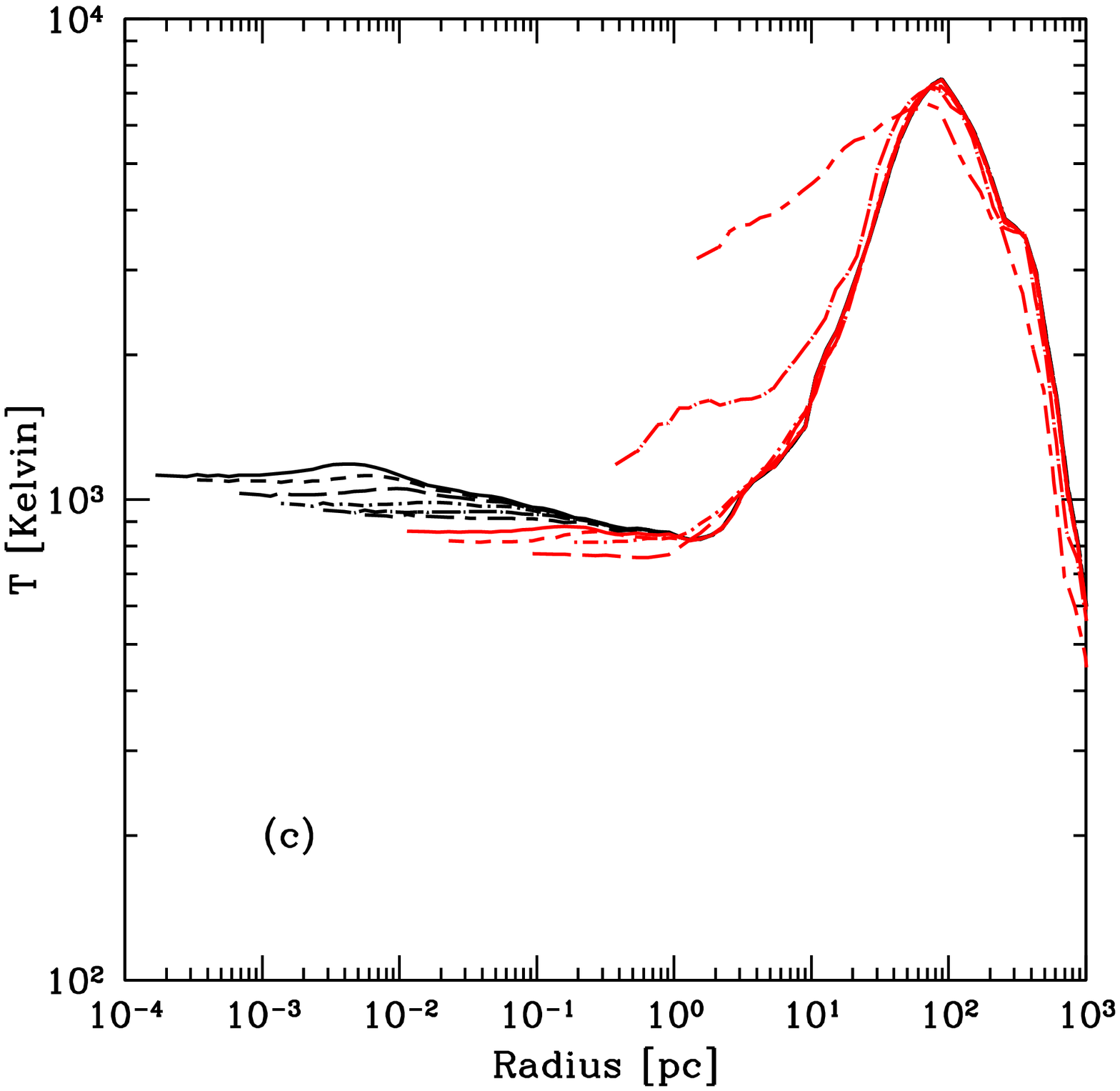}
\includegraphics[width=0.3\textwidth]{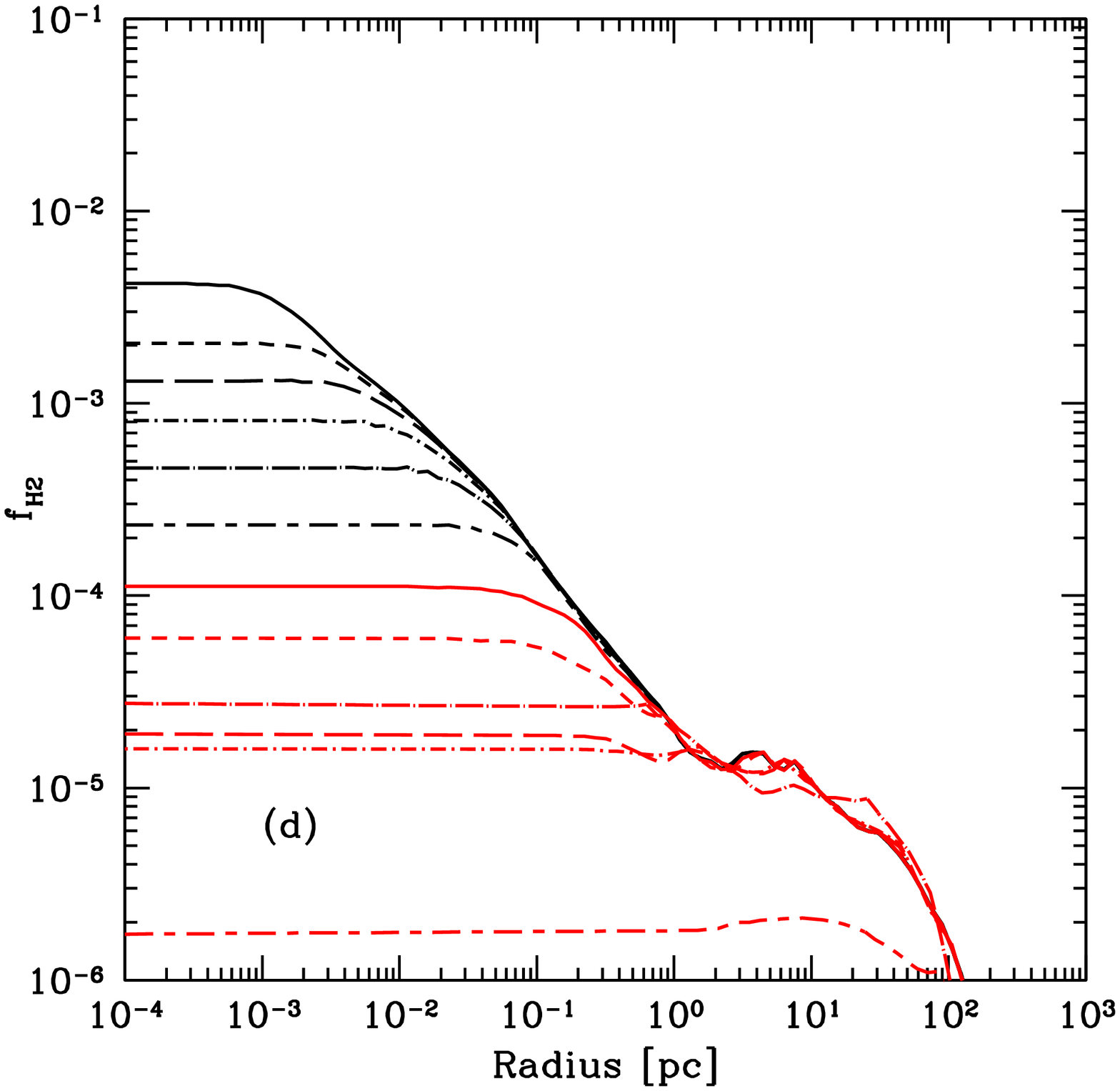}
\includegraphics[width=0.3\textwidth]{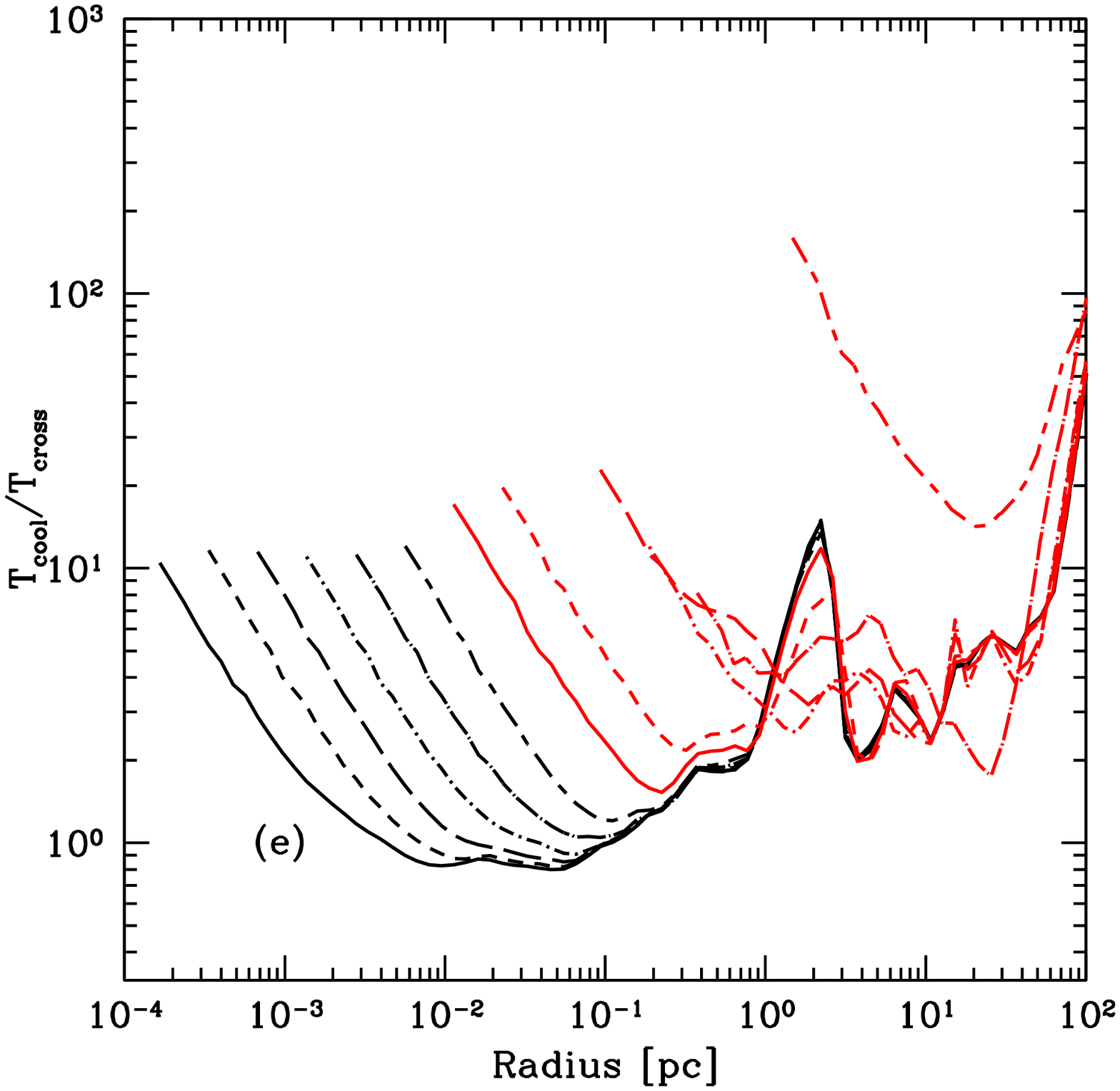}
\includegraphics[width=0.3\textwidth]{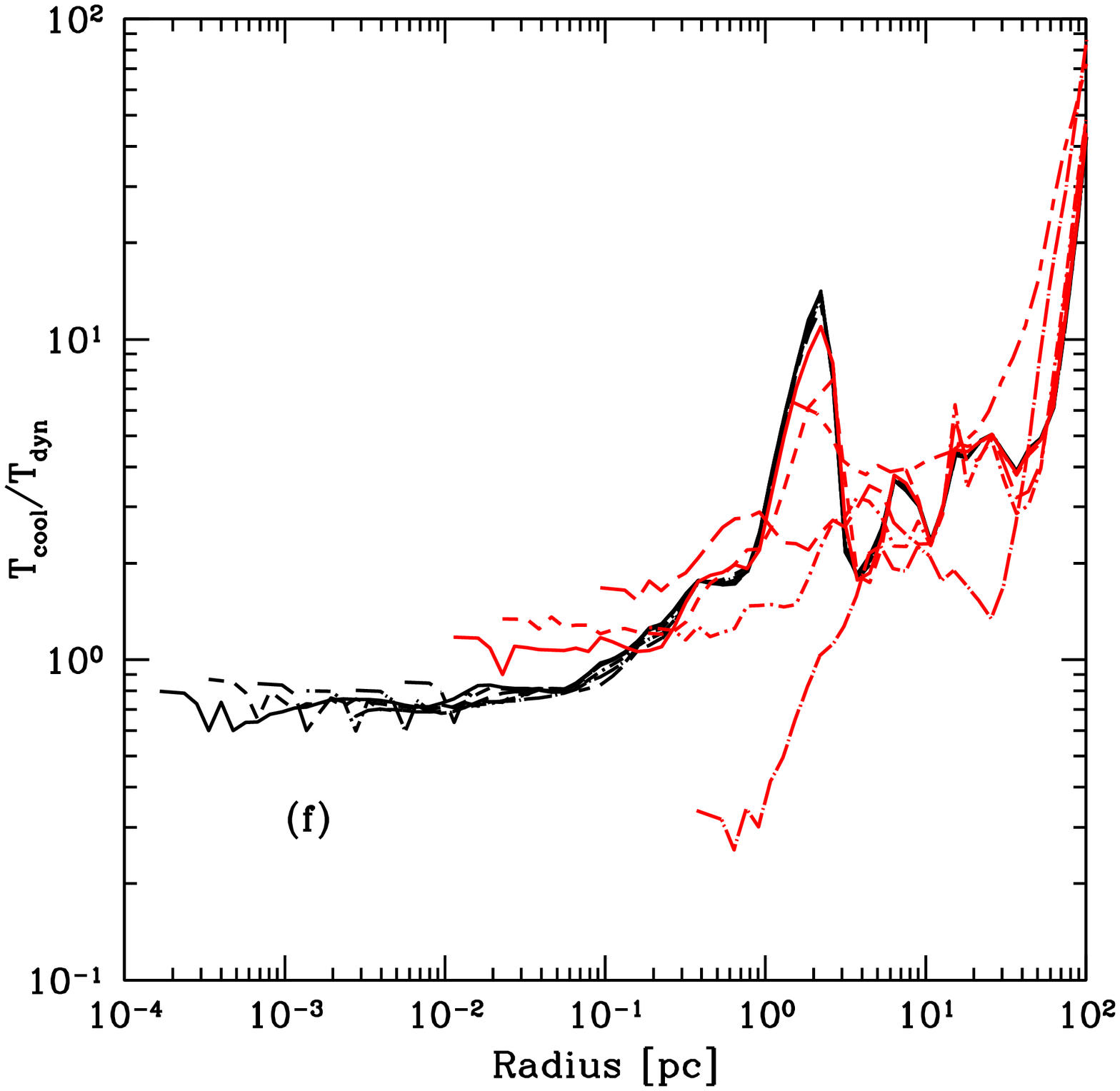}
\end{center}
\caption{
Evolution of several spherically-averaged baryon quantities as a function
of time for the J$_{21} = 1$ simulation.
Panel (a):  Number density as a function of radius.
Panel (b):  Enclosed gas mass as a function of radius.
Panel (c):  Temperature as a function of radius.
Panel (d):  $H_2$ fraction as a function of radius.
Panel (e):  Ratio of cooling time to sound crossing time as a function of radius.
Panel (f):  Ratio of cooling time to dynamical time as a function of radius.
All quantities except enclosed gas mass are mass-weighted.
Line types and weights correspond to different times, as follows.
Red short dashed-long dashed line:  $z =  18.384605$, $t = 1.995 \times 10^8$ years.  
Red dot long-dashed line:  $z = 17.737221 $, $\Delta t = 5.2141 \times 10^6$ years.  
Red dot short-dashed line:  $z = 17.394373$, $\Delta t =  1.7253 \times 10^6$ years.  
Red long-dashed line:  $z = 17.373929$, $\Delta t = 3.6030 \times 10^5$ years.  
Red short-dashed line:  $z = 17.330286$, $\Delta t = 7.7255 \times 10^5$ years.  
Red solid line:  $z = 17.325077$, $\Delta t = 9.25088 \times 10^4$ years.  
Black short dashed-long dashed line:  $z = 17.323130$, $\Delta t = 3.4610 \times 10^4$ years.  
Black dot long-dashed line:  $z = 17.322398$, $\Delta t = 1.3001 \times 10^4$ years.  
Black dot short-dashed line:  $z = 17.322050$, $\Delta t = 6.1828 \times 10^3$ years.  
Black long-dashed line:  $z = 17.321880$, $\Delta t = 3.0262 \times 10^3$ years.  
Black short-dashed line:  $z = 17.321796 $, $\Delta t = 1.4813 \times 10^3$ years.  
Black solid line:  $z = 17.321757 $, $\Delta t = 7.0496 \times 10^2$ years.  
}
\label{fig.radprof_jlw1em21.1}
\end{figure}

Figures~\ref{fig.radprof_jlw0.1} and~\ref{fig.radprof_jlw1em21.1} show the temporal 
evolution of several spherically-averaged baryon quantities for the 
J$_{21} = 0$ and $1$ calculations, 
respectively.  These plots show baryon number density, enclosed baryon mass, 
temperature, H$_2$ fraction, ratio of cooling time to sound crossing time, and
ratio of cooling time to dynamical time, as a function of radius, evolving from a
number density of $\simeq 10$~cm$^{-3}$ to $\simeq 10^{10}$~cm$^{-3}$ in both calculations.
Some qualitative commonalities are obvious between the two calculations.  Both halos
steadily grow in central density (Panel (a)), which is coupled to the growing molecular
hydrogen fraction in the central region of the halo (Panel (d)).  Both halos collapse quasi-statically, as 
shown in panels (e) and (f).  Also, the temperature profiles (as shown in Panel (c)) are 
broadly similar, in that the gas in the halo central region cools to some minimum value at r $\sim 1-2$ pc
(corresponding to n $\sim 10^4$~cm$^{-3}$ in both simulations) and then creeps steadily upward
as gas collapses to higher densities.

Though there are qualitative similarities between the evolution of the two calculations, detailed 
examination shows some significant quantitative differences.  The time that the halo in the J$_{21} = 0$
simulation takes to evolve from a central density of $n \simeq 10$~cm$^{-3}$ to 
$10^{10}$~cm$^{-3}$ is $\simeq 24$~Myr, with 21 Myr of that being the time required to 
evolve from  $n \simeq 10$ to n $\simeq 10^4$ cm$^{-3}$.  The time required for 
the J$_{21} = 1$ calculation 
to evolve from $n \simeq 10$~cm$^{-3}$ to $10^{10}$~cm$^{-3}$ is slightly shorter,
$\simeq$ 17 Myrs, with 8 Myrs required to evolve to n $\simeq 10^{4}$~cm$^{-3}$.
The temperature evolution is also somewhat different -- the J$_{21} = 0$ simulation
has a minimum temperature at r $\sim 1$ pc of 200 K, with n $\simeq 10^4$$^{-3}$
and a molecular hydrogen fraction of $\simeq 10^{-3}$, and roughly 1000 M$_\odot$
of enclosed gas.  As the gas within this radius evolves to higher densities, the 
temperature increases by a factor of a few, peaking at T $\sim 500$ K at 
n~$\sim 10^{10}$~cm$^{-3}$, with f$_{H_2} \simeq 0.02$.  The gas cooling time 
within r $\sim 1$ pc in the J$_{21} = 0$ simulation is always larger than the sound 
crossing and dynamical times by at least a factor of two, implying a quasistatic
gas contraction at all times.

The J$_{21} = 1$ simulation evolves somewhat differently.  The collapsing
gas reaches a minimum temperature of $\sim 800$~K at r $\sim 2$ pc, with a baryon
density of $\sim 2 \times 10^3$~cm$^{-3}$, an enclosed gas mass
of roughly $10^4$ M$_\odot$, and a molecular hydrogen fraction of $\sim 10^{-5}$,
two orders of magnitude below that in the J$_{21} = 0$ calculation at the
equivalent temperature minimum.  The low H$_2$ fraction is due to the high
LW background radiation flux.  As the halo evolves to higher densities, the gas
temperature and H$_2$ fraction also creep upwards.  The temperature of the gas when
the peak reaches roughly $10^{10}$~cm$^{-3}$ is $\sim 1200$ K, with a H$_2$ fraction of 
$\sim 4 \times 10^{-3}$.  The cooling time at r $\sim 2$ pc is an order of magnitude 
higher than the sound crossing or dynamical times, implying an extremely slow contraction
of the gas at that radius.  As the central density increases, however, the cooling 
time becomes lower, with T$_{cool}$/T$_{dyn} \sim 0.7$ and T$_{cool}$/T$_{cross}$
dipping to $\sim 0.8$ at its minimum, but the latter increasing toward the 
central density peak to a ratio of $\sim 10$.  This implies a more rapid contraction
of gas in the halo center in the simulation with a higher UV background, despite the lack of
molecular hydrogen.  It is also worth noticing that the J$_{21} = 1$ calculation
has H$_2$ fractions at n $\sim 10^4$ and $10^8$~cm$^{-3}$ of $\simeq 2 \times 10^{-5}$
and $8 \times 10^{-4}$, respectively, as compared to f$_{H_2} \simeq 10^{-3}$ and
$2 \times 10^{-3}$ at the same densities in the J$_{21} = 0$ calculation.

\clearpage
%%%%%%%%%%%%%%%%%%% FIGURE %%%%%%%%%%%%%%
%%%%%%%%%%%%%%%%%%%%%%%%%%%%%%%%%%%%%%%%%

%%%%%%%%%%%%%%%%%%% FIGURE %%%%%%%%%%%%%%
%%%%%%%%%%%%%%%%%%%%%%%%%%%%%%%%%%%%%%%%%
\clearpage

%%%%%%%%%%%%%%%%%%%%%%%%%%%%%%%%%%%%%%%%%%%%%%%%%%%%%%%%%%%%%%%%%%%%%%%%%%%%
\section{Neglected Physics and Possible Numerical issues}\label{sect.issues}

In this paper we have examined aspects of the formation of Population III 
stars in the presence
of a soft ultraviolet background, using an adaptive mesh refinement cosmological 
structure formation code. Given the nature of the tool and current limitations in 
computing power, some physics were neglected, and possible numerical issues may 
arise.  We discuss this here.

The primordial chemistry model used in these calculations ignores the effects of 
deuterium, lithium, and the various molecules that form between these elements 
and ordinary hydrogen.  Deuterium and lithium have been shown to be unimportant 
in the temperature and density regimes that we have examined in this paper 
\markcite{1998A&A...335..403G,2002P&SS...50.1197G,2005MNRAS.361..850L}({Galli} \& {Palla} 1998, 2002; {Lipovka}, {N{\'u}{\~n}ez-L{\'o}pez}, \&  {Avila-Reese} 2005).  
However, it is possible that they may be relevant 
in other situations of importance to Population III star formation -- in particular, 
regions which have been ionized to very high electron fractions may experience 
significant cooling from the HD molecule, which due to its permanent dipole moment
makes it more than 100 times more effective as a cooling agent than molecular 
hydrogen (per molecule), and has the potential to cool gas down to approximately 
the temperature of the cosmic microwave background, which scales with redshift 
as $T_{cmb}(z) = 2.73 \times (1+z)$~K
~\markcite{2000MNRAS.314..753F,2002P&SS...50.1197G,2005MNRAS.361..850L}({Flower} {et~al.} 2000; {Galli} \& {Palla} 2002; {Lipovka} {et~al.} 2005).
This gives a minimum baryon temperature of approximately $55$ Kelvin at $z=20$ and 
could reduce the minimum accretion rate onto a primordial protostar significantly.
Lithium, while in principle an effective coolant as LiH, is safely ignored, since only
a tiny fraction of lithium is converted into LiH~\markcite{2005PASJ...57..951M}({Mizusawa}, {Omukai}, \&  {Nishi} 2005).
A final chemical process that is omitted from our calculation is heating 
caused by molecular hydrogen formation at high densities (n $\ga 10^8$~cm$^{-3}$),
the inclusion of which may result in differences in the temperature evolution
of the gas in the highest density regimes considered here.  This does not 
significantly affect the conclusions of our paper, as the significant differences
we observe between simulations occur at densities much lower than 
$\sim 10^8$~cm$^{-3}$.

Self-shielding of the photodissociating background by molecular hydrogen in the high-density
gas is ignored in these calculations.  Though this effect could 
in principle be important, the actual column
densities of molecular hydrogen are typically far too small to actually block the soft UV flux.  
According to
\markcite{2001ApJ...548..509M}{Machacek} {et~al.} (2001) (and references therein), a column density of 
$5 \times 10^{14}$~cm$^{-2}$ is enough for 
shielding to become important.  However,
this was derived for a static distribution of H$_2$, while the LW band consists of 
hundreds of individual lines
whose width in this case is dominated by Doppler broadening. It is useful to note in 
this case that the average line 
width is $\sim 2$~km/s and the RMS baryon velocity in our calculations are $\sim 4$~km/s. 
In order for self-shielding to be important in the case of a turbulent medium, the column 
density must be much higher.
Typical maximum H$_2$ column densities in our calculations are on the order of 
$10^{16}$~cm$^{-2}$, but these
occur late in the collapse of the halo core, and in the highest density regions the cooling and
H$_2$ production times are much shorter than the photodissociation time scale, at which point 
self-shielding becomes unimportant.  It is worth noting that there are some regimes where
self-shileding can be critical.  \markcite{2007ApJ...659..908S}{Susa} (2007) finds that
self-shielding can strongly affect the evolution of a collapsing primordial halo which is being
illuminated with photodissociating flux by a neighboring star, though the situation is somewhat 
idealized.
Additionally, we do not consider the more complicated effects relating to Population III
stars which form in halos in the cosmic neighborhoods where previous generations of stars have
existed.  This allows us to ignore complex radiative, chemical and dynamical effects that would
vastly complicate our 
calculations~\markcite{2005ApJ...628L...5O,2006astro.ph.12254J,2006astro.ph.10819Y,2007arXiv0705.3048G,
2007ApJ...659L..87A,2007MNRAS.375..881A}({O'Shea} {et~al.} 2005a; {Johnson}, {Greif}, \&  {Bromm} 2006; {Yoshida} {et~al.} 2006a; {Greif} {et~al.} 2007; {Abel}, {Wise}, \&  {Bryan} 2007; {Ahn} \& {Shapiro} 2007).

A further effect that is ignored in this paper is $H^-$ photodetachment.  This could in principle
be a significant effect -- $H^-$ is the catalyst for molecular hydrogen formation in the 
dominant \h2 formation channel at densities $\la 10^8$~cm$^{-3}$, and it can be detached by
photons with energies $h \nu \ga 0.75$~eV. Photons with this energy would be produced in
great numbers by the same massive Population III stars that we assume are producing the 
molecular hydrogen photodissociating background.  However, as shown by \markcite{2001ApJ...548..509M}{Machacek} {et~al.} (2001),
the rate of photodetachment is orders of magnitude smaller than the rate of $H^-$ formation
at densities comparable to that found in the centers of the halos examined in this work,
and thus the process of $H^-$ photodetachment can be safely ignored.

In this paper, we examine the effects of molecular hydrogen dissociating backgrounds which range in 
strength from J$_{LW} = 10^{-24}$ to $10^{-21}$~ergs$^{-1}$~cm$^{-2}$~Hz$^{-1}$~sr$^{-1}$ in the LW
band ($11.18 - 13.6$ eV).  This range is in good agreement with the expected range of photodissociating
backgrounds predicted by~\markcite{2005ApJ...629..615W}{Wise} \& {Abel} (2005), but it is still finite.  Examination of our
simulations show that the case with the lowest UV background (J$_{21} = 10^{-3}$) is almost identical
to the ``control'' (J$_{21} = 0$) calculation, justifying our choice of minimum value.  Our upper value
is consistent with Wise \& Abel, and examination of Figures~\ref{fig.halovals1} and 
\ref{fig.radprof_zcoll.1}--\ref{fig.radprof_zcoll.3}
suggests that increases in the strength of the UV background (within reasonable values) will 
result in a further delay in halo
collapse, but no major change in the mode of star formation observed.  This is an effect of the halo
properties -- a small amount of molecular hydrogen will always exist in these halos, allowing
the gas at the center to cool and contract quasi-statically to higher densities.
Once the gas has collapsed to very high densities, rapid H$_2$ formation via the 3-body process will
occur, essentially independent of the strength of the soft UV background, and the
gas will be able to cool down to $\simeq 200$ K very quickly.

The effects of magnetic fields are completely ignored in the simulations discussed in this work.  
Magnetic fields are  
discussed in detail in Paper I, but, to summarize, a fairly high seed magnetic field is needed to 
be dynamically
significant at the relatively low densities we explore in this work.  The possible importance of magnetic
fields has been explored in analytic and semi-analytic work 
\markcite{TanMcKee2004,2006MNRAS.371..444S,2007arXiv0704.1853M}({Tan} \& {McKee} 2004; {Silk} \& {Langer} 2006; {Maki} \& {Susa} 2007).
We will examine the possible evolution of magnetic fields within the context of 
cosmological AMR simulations in a
later paper (O'Shea \& Turk 2007, in preparation).

The simulations presented in this paper are generated assuming a cosmology
that is somewhat different than the currently-favored WMAP Year III ``best-fit''
model~\markcite{2006astro.ph..3449S}({Spergel} {et~al.} 2006).  Most importantly, our value of $\sigma_8$ 
is 0.9, which is significantly higher than the WMAP value of 0.761.  The general
effect of a higher $\sigma_8$ is to cause structure formation to take place earlier,
and thus at a given redshift one would expect significantly more halos in our chosen
cosmological model than in the WMAP Year III model.  However, the evolution of any
single cosmological halo, such as the one examined in this work, is not particularly
affected by this parameter in the sense that we are not examining halo stastical
properties.  In addition to $\sigma_8$, the ratio of $\Omega_b/\Omega_m$ in our
simulations is 0.1337, while it is $0.1746$ in the WMAP Year III cosmology.
This may result in some small quantitative differences in, e.g., the redshift of 
halo collapse, but should not significantly affect our results.

We direct the reader to Paper I for a detailed discussion of other possible numerical issues, 
such as the generation of cosmological initial conditions, the assumption that the cold dark 
matter model is correct, the choice of halo in our simulations, and the size of the simulation 
volumes used.

\section{Discussion}\label{discuss}

This paper explores the formation of Population III stars in simulations with a constant 
soft UV background.  Our results agree well qualitatively with that of \markcite{2001ApJ...548..509M}{Machacek} {et~al.} (2001);
 we both find that a soft UV background can delay the formation of molecular hydogen, and thus the
cooling and collapse, of small ($\sim 10^6$~M$_\odot$) cosmological halos in which Population III 
stars form. We also find that increasing the soft UV background  increases the minimum halo mass 
required for a halo to collapse (in a way similar to that of smoothing the dark matter 
power spectrum at small scales -- see \markcite{oshea_wdm}{O'Shea} \& {Norman} (2006) for a discussion).  Machacek et al. derived
a mass threshold for collapse as a function of the LW background flux that agrees well with our 
simulations, though the halo masses in our calculation are significantly higher.  This is due to the halo that 
we examine being a ``typical'' halo rather than at the threshold mass for star formation.  Presumably, if 
we performed these calculations using many halos in a range of cosmological realizations, we would find
a minimum halo mass that is somewhat lower than the masses seen in our calculations.
Our work is a significant improvement upon that of Machacek et al. in some respects, as our 
simulations are much more 
highly resolved and we examine the evolution of a single halo over a much wider range of 
soft UV background fluxes.  One drawback of our work compared to Machacek et al. is that we examine the
evolution of a single halo, albeit with a broad range of UV backgrounds.

Our work is similar to the calculations in~\markcite{2003ApJ...592..645Y}{Yoshida} {et~al.} (2003) which examine
Population III star formation in the presence of a photodissociating ultraviolet background.  
They find 
that for values of J$_{LW}$ above $10^{-23}$~erg s$^{-1}$~cm$^{-2}$~Hz$^{-1}$~sr$^{-1}$, hydrogen
molecules are rapidly dissociated and gas cooling is inefficient, implying that
the collapse of gas in the halo center may be delayed until atomic line cooling dominates, in halos with
T$_{vir} \ga 10^4$~K.  While we see delays in halo collapse, we do see that a tiny
amount of \h2 can form even at high FUV background strengths, eventually
allowing gas to cool and collapse to high densities without ever reaching temperatures 
at which atomic line cooling would be effective (T $\ga 10^4$~K).
 Yoshida et al. make this statement based
on a set of simulations using two different values of J$_{LW}$ ($10^{-23}$ and $10^{-22}$), which
bracket the ``break'' in properties seen in our results.  The simulations used to obtain this result
 had SPH particles with m$_{gas} = 100.0$~h$^{-1}$~M$_\odot$ and thus an effective mass 
resolution comparable
to the mass of gas within the quasistatically-collapsing central region of a halo.  Our mass resolution is more than
two orders of magnitude higher in the center of the cosmological halos that we examine.
The differences between our result and theirs is likely due to the significant differences in
resolution and may in fact be consistent once resolution is taken into account.

We see a strong relationship between accretion rates onto the protostellar cloud
and the strength of the photodissociating background.  Calculations with higher 
ultraviolet background strengths typically have a larger spherically-averaged baryon accretion rate
onto the evolving protostellar cloud.  This is clearly due to variation in halo central
temperature relating to the amount of molecular hydrogen existing in the center of the halo at the
epoch of collapse.  Accretion onto the evolving protostellar cloud is subsonic, and
thus regulated by the local sound speed (which scales as T$^{0.5}$).
This implies some relationship 
between the strength of the photodissociating background and the final stellar mass of 
the primordial star -- however, the details of this relationship are unclear, and depends
on many factors.  For example,~\markcite{2003ApJ...589..677O}{Omukai} \& {Palla} (2003) suggest that an increase in accretion
rate above $\dot{m} \simeq 4 \times 10^{-3}$~M$_\odot$/yr 
will actually result in a \emph{decrease} in the final mass of the star due to radiative 
feedback from the evolving protostar.  However, their results
use one-dimenstional, fairly idealized models, and geometrical effects may be important.  This is
explored in~\markcite{TanMcKee2004}{Tan} \& {McKee} (2004), who use a combination of analytic and semi-analytic models of 
the evolving system.  They suggest that accretion onto the protostar is highly non-spherical, 
and is in fact mediated by an accretion disk.  They argue that a larger accretion rate onto the disk
will lead to a larger star, though with some limitations.  Finally, \markcite{2001ApJ...546..635O}{Omukai} (2001) 
and \markcite{2003ApJ...599..746O}{Omukai} \& {Yoshii} (2003) study the IMF of 
stars that form in T$_{vir} \ga 10^4$~K metal-free protogalaxies in the presence of an H$_2$ photodissociating
background.  They conclude that photodissociation actually decreases the Jeans mass of the gas at high
densities, and thus reduces the fragmentation mass scale of the clouds and presumably the stellar mass.
They use one-dimensional, spherically symmetric simulations, however.  We can explore this issue in 
more detail in later calculations, when appropriate models for shielding, cooling via $H^-$, and 
more advanced chemistry have been implemented into Enzo.  This will be examined in a later paper.
It is worth noting that the range of accretion rates observed in this situation are within
the range of rates seen in the $\Lambda$CDM simulations discussed in Paper I.  This implies that the 
mass ranges of the resulting stars will not be significantly different than in the 
``standard'' Population 
III star formation scenario.

We see no fragmentation of the high-density baryon core, up 
to a baryon number density of $n \simeq 10^{10}$~cm$^{-3}$, for the entire range of simulations
explored in this paper.  This is due to a combination of effects, but predominantly the poor
cooling properties of molecular hydrogen.  The gas is relatively hot ($\sim 1000$ Kelvin) 
and thus has a high sound speed, which helps to damp out perturbations in the halo center which would
otherwise result in multiple fragments.  This effect is exacerbated in simulations with
high ultraviolet background strengths, as the temperatures (and thus the sound speeds)
are generally higher.
  This result implies that there is no fundamental change in the mode of Population III 
star formation as halos grow, and that more massive ``proto-galactic'' halos, with T$_{vir}
\ga 10^4$~K, will continue to form a single massive star per halo.  These halos will have 
a much larger binding energy than the smaller, M $\sim 10^5-10^6$~M$_\odot$ halos which have 
traditionally been examined by numerical simulations, and implies that multiple generations of star formation
may be able to take place in a single halo.  This leads to the possibility of ``self-enrichment,'' where
a single Population III star enriches the high-density center of the halo to metallicities high
enough to change the cooling properties of the gas.  This could cause a strong change in the IMF 
\markcite{2006ApJ...643...26S,2007ApJ...661L...5S}({Santoro} \& {Shull} 2006; {Smith} \& {Sigurdsson} 2007), and these objects could make a significant contribution
to the reionization history of the universe~\markcite{2003ApJ...586....1M,2006MNRAS.373..128G}({Mackey}, {Bromm}, \&  {Hernquist} 2003; {Greif} \& {Bromm} 2006).  We will examine whether the high-density gas in the centers
of these more massive halos fragments in a later work.

The values of accretion rates onto the evolving protostellar cloud which are
observed with increasing photodissociating
background flux (which are always comparable to those observed in Paper I), in addition to the 
apparent lack of fragmentation of gas in the halo center up to
densities of n$_h \sim 10^{10}$~cm$^{-3}$, implies that there is no major
change in the mass range of Population III stars  as halo masses increase to T$_{vir} \sim 10^4$~K.  
This is in mild disagreement with previous semianalytic results by \markcite{2002ApJ...569..558O}{Oh} \& {Haiman} (2002) and 
\markcite{2003ApJ...586....1M}{Mackey} {et~al.} (2003).  This is apparently
because previous authors assumed that \h2 cooling in ``second generation'' (T$_{vir} \ga 10^4$~K) objects
is unimportant, while we observe that a small \h2 fraction, which forms in equilibrium with the UV
background, can cool gas effectively to temperatures of $\simeq 1000$~K.  We do
agree with these authors that the majority of primordial star formation will take place in objects with
T$_{vir} \sim 10^4$~K, the so-called ``Second generation'' objects~\markcite{2006MNRAS.368.1301M}({MacIntyre}, {Santoro}, \&  {Thomas} 2006)

Our work appears to contradict the results of \markcite{2006MNRAS.373..128G}{Greif} \& {Bromm} (2006), who use analytic and semianalytic
techniques to follow the collapse of gas within halos of T$_{vir} > 10^4$.  They find that, 
in their model, gas cools
to $\sim 8000$~K via atomic line cooling, and then contracts nearly isothermally to high densities,
allowing a molecular hydrogen fraction of f$_{H2} \sim 0.001$ to build up independent of local
density and temperature.  The gas in the center of the halo then cools and allows the gas to fragment
on scales of $\sim 20$~M$_\odot$.  We do not see this mode of star formation, possibly because
we never generate the significantly ionized halos with T$ > 10^4$~K upon which their scenario depends.
Rather, we simulate halos that in the cases of J$_{21} \geq 0.1$ are built up to T$_{vir} \sim 10^4$~K
via mergers, and the gas is never ionized to a significant degree.  One possible reason for
the observed differences is that we ignore the formation of deuterium hydride and its effects on the 
cooling properties of primordial gas.  It is possible that, in some contexts, the inclusion of HD may
result in enhanced fragmentation.  This will be examined in future studies.  The scenario described by Oh \& Haiman may still in fact be able
to occur, and deserves further detailed numerical study to determine the fate of the collapsing gas and
the implications of the (presumably) extremely high accretion rates onto the forming protostellar clouds.

Our results are similar to those shown by \markcite{2007arXiv0707.2059W}{Wise} \& {Abel} (2007),
who perform similar simulations examining the collapse of gas in cosmological halos in the presence 
of a photodissociating background.  They also find that, regardless of the strength of the UV background, 
the collapse of gas in the centers of the cosmological halos in question still occurs via
H$_2$ cooling.  This holds even for quite extreme examples, such as when the original electron
fraction is set to unphysically low levels (which should suppress H$_2$ formation, at 
least temporarily).

%%%%%%%%%%%%%%%%%%%%%%%%%%%%%%%%%%%%%%%%%%%%%%%%%%%%%%%%%%%%%%%%%%%%%%%%%%%%
\section{Summary}\label{summary}

In this paper we have performed a suite of high dynamical range (L$/\Delta x \sim 5 \times 10^8$) 3D adaptive
mesh cosmological simulations of the formation of Population III
stars in a $\Lambda$CDM universe in the presence of a molecular
hydrogen photodissocating (``Lyman-Werner'') ultraviolet
background.
The purpose of these calculations is to understand the effect that the soft
ultraviolet background has on the evolution of the gas in collapsing cosmological halos 
and to determine possible effects on the forming protostellar cloud.  These simulations
are all of a single cosmological realization, but with varied ultraviolet background strengths.
Our principal results are as follows:

1.  Our calculations show that, as the flux of ultraviolet radiation in the Lyman-Werner
band is increased, Population III star formation in a given cosmological halo 
is delayed to later times and, as a result, an increase in the virial mass of this
halo at the onset of baryon collapse.  This is in good
agreement with previous work by~\markcite{2001ApJ...548..509M}{Machacek} {et~al.} (2001) and~\markcite{2003ApJ...592..645Y}{Yoshida} {et~al.} (2003).

2.  We find that, contrary to previous work, the formation of primordial stars is never completely suppressed, 
regardless
of the strength of the UV background.  A small amount of molecular
hydrogen always exists in cosmological halos, and allows cooling and halo collapse
in gas which is bathed in a strong photodissociating background but has not
reached $10^4$~K.  The previously-suggested mode of star formation in  ``second generation'' halos, where
the collapse of gas to high density is completely suppressed until cooling can take place via atomic
hydrogen line transitions, is never observed in our calculations, which have
photodissociating background strengths up to J$_{21} = 1$.

3.  Though the molecular hydrogen fraction in the centers of halos which are bathed in strong 
(J$_{21} \geq 0.1$) Lyman-Werner radiation is very small at early times (f$_{H2} \sim 10^{-6}-10^{-5}$), the
gas at the center of the halo can still collapse due to the efficient cooling of molecular hydrogen at
$2-5 \times 10^3$~K, and to the extended time these halos require to collapse compared to
halos in the presence of much smaller UV backgrounds. 

4.  We observe that the estimated accretion rate onto the forming protostellar
cloud varies strongly as a function of J$_{LW}$, with simulations that have
a stronger ultraviolet background having higher accretion rates.  This is 
 a function of the suppression of molecular hydrogen formation
(and thus suppression of effective cooling) by the photodissociating background
and of the higher virial temperatures of these halos when the gas finally collapses.

5.  Only a single clump is formed at the center of each collapsing cosmological 
halo, regardless of the strength of the photodissociating background, up to
a baryon number density of $n \sim 10^{10}$~cm$^{-3}$.  This implies that, as in
the more commonly-studied mode of star formation, we will find only a single star
per halo even in objects which are massive enough that T$_{vir} \ga 10^4$~K.

%%%%%%%%%%%%%%%%%%%%%%%%%%%%%%%%%%%%%%%%%%%%%%%%%%%%%%%%%%%%%%%%%%%%%%%%%%%%
\acknowledgments{
B.W.O. would like to thank Tom Abel, Greg Bryan, Simon Glover, Thomas Greif,
Matthew Turk, and John Wise for 
useful discussions.  This work supported in part by NASA
grant NAG5-12140 and NSF grant AST-0307690. 
B.W.O. carried out this work under the auspices of the
National Nuclear Security Administration of the
U.S. Department of Energy at Los Alamos National
Laboratory under Contract No. DE-AC52-06NA25396, and was 
supported by a LANL Director's Postdoctoral Fellowship (DOE LDRD grant 
20051325PRD4).
The simulations were performed at SDSC and NCSA with computing time provided by 
NRAC allocation MCA98N020.  
}

%%%%%%%%%%%%%%%%%%%%%%%%%%%%%%%%%%%%%%%%%%%%%%%%%%%%%%%%%%%%%%%%%%%%%%%%%%%%
%\bibliographystyle{apj}
\bibliography{}  % looks in ms.bib for bibliography info

\end{document}